\documentclass[aps,prx,twocolumn,superscriptaddress]{revtex4-1}

\usepackage{bm}
\usepackage{braket}
\usepackage{amsmath}
\usepackage{mathtools}
\usepackage{dsfont}

\begin{document}

\preprint{}

\title{Efficient multiphoton sampling of molecular vibronic spectra on a superconducting bosonic processor}

\author{Christopher S. Wang}
\email{christopher.wang@yale.edu}
\author{Jacob C. Curtis}
\author{Brian J. Lester}
\author{Yaxing Zhang}
\author{Yvonne Y. Gao}
\affiliation{Departments of Physics and Applied Physics, Yale University, New Haven, CT 06511, USA.}
\affiliation{Yale Quantum Institute, Yale University, New Haven, CT 06520, USA.}
\author{Jessica Freeze}
\author{Victor S. Batista}
\author{Patrick H. Vaccaro}
\affiliation{Department of Chemistry, Yale University, New Haven, CT 06511, USA.}
\author{Isaac L. Chuang}
\affiliation{Department of Physics, Center for Ultracold Atoms, and Research Laboratory of Electronics, Massachusetts Institute of Technology, Cambridge, MA 02139, USA.}
\author{Luigi Frunzio}
\author{Liang Jiang}
\altaffiliation{current address: Pritzker School of Molecular Engineering, University of Chicago, Chicago, IL 60637, USA.}
\author{S. M. Girvin}
\author{Robert J. Schoelkopf}
\email{robert.schoelkopf@yale.edu}
\affiliation{Departments of Physics and Applied Physics, Yale University, New Haven, CT 06511, USA.}
\affiliation{Yale Quantum Institute, Yale University, New Haven, CT 06520, USA.}
\date{\today}

\begin{abstract}
The efficient simulation of quantum systems is a primary motivating factor for developing controllable quantum machines. For addressing systems with underlying bosonic structure, it is advantageous to utilize a naturally bosonic platform. Optical photons passing through linear networks may be configured to perform quantum simulation tasks, but the efficient preparation and detection of multiphoton quantum states of light in linear optical systems are challenging. Here, we experimentally implement a boson sampling protocol for simulating molecular vibronic spectra [Nature Photonics $\textbf{9}$, 615 (2015)] in a two-mode superconducting device. In addition to enacting the requisite set of Gaussian operations across both modes, we fulfill the scalability requirement by demonstrating, for the first time in any platform, a high-fidelity single-shot photon number resolving detection scheme capable of resolving up to 15 photons per mode. Furthermore, we exercise the capability of synthesizing non-Gaussian input states to simulate spectra of molecular ensembles in vibrational excited states. We show the re-programmability of our implementation by extracting the spectra of photoelectron processes in H$_2$O, O$_3$, NO$_2$, and SO$_2$. The capabilities highlighted in this work establish the superconducting architecture as a promising platform for bosonic simulations, and by combining them with tools such as Kerr interactions and engineered dissipation, enable the simulation of a wider class of bosonic systems.
\end{abstract}

\maketitle

\section{Introduction}

Simulation of quantum systems with quantum hardware offers a promising route towards understanding the complex properties of those systems that lie beyond the computational power of classical computers \cite{Feynman1982}. A particularly efficient approach is one that utilizes the natural properties of the quantum hardware to simulate physical systems that share those properties. This approach serves as the foundation for bosonic quantum simulation, where the natural statistics and interference between bosonic excitations are directly exploited. A prominent example of this is the manipulation of bosonic atoms in an optical lattice to explore various types of many-body physics \cite{Bakr2009, Tai2017, Bernien2017}.

Boson sampling is another example of a computationally challenging task that can be performed by manipulating bosonic excitations \cite{Aaronson2013, Hamilton2017}. Conventionally described in the context of linear optics, boson sampling, in its many forms, involves the single photon detection of nonclassical states of light passing through a linear interferometric network. Current technologies for optical waveguides allow the creation of complex interferometers across many modes. An outstanding challenge in the optical domain, however, is generating and detecting nonclassical states of light with high efficiencies.

In the microwave domain, nonlinearities provided by Josephson junctions enable the powerful preparation and flexible quantum nondemolition (QND) measurement of quantum states of light in superconducting circuits \cite{Schuster2007, Johnson2010}. The ability to performing QND photon number measurements not only enables the high fidelity measurement of bosonic qubits via repeated detections \cite{Hann2018, Elder2019}, but also enables the construction of more complex measurement operators. The circuit QED platform has also demonstrated universal control over individual bosonic modes as well as robust beamsplitter operations between separate modes \cite{Heeres2017, Gao2018}. These capabilities motivate an alternative approach to performing both boson sampling \cite{Peropadre2016} and bosonic quantum simulation protocols using superconducting circuits.

An instance of bosonic quantum simulation that maps onto a generalized boson sampling problem is obtaining molecular vibronic spectra associated with electronic transitions \cite{Huh2015}. The algorithm requires, in addition to beamsplitters, single-mode displacement and squeezing operations. Furthermore, the output will generally be a multimode multiphoton state, thus requiring a series of photon number resolving detectors at the output. Experimental imperfections in the controls, photon loss, and detector inefficiency has made a linear optical implementation of this algorithm challenging \cite{Clements2018}. Recent work leveraging the bosonic nature of two phonon modes of a single trapped ion has been successful in generating more accurate spectra \cite{Shen2018}; However, an efficient detection scheme capable of directly sampling from the exponentially growing bosonic Hilbert space remains elusive.

Here, we experimentally implement a two-mode superconducting bosonic processor with a full set of controls that enables the scalable simulation of molecular vibronic spectra. The processor combines arbitrary (Gaussian and non-Gaussian) state preparation and a universal set of Gaussian operations enabled by four-wave mixing of a Josephson potential. Most importantly, we implement a single-shot QND photon number resolving detection scheme capable of resolving up to $n_\textrm{max}$ = 15 photons per mode. This detection scheme, when operated without errors on a multiphoton distribution bounded within $n_\textrm{max}$ per mode, extracts the maximum possible amount of information from the underlying spectra per run of the experiment.

\section{Bosonic algorithm for Franck Condon factors}

The mapping of molecular vibronic spectra onto a bosonic simulation framework can be understood by considering the nature of vibrational dynamics that accompanies an electronic transition. In keeping with the tenets of the adiabatic Born-Oppenheimer approximation, the presumed separability between electronic and nuclear degrees of freedom results in distinct electronic states, each of which forms a potential-energy surface (PES) that supports a distinct manifold of vibrational eigenstates. The normal modes of vibrations for a given electronic state are obtained by expanding the PES in powers of displacement coordinates referenced to the minimum-energy (equilibrium) configuration and retaining only up to quadratic terms. Under this harmonic approximation, the corresponding transformation of the set of creation and annihilation operators $\hat{\textbf{a}} = (\hat{a}_1,...,\hat{a}_N)$ for $N$ vibrational modes, may be expressed using the Doktorov transformation \cite{Doktorov1977}:
\begin{gather}
\hat{\textbf{a}} \rightarrow \hat{U}_\textrm{Dok}\hat{\textbf{a}}\hat{U}^\dagger_\textrm{Dok} \\
\hat{U}_\textrm{Dok} = \hat{\boldsymbol{D}}(\boldsymbol{\alpha})\hat{\boldsymbol{S}}^\dagger(\boldsymbol{\zeta'})\hat{\boldsymbol{R}}(U)\hat{\boldsymbol{S}}(\boldsymbol{\zeta})
\end{gather}
where 
\begin{align}
\hat{\boldsymbol{D}}(\boldsymbol{\alpha}) & = \hat{D}(\alpha_1)\otimes\hat{D}(\alpha_2)\otimes ... \otimes\hat{D}(\alpha_N) \\
\hat{\boldsymbol{S}}^{(\dagger)}(\boldsymbol{\zeta^{(')}}) & = \hat{S}^{(\dagger)}(\zeta^{(')}_1)\otimes\hat{S}^{(\dagger)}(\zeta^{(')}_2)\otimes ... \otimes\hat{S}^{(\dagger)}(\zeta^{(')}_N)
\end{align}
correspond to $N$-dimensional vectors of single-mode displacement and squeezing operations, respectively. $\hat{\boldsymbol{R}}(U)$ is an $N$-mode rotation operator corresponding to a $N \times N$ rotation matrix $U$ which can be decomposed into a product of two-mode beamsplitter operations (see supplementary text I) \cite{Reck1994}. For $N$ = 2, $U$ is a two-dimensional rotation matrix parameterized by a single angle $\theta$. The set of dimensionless Doktorov parameters $\boldsymbol{\alpha}=(\alpha_1,...,\alpha_N), \boldsymbol{\zeta}=(\zeta_1,...,\zeta_N), \boldsymbol{\zeta'}=(\zeta'_1,...,\zeta'_N)$ and $U$ originate from molecular structural information in the different electronic configurations (see supplementary text I). Applying $\hat{U}_\textrm{Dok}$ to an initial state $\ket{\psi_0}$ of the bosonic processor directly emulates the physical process of a pre-transition molecular vibrational state experiencing a sudden change in the vibrational PES and being expressed in the post-transition vibrational basis. This projection gives rise to the Franck-Condon factors (FCFs) defined as the vibrational overlap integrals of an initial pre-transition vibrational eigenstate, $\ket{\vec{n}} = \ket{n,m,...}$, with a final post-transition vibrational eigenstate, $\ket{\vec{n}'} = \ket{n',m',...}$:
\begin{equation}
\text{FCF}_{\vec{n},\vec{n}'} = |\bra{\vec{n}'}\hat{U}_\textrm{Dok}\ket{\vec{n}}|^2
\end{equation}
In spectroscopic experiments, the validity of this ``sudden approximation" depends on the vastly different energy and time scales that typically characterize electronic and nuclear motions (e.g., $\gg$10000 cm$^{-1}$ vs. $\sim$1000 cm$^{-1}$ and $\ll 10^{-18}$ s vs. $\geq 10^{-15}$ s, respectively). By further assuming that the transition electric dipole moment does not depend on nuclear coordinates (i.e., the Condon approximation), the relative intensities of features appearing in vibrationally resolved absorption and emission spectra will be directly proportional to the corresponding FCFs. Indeed, the practical importance of these quantities often stems from the structural and dynamical insights that they can provide about excited electronic states, information that can be challenging to obtain through other conventional spectroscopic means.

\begin{figure*}[t]
  \includegraphics[width=\linewidth]{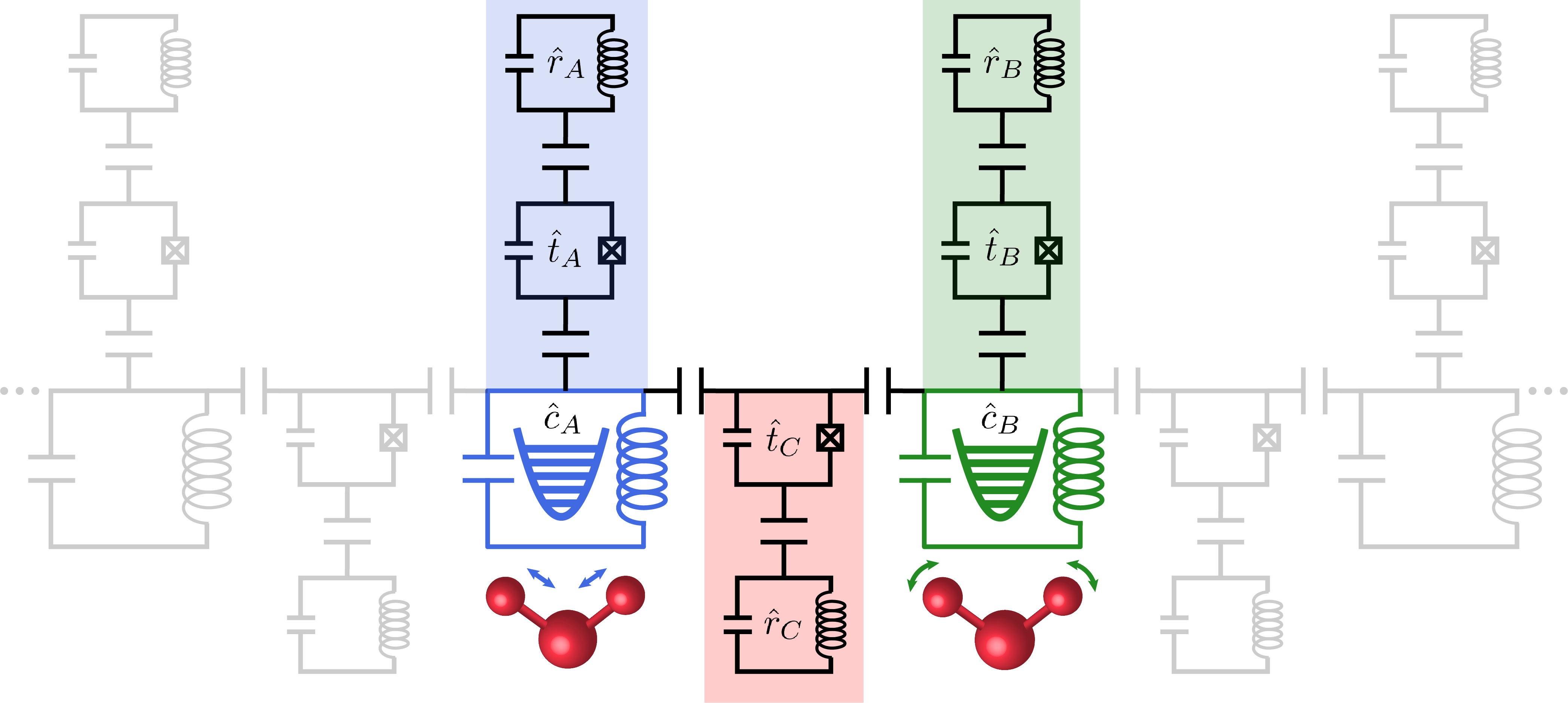}
  \caption{\textbf{Circuit schematic of the superconducting bosonic processor.} The device consists of two microwave cavity modes (blue, $\hat{c}_A$ and green, $\hat{c}_B$) which represent the symmetric-stretching and bending modes of a triatomic molecule in the $C_{2v}$ point group, respectively. Ancilla measurement and control modules (shaded in blue and green) consisting of a transmon qubit $(\hat{t}_A,\hat{t}_B)$ and readout resonator $(\hat{r}_A,\hat{r}_B)$ couple to each cavity mode for state preparation and measurement. The coupler module (shaded in red) consists of a coupler transmon $\hat{t}_C$ and readout resonator $\hat{r}_C$. The coupler transmon is used to facilitate bilinear Gaussian operations on the two cavity modes through four-wave mixing, and the readout resonator is used for both characterization and post-selection. This configuration is extensible to a general linear array of $N$ cavity modes with nearest-neighbor coupler modules and $N$ ancillary modules, as depicted in light grey.}
  \label{fig:fig1}
\end{figure*}

\section{Experimental implementation}

Our superconducting processor is designed to manipulate the bosonic modes of two microwave cavities, $\hat{c}_A$ and $\hat{c}_B$ (Fig. 1). In our design, a coupler transmon $\hat{t}_C$ dispersively couples to both cavities, enabling beamsplitter \cite{Gao2018} and squeezing operations through driven four-wave mixing processes. The coupler transmon is also dispersively coupled to a readout resonator $\hat{r}_C$. Displacement operations on the cavity modes are performed via resonant drives through local coupling ports. Additional ancillary transmon-readout systems $\{\hat{t}_A,\hat{r}_A\}$ and $\{\hat{t}_B,\hat{r}_B\}$ are inserted into the device and couple to each cavity for the purposes of state preparation and tomography.

The goal of the present simulation is to emulate the transformation of a molecular vibrational state due to an electronic transition using the photonic state of the quantum processor. The processor is first initialized in a state $\ket{\psi_0}$ corresponding to the pre-transition molecular vibrational state of interest. A vacuum state of both bosonic modes is prepared through feedback cooling protocols, and Fock states $\ket{n,m}$ are initialized using optimal control techniques \cite{Heeres2017} (see supplementary text IV). The ability to reliably synthesize arbitrary Fock states translates to the powerful capability of simulating FCFs starting from vibrationally excited states, a task which is challenging in most other bosonic simulators. The Doktorov transformation is then applied, producing a basis change to that of the post-transition vibrational Hamiltonian. In our case where $N$ = 2, the rotation operator corresponds directly to enacting a single beamsplitter. Both the single-mode squeezing and beamsplitter operations utilize the four-wave mixing capabilities of the coupler transmon. Two pump tones that fulfill the appropriate frequency matching condition ($\omega_1 + \omega_2 = 2\omega_{A/B}$ for the squeezing operation, and $\omega_2 - \omega_1 = \omega_B - \omega_A$ for the beamsplitter operation, where $\omega_{1/2}$ are the two pump frequencies and $\omega_{A/B}$ are the cavity frequencies) are sent through a port that primarily couples to the coupler transmon, which enacts the desired Hamiltonians \cite{Gao2018}:
\begin{align}
\hat{H}_\textrm{BS}/\hbar & = g_\textrm{BS}(t)(e^{i\varphi}\hat{c}_A\hat{c}^\dagger_B + e^{-i\varphi}\hat{c}^\dagger_A\hat{c}_B) \\
\hat{H}_{\textrm{sq},i}/\hbar & = g_{\textrm{sq},i}(t)(e^{i\phi_i}\hat{c}^2_i + e^{-i\phi_i}\hat{c}^{\dagger 2}_i)
\end{align}
where $i \in \{A,B\}$. Importantly, the phase of the operations $\{\varphi, \phi_A, \phi_B\}$ implemented by these Hamiltonians is controlled by the phase of the pump tones, allowing the microwave control system to generate the correct family of beamsplitter and squeezing operations for performing $\hat{U}_\textrm{Dok}$. A set of transmon measurements is then carried out for the purpose of post-selecting the final data on measuring all transmons in their ground states. This verification step primarily aims to reject heating events of the transmons out of their ground state; the heating of these ancillae otherwise dephases the cavities while coupler heating effectively halts the pumped operations by shifting the requisite frequency matching conditions. This post-selection also serves to ensure that the ancillae begin in their ground states for the subsequent measurement of the cavities. In our experiment, we reject $5 - 10\%$ of the data depending on which transformation is simulated. Finally, averaging many measurements of the cavities in a photon number basis, $\{n',m'\}$, give the desired FCFs. The full set of controls is detailed in Table 1; the relatively small error rates due to photon loss provide a sense of scale for attainable circuit depths while maintaining high fidelity performance.

\renewcommand{\arraystretch}{1.6}
\begin{table*}[]
\setlength{\tabcolsep}{10pt}
\begin{tabular}{|l|l|l|}
\hline
 & \multicolumn{1}{|c|}{\textbf{Interaction}} & \multicolumn{1}{|c|}{\textbf{Error Rate}} \\ \hline
 \textbf{State preparation} &  &  \\ \hline
\quad Optimal Control  & $\hat{H}_\textrm{drive}(t) = \varepsilon^*_i(t)\hat{c}_i + \varepsilon_i(t)\hat{c}^\dagger_i + \epsilon^*_i(t)\hat{t}_i + \epsilon_i(t)\hat{t}^\dagger_i$  &  $\kappa_i \tau_{\textrm{prep},i} \sim 10^{-3} - 10^{-2}$ \\ \hline
\textbf{Operations}    &  &  \\ \hline
\quad Displacement     & $\tilde{\varepsilon}^*_i(t)\hat{c}_i + \tilde{\varepsilon}_i(t)\hat{c}^\dagger_i$ & $\kappa_i \tau_{\textrm{disp},i} \sim 10^{-4}$ \\ \hline
\quad Squeezing        & $g_{\textrm{sq},i}(t)(e^{-i\phi_i}\hat{c}^2_i + e^{i\phi_i}\hat{c}^{\dagger 2}_i)$ & $\kappa_i / g_{\textrm{sq},i} \sim 5 \times 10^{-2}$ \\ \hline
\quad Beamsplitter     & $g_\textrm{BS}(t)(e^{i\varphi}\hat{c}_A\hat{c}^\dagger_B + e^{-i\varphi}\hat{c}^\dagger_A\hat{c}_B)$ & $\kappa_i\tau_\textrm{BS} \sim 10^{-2}$ \\ \hline
\textbf{Measurement}   &  &  \\ \hline
\quad Single-bit extraction  & $\hat{M}_0 = \ket{n',m'}\bra{n',m'} \quad \hat{M}_1 = \hat{\mathds{1}} - \hat{M}_0$ & $\kappa_i\tau_\textrm{meas} \sim 10^{-3} - 10^{-2}$ \\ \hline
\quad Sampling         & $\ket{n'} = \ket{b_3,b_2,b_1,b_0}$ & $\kappa_i\tau_\textrm{meas} \sim 10^{-2} - 10^{-1}$ \\ \hline
\end{tabular}
\caption{\textbf{Full set of controls of the bosonic processor.} The subscripts $i \in \{A,B\}$ correspond to operations on each cavity and their respective ancillary modules. $\kappa_A$ and $\kappa_B$ are the intrinsic linewidths of the cavity modes. Optimal control pulses that utilize both resonant cavity and ancilla drives with complex envelopes, $\varepsilon_i(t)$ and $\epsilon_i(t)$, respectively, generate a desired initial Fock state (see supplementary text IV). $\ket{n',m'}$ denotes a particular joint photon number to be measured in the cavities, and $\{b_i\}$ represent the bits associated with the binary decomposition of the photon number. For example, $\ket{5} = \ket{0101}$. The third column presents an estimate of the expected limits on fidelity due to photon loss for the hardware in this current implementation of the simulator.}
\end{table*}

\section{Measurement Protocols}

Two complementary measurement schemes are used to extract FCFs from the final state of the processor, both of which utilize the dispersive coupling of each microwave cavity to its ancillary transmon-readout system (Fig. 2a): $\hat{H}_\textrm{int}/\hbar = -\sum_{i\in\{A,B\}}\chi_i\hat{c}^\dagger_i\hat{c}_i\hat{t}^\dagger_i\hat{t}_i$, where the dispersive interaction strengths in our experiment are $\chi_A = 2\pi \times 748$ kHz and $\chi_B = 2\pi \times 1240$ kHz. Fundamentally, the difference between these two measurement schemes arises from the ability of the latter to extract more than one bit of information on a given single shot of the experiment.

\begin{figure*}[t]
	\includegraphics[scale = 0.35]{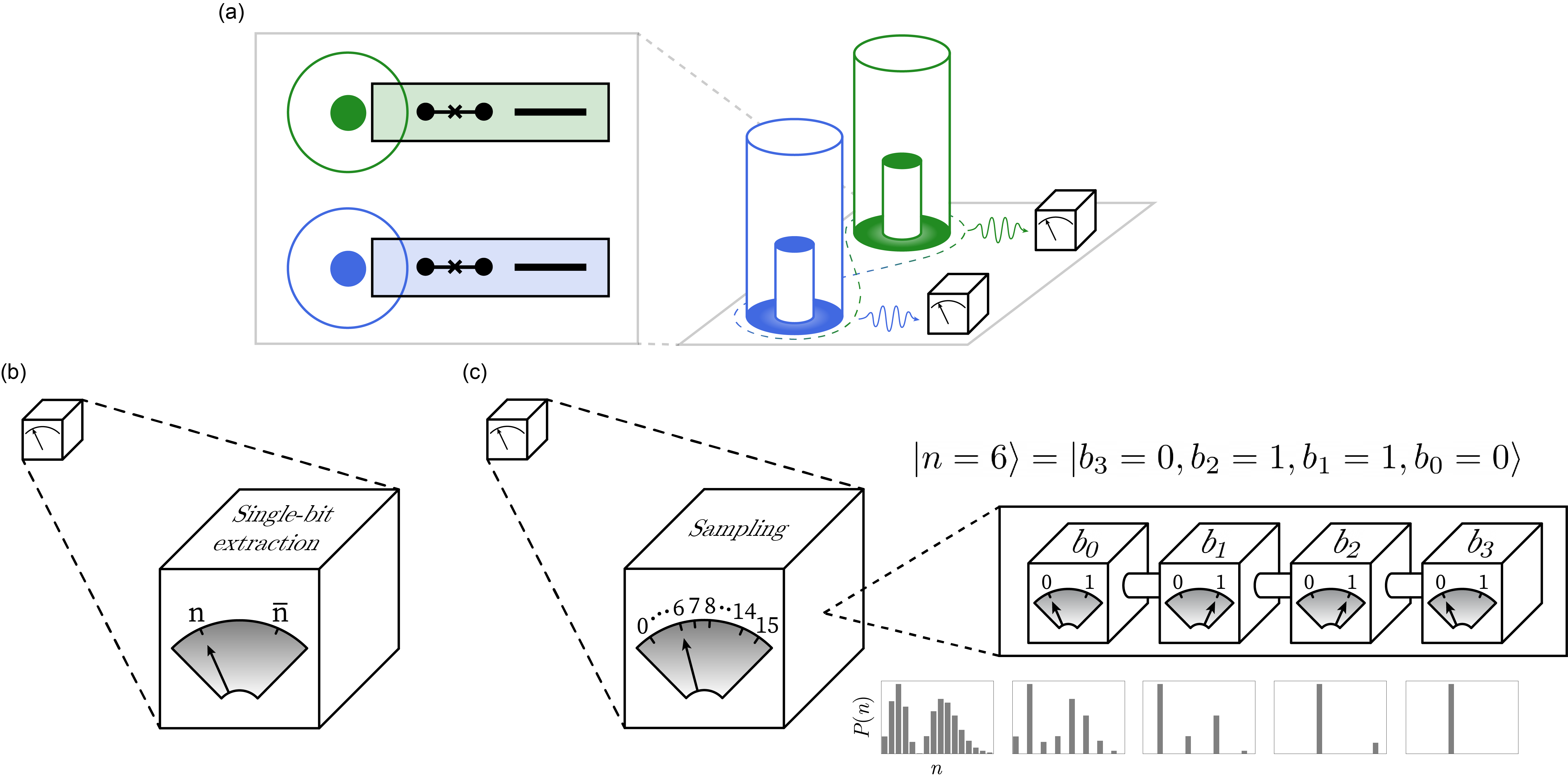}
	\caption{\textbf{QND measurements of cavity photon number.} (a) Depiction of two 3D $\lambda/4$ co-axial cavities dispersively coupled to individual ancillary modules (outlined in black), each consisting of an ancilla qubit and a readout resonator. From the point of view of the measurement, the ancillary modules serve as reconfigurable black boxes used to detect the photon number in each cavity. The dotted trace serves to illustrate that the state of the two cavities can generally be entangled. (b) \textit{Single-bit extraction.} In this scheme, on a given run of the experiment, each ancilla is excited conditioned on a pre-determined photon number $n$ in its respective cavity. (c) \textit{Sampling.} Here, instead of having a binary output, each detection module serves as a photon number resolving detector. Sequential QND measurements of the operators associated with the first four bits of each photon number's binary decomposition resolve up to 15 photons per mode. For a general state of the cavity, each measurement projects the state into the eigenspace of the outcome, ultimately projecting out a single Fock state with its corresponding probability. In the schematic, a sequence sampling the 6 photon component ($\ket{6} = \ket{0110}$) from a displaced Fock state is shown.}
	\label{fig:fig2}
\end{figure*}

\subsection{Single-bit extraction}

The first scheme (Fig. 2b)  maps a given joint cavity photon number population $\{n',m'\}$ onto the joint state of the two transmons via state-selective $\pi$ pulses. These pulses have frequencies $\omega_{t_A} = \omega^0_{t_A} - n'\chi_A + (n'^2 - n')\frac{\chi'_A}{2}$ and $\omega_{t_B} = \omega^0_{t_B} - m'\chi_B + (m'^2 - m')\frac{\chi'_B}{2}$ where the small second-order dispersive shift is also taken into account: $\hat{H}'_\textrm{int}/\hbar = \sum_{i\in\{A,B\}}\frac{\chi'_i}{2}\hat{c}^\dagger_i\hat{c}^\dagger_i\hat{c}_i\hat{c}_i\hat{t}^\dagger_i\hat{t}_i$. In our experiment, $\chi'_A = 2\pi \times 1.31$ kHz and $\chi'_B = 2\pi \times 1.35$ kHz. The pulses are applied with a Gaussian envelope truncated at $\pm 2\sigma_t$ such that the bandwidth is approximately $\sigma_f = 1/(2\pi\sigma_t)$, where $\sigma_t$ is the standard deviation of the pulse in time. The selectivity is defined as the probability of exciting the ancilla given occupation in an adjacent photon number state of interest. Pulses with $\sigma_t$ = 1 $\mu$s for both ancillas are used in the experiment, which give selectivities above 99.9$\%$ and implement the following mapping for a general state of the two cavities $\ket{\psi} = \sum_{i,j}c_{ij}\ket{i,j}$ for a chosen set of photon numbers $\{n',m'\}$ to probe:
\begin{align}
& \sum_{i,j}c_{ij}\ket{i,j}\otimes\ket{g,g} \rightarrow \sum_{i\neq n', j\neq m'}c_{ij}\ket{i,j}\otimes\ket{g,g} \nonumber \\ + & \sum_{i\neq n'}c_{im'}\ket{i,m'}\otimes\ket{g,e} +  \sum_{j\neq m'}c_{n'j}\ket{n',j}\otimes\ket{e,g} \nonumber \\ & + c_{n'm'}\ket{n',m'}\otimes\ket{e,e}
\end{align}
where $\ket{g}$ and $\ket{e}$ are the ground and first excited state of each ancilla transmon. The transmons are then individually read out using standard dispersive techniques, and the results are correlated on a shot-by-shot basis to extract a single bit of information for each joint photon number state probed. We thus call this measurement scheme ``single-bit extraction."  

\subsection{Photon number resolved sampling}

Extracting FCFs using the ``single-bit extraction" scheme, however, is not scalable. The bosonic Hilbert space grows exponentially with the number of modes $N$ as $n^N_\textrm{max}$, where $n_\textrm{max}$ is the maximum number of excitations considered for each mode. Thus, any measurement protocol that only extracts a single bit of information from the underlying distribution at a time must necessarily query the exponentially growing number of final states. This is the case for a recent implementation of this simulation using two phonon modes of a single trapped ion \cite{Shen2018}.

The second measurement scheme circumvents this problem and implements single-shot photon number resolving detection (Fig. 2c). At the heart of this technique is the fact that the following Hamiltonian:
\begin{align}
\hat{H} = -\chi\hat{c}^\dagger\hat{c}\hat{t}^\dagger\hat{t} + \epsilon^*(t)\hat{t} + \epsilon(t)\hat{t}^\dagger
\end{align}
enables the QND mapping of any binary valued operator of the cavity Hilbert space onto the state of the transmon by numerically optimizing an appropriate waveform $\epsilon(t)$ that is applied to only the transmon. Through this capability, we synthesize pulses $\epsilon_k(t)$, $k \in \{0, 1, 2, 3\}$ that excite the transmon from its ground state $\ket{g}$ to the excited state $\ket{e}$ conditioned on the cavity state's projection onto a series of parity subspaces that represent a binary decomposition $b_k$ of the photon number $\ket{n} = \ket{b_3b_2b_1b_0}$:
\begin{equation}
(\hat{P}_k)_{ij} = \left\{
\begin{array}{ll}
      0 & \textrm{if} \quad i \neq j \\
      1 - 2\bigg( \lfloor\frac{i}{2^k}\rfloor\quad (\textrm{mod} \, 2)\bigg) & \textrm{if} \quad i = j \\
\end{array} 
\right.
\end{equation}
with eigenvalues $\pm$1. In our experiment, we synthesize pulses each with a duration between $800 - 1200$ ns that explicitly designate the above parity operators over the Hilbert space of each cavity up to $n_{max} = 15$.

The QND nature of each of these mapping pulses on the cavity state ensures that the pulses can be applied sequentially (with transmon measurements following each pulse) to project an initial cavity state $\ket{\psi} = \sum_{n=0}^{15}c_n\ket{n}$ with bounded support within $n_{max} = 15$ into a definite Fock state $\ket{n}$ with a probability $|c_n|^2$. In order to minimize errors due to decoherence when the transmon is excited, after the $k^\textrm{th}$ bit is mapped onto the transmon and measured, the transmon is reset to its ground state using real-time feed-forward control to prepare for the measurement of the $(k+1)^\textrm{th}$ bit of the same cavity state. This protocol is performed simultaneously on each cavity, which returns a sample from the underlying joint photon number distribution on a shot-by-shot basis. This measurement scheme thus implements bosonic sampling of the final state distribution, which we will simply call ``sampling." These binary detectors optimally resolve the photon number in the cavities in $N\textrm{log}_2(n_\textrm{max})$ measurements, but are prone to more errors due to transmon decoherence that are, in principle, correlated bit-to-bit. We leave the task of characterizing these errors in full detail, optimizing the technique, and potentially applying de-convolution methods \cite{Nachman2019} to improve the accuracy of the sampled distribution as the subject of future work.

\begin{figure}
	\includegraphics[scale = 0.5]{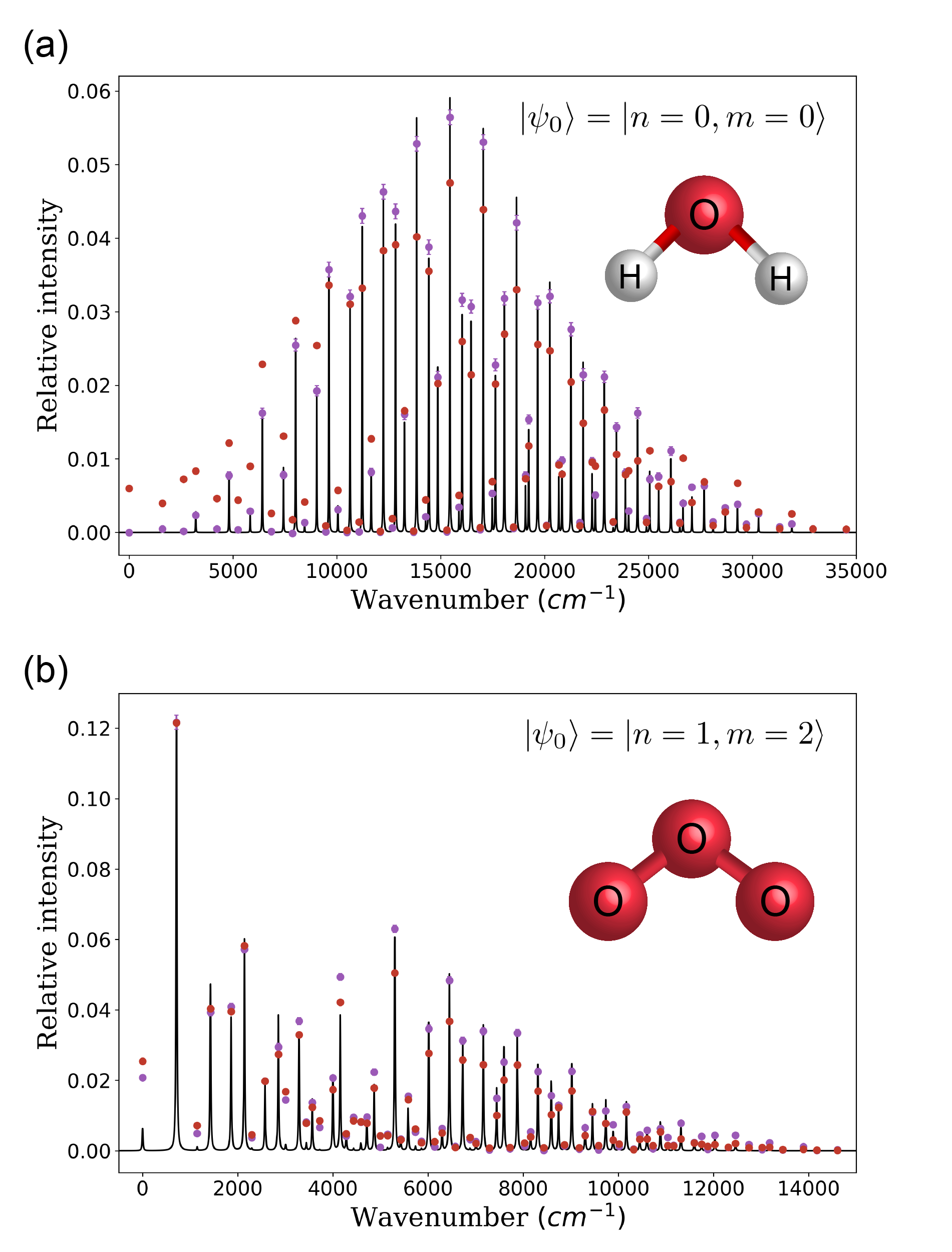}
	\caption{\textbf{Experimental Franck-Condon factors.} Measured data for (a) photoionization of water to the $(\tilde{B}^2B_2)$ excited state of the cation H$_2$O $\xrightarrow{h\nu}$ H$_2$O$^+ (\tilde{B}^2$B$_2)$ + e$^-$ starting in the vacuum (vibrationless, $n$ = 0, $m$ = 0) state and (b) the photodetachment of the ozone anion to the ground state of the neutral species O$^-_3 \xrightarrow{h\nu}$ O$_3$ + e$^-$ starting from a vibrational eigenstate possessing one quantum of symmetric-stretching and two quanta of bending excitation ($n$ = 1, $m$ = 2). The abscissa scale corresponds to vibrational term values (energies in cm$^{-1}$) within the final (post-transition) electronic PES calculated from the harmonic frequencies for symmetric-stretching and bending degrees of freedom: $\tilde{\nu} = n'\tilde{\nu}'_\textrm{stretch} + m'\tilde{\nu}'_\textrm{bend}$. Solid lines depict theoretical FCFs, artificially broadened with Lorentzian profiles (10 cm$^{-1}$ FWHM). Circles represent experimental data using the ``single-bit extraction" (purple) and ``sampling" (red) measurement schemes explained in the main text; statistical error bars for the latter measurement are not visible on this scale (see supplementary text VI). Systematic errors associated with transmon decoherence during the selective $\pi$ pulses are corrected for (see supplementary text V). Additional errors are present in the sampled values, owing to decoherence effects during the binary decomposition measurement chain.}
	\label{fig:fig3}
\end{figure}

\section{Simulated photoelectron spectra}

The full set of FCFs for a given electronic transition can provide, under the Condon approximation, the relative intensities of vibronic progressions appearing in corresponding photoelectron spectra. We simulate four different photoelectron processes starting in various vibrational initial states $\ket{\psi_0} = \ket{n,m}$: H$_2$O $\xrightarrow{h\nu}$ H$_2$O$^+ (\tilde{B}^2$B$_2)$ + e$^-$, O$^-_3 \xrightarrow{h\nu}$ O$_3$ + e$^-$, NO$^-_2 \xrightarrow{h\nu}$ NO$_2$ + e$^-$, and SO$_2 \xrightarrow{h\nu}$ SO$^+_2$ + e$^-$. Experimental results for the first two processes starting in $\ket{\psi_0} = \ket{0,0}$ and $\ket{\psi_0} = \ket{1,2}$, respectively, are presented in Fig. 3 (see supplementary text VI for additional data). The particular electronic state of the water cation considered here $(\tilde{B}^2\textrm{B}_2)$ is the second excited state of doublet spin multiplicity, with the attendant electronic wave function having B$_2$ symmetry. The other four processes consider transitions beginning and ending in electronic ground states. All of the triatomic molecules targeted here retain $C_{2v}$ point-group symmetry for their equilibrium configurations in both the pre-transition and post-transition electronic states, thereby ensuring that analyses can be restricted to the two-dimensional subspace of the symmetric-stretching and bending modes. 

A figure of merit for quantifying the quality of the quantum simulation is the distance $D = \frac{1}{2}\sum_{i=0}^{n_\textrm{max}}\sum_{j=0}^{n_\textrm{max}}|p^\textrm{meas}_{ij} - p^\textrm{ideal}_{ij}|$ between the measured probabilities $\{p^\textrm{meas}_{ij}\}$ and the ideal distribution $\{p^\textrm{ideal}_{ij}\}$. The distances for the two simulated processes in Fig. 3 are D = 0.049 (H$_2$O) and 0.105 (O$_3$) for the ``single-bit extraction" scheme. This distance metric is accompanied with a success probability due to post-selection of transmon heating events, which are 95$\%$ and 93$\%$ for the aforementioned simulations. The heating events are dominated by the coupler transmon; the dynamics of a driven Josephson element for engineered bilinear operations presents a multi-dimensional optimization problem that seeks to maximize the desired interaction rates while minimizing induced decoherence and dissipation rates \cite{Zhang2019}. Self-Kerr Hamiltonian terms of the form $\hat{H}_\textrm{Kerr} / \hbar = -\sum_{i \in \{A,B\}} \frac{K_i}{2}\hat{c}^\dagger_i\hat{c}^\dagger_i\hat{c}_i\hat{c}_i$, photon loss, and imperfect state preparation account for errors that are undetected by post-selection. The first two effects are captured through full time-domain master equation simulations (see supplementary text VI). The magnitudes of these two errors depends on the molecular process; each corresponding Doktorov transformation will have different squeezing and rotation parameters thus leading to varying lengths of the pumped operations. Simulations of shorter length circuits will therefore have lower error rates. Additionally, errors due to self-Kerr interactions of the cavities are larger for higher photon number states.

\section{Resource requirements and Scalability}

The central advantages of simulating the transformation of a bosonic Hamiltonian using a bosonic system lie in both the native encoding and the efficient decomposition of the Doktorov transformation into Gaussian operations. A recent proposal for obtaining Franck-Condon factors on a conventionally envisioned spin-$\frac{1}{2}$ quantum computer \cite{Sawaya2019} needs to map the problem onto qubits and a univeral gate set. This first requires encoding the Hilbert space of size $n^N_\textrm{max}$ onto $n_q = N\textrm{log}_2(n_\textrm{max})$ qubits. The choice of $n_\textrm{max}$ is dependent on the initial state as well as the magnitude of the displacement and squeezing; both operations can produce states with large photon numbers. Using quantum signal processing \cite{Low2017}, the approximate number of gates $n_g$ then needed to implement $\hat{U}_\textrm{Dok}$ using a universal qubit gate set to within an error $\varepsilon$ is $n_g = O(N^2n^2_\textrm{max}\textrm{log}^3(1/\varepsilon))$. For our experiment with $N$ = 2 modes, taking $n_\textrm{max}$ = 16 and desiring an error $\varepsilon = 5 \times 10^{-2}$, this translates to $n_q$ = 8 qubits and $n_g = O(10^3)$ gates. The coherence requirements for performing such a computation this way is thus relatively demanding and exceeds the capabilities of current technologies. The advantage of the qubit-based algorithm, however, is the ability to systematically incorporate anharmonicities in the PES, a task which still needs to be theoretically investigated for the bosonic implementation. By comparison, our native bosonic simulator containing $N$ modes simply requires $2N$ squeezing operations, $N$ displacement operations, and a maximum of $N(N-1)/2$ nearest-neighbor beamsplitter operations \cite{Reck1994,Cybenko2001}. This translates to a total of $O(N^2)$ operations and a corresponding circuit depth of $O(N)$ when non-overlapping beamsplitters are applied simultaneously.

\section{Discussion and Outlook}

The superconducting platform demonstrated here is capable of successfully integrating all of the necessary components for performing a high-fidelity, scalable implementation of a practical computational task of interest. Looking ahead, there exist concrete steps toward scaling up, improving performance, and mitigating sources of error. High-Q superconducting modules controlling up to three modes have been experimentally demonstrated \cite{Chou2018}, which would allow simulations to encompass nonlinear triatomic molecules of $C_s$ symmetry. In general, a molecule composed of $M$ atoms will have $3M - 6$ vibrational degrees of freedom ($3M - 5$ for linear species), which sets the requirement for the number of modes needed in the simulator. A linear array of cavity modes (Fig. 1) is sufficient for preserving the efficiency of the implementation as discussed in Section VI. The photon loss rates of the microwave cavities in the device presented here can also be improved \cite{Reagor2016,Romanenko2018}. Finally, the self-Kerr Hamiltonian terms of the cavity modes may be cancelled with extra off-resonant pump tones, or three-wave mixing methods that avoid self-Kerr nonlinearities altogether may be used \cite{Frattini2017}.

The capabilities of the bosonic processor extend beyond the simulation shown for the estimation of FCFs. For applications in quantum chemistry, the full set of Gaussian operations enables the simulation of time-domain vibrational dynamics, potentially probing coherent evolution after a vibronic transition. Beyond the Gaussian framework, however, there also exist possibilities to utilize programmable self-Kerr nonlinearities to probe anharmonic effects in molecular dynamics \cite{Lloyd1999}. More generally, for condensed matter many-body systems, the tools utilized here enable the simulation of tight-binding lattice Hamiltonians, and the inclusion of controllable self-Kerr interactions extends the scope to extended Bose-Hubbard models \cite{Ma2019, Yanay2019}. The ability to use Josephson nonlinearities to manipulate microwave photons in these rich and diverse ways opens up promising avenues for the simulation of bosonic quantum systems in a superconducting platform.

\begin{acknowledgments}
We thank Salvatore Elder, Hannah Lant, Zachary N. Vealey, Lidor Foguel, Kenneth A. Jung, and Shruti Puri for helpful discussions. This research was supported by the U.S. Army Research Office (W911NF-18-1-0212, W911NF-16-1-0349). We gratefully acknowledge support of the U.S. National Science Foundation under grants CHE-1900160 (VSB), CHE-1464957 (PHV), and DMR-1609326 (SMG), the NSF Center for Ultracold Atoms (ILC), and the Packard Foundation (LJ). Fabrication facilities use was supported by the Yale Institute for Nanoscience and Quantum Engineering (YINQE) and the Yale SEAS clean room.
\end{acknowledgments}


\begin{thebibliography}{9}
\bibitem{Feynman1982}
R. P. Feynman. Simulating physics with computers. \textit{Int J Theor Phys.} 21, 467–488 (1982).

\bibitem{Bakr2009}
W. S. Bakr, J. I. Gillen, A. Peng, S. Fölling, M. Greiner. A quantum gas microscope for detecting single atoms in a Hubbard-regime optical lattice. \textit{Nature}. 462, 74–77 (2009)

\bibitem{Tai2017}
M. E. Tai, A. Lukin, M. Rispoli, R. Schittko, T. Menke, D. Borgnia, P. M. Preiss, F. Grusdt, A. M. Kaufman, M. Greiner. Microscopy of the interacting Harper-Hofstadter model in the two-body limit. \textit{Nature}. 546, 519-523 (2017).

\bibitem{Bernien2017}
H. Bernien, S. Schwartz, A. Keesling, H. Levine, A. Omran, H. Pichler, S. Choi, A. S. Zibrov, M. Endres, M. Greiner, V. Vuletic, M. D. Lukin. Probing many-body dynamics on a 51-atom quantum simulator. \textit{Nature}. 551, 579-584 (2017).

\bibitem{Aaronson2013}
S. Aaronson and A. Arkhipov. The Computational Complexity of Linear Optics. \textit{Theory of Computing} 9, 143–252 (2013).

\bibitem{Hamilton2017}
C. S. Hamilton, R. Kruse, L. Sansoni, S. Barkhofen, C. Silberhorn, I. Jex. Gaussian Boson Sampling. \textit{Phys. Rev. Lett.} 119, 170501 (2017).

\bibitem{Schuster2007}
D. I. Schuster, A. A. Houck, J. A. Schreier, A. Wallraff, J. M. Gambetta, A. Blais, L. Frunzio, J. Majer, B. Johnson, M. H. Devoret, R. J. Schoelkopf. Resolving photon number states in a superconducting circuit. \textit{Nature.} 445, 515–518 (2007).

\bibitem{Johnson2010}
B. R. Johnson, M. D. Reed, A. A. Houck, D. I. Schuster, L. S. Bishop, E. Ginossar, J. M. Gambetta, L. DiCarlo, L. Frunzio, S. M. Girvin, R. J. Schoelkopf. Quantum non-demolition detection of single microwave photons in a circuit. \textit{Nature Physics.} 6, 663–667 (2010).

\bibitem{Hann2018}
C. T. Hann, S. S. Elder, C. S. Wang, K. Chou, R. J. Schoelkopf, L. Jiang. Robust readout of bosonic qubits in the dispersive coupling regime. \textit{Phys. Rev. A} 98, 022305 (2018).

\bibitem{Elder2019}
S. S. Elder, C. S. Wang, P. Reinhold, C. T. Hann, K. S. Chou, B. J. Lester, S. Rosenblum, L. Frunzio, L. Jiang, R. J. Schoelkopf. High-fidelity measurement of qubits encoded in multilevel superconducting circuits. \textit{Phys. Rev. X} (Accepted) (2019).

\bibitem{Heeres2017}
R. W. Heeres, P. Reinhold, N. Ofek, L. Frunzio, L. Jiang, M. H. Devoret, R. J. Schoelkopf. Implementing a universal gate set on a logical qubit encoded in an oscillator. \textit{Nature Communications.} 8, 94 (2017).

\bibitem{Gao2018}
Y. Y. Gao, B. J. Lester, Y. Zhang, C. Wang, S. Rosenblum, L. Frunzio, L. Jiang, S. M. Girvin, R. J. Schoelkopf. Programmable Interference between Two Microwave Quantum Memories. \textit{Phys. Rev. X.} 8, 021073 (2018).

\bibitem{Peropadre2016}
B. Peropadre, G. G. Guerreschi, J. Huh, A. Aspuru-Guzik. Proposal for Microwave Boson Sampling. \textit{Phys. Rev. Lett.} 117, 140505 (2016).

\bibitem{Huh2015}
J. Huh, G. G. Guerreschi, B. Peropadre, J. R. McClean, A. Aspuru-Guzik. Boson sampling for molecular vibronic spectra. \textit{Nature Photonics.} 9, 615–620 (2015).

\bibitem{Clements2018}
W. R. Clements, J. J. Renema, A. Eckstein, A. A. Valido, A. Lita, T. Gerrits, S. W. Nam, W. S. Kolthammer, J. Huh, I. A. Walmsley. Approximating vibronic spectroscopy with imperfect quantum optics. \textit{J. Phys. B: At. Mol. Opt. Phys.} 51, 245503 (2018).

\bibitem{Shen2018}
Y. Shen, Y. Lu, K. Zhang, J. Zhang, S. Zhang, J. Huh, K. Kim. Quantum optical emulation of molecular vibronic spectroscopy using a trapped-ion device. \textit{Chem. Sci.} 9, 836–840 (2018).

\bibitem{Doktorov1977}
E. V. Doktorov, I. A. Malkin, V. I. Man’ko. Dynamical symmetry of vibronic transitions in polyatomic molecules and the Franck-Condon principle. \textit{Journal of Molecular Spectroscopy.} 64, 302–326 (1977).

\bibitem{Reck1994}
M. Reck, A. Zeilinger, H. J. Bernstein, P. Bertani. Experimental realization of any discrete unitary operator. \textit{Phys. Rev. Lett.} 73, 58-61 (1994).

\bibitem{Nachman2019}
B. Nachman, M. Urbanek, W. A. de Jong, C. W. Bauer. Unfolding quantum computer readout noise. \textit{arXiv:1910.01969 [quant-ph]} (2019) (available at http://arxiv.org/abs/1910.01969).

\bibitem{Zhang2019}
Y. Zhang, B. J. Lester, Y. Y. Gao, L. Jiang, R. J. Schoelkopf, S. M. Girvin. Engineering bilinear mode coupling in circuit QED: Theory and experiment. \textit{Phys. Rev. A.} 99, 012314 (2019).

\bibitem{Sawaya2019}
N. P. D. Sawaya and J. Huh. Quantum Algorithm for Calculating Molecular Vibronic Spectra. \textit{J. Phys. Chem. Lett.} 10, 3586 (2019).

\bibitem{Low2017}
G. H. Low and I. L. Chuang. Optimal Hamiltonian Simulation by Quantum Signal Processing. \textit{Phys. Rev. Lett.} 118, 010501 (2017).

\bibitem{Cybenko2001}
George Cybenko. Reducing Quantum Computations to Elementary Unitary Operations, \textit{Computing in Science $\&$ Engineering} 3, 27 (2001).

\bibitem{Chou2018}
K. S. Chou, J. Z. Blumoff, C. S. Wang, P. Reinhold, C. J. Axline, Y. Y. Gao, L. Frunzio, M. H. Devoret, L. Jiang, R. J. Schoelkopf. Deterministic teleportation of a quantum gate between two logical qubits. \textit{Nature.} 561, 368 (2018).

\bibitem{Reagor2016}
M. Reagor, W. Pfaff, C. Axline, R. W. Heeres, N. Ofek, K. Sliwa, E. Holland, C. Wang, J. Blumoff, K. Chou, M. J. Hatridge, L. Frunzio, M. H. Devoret, L. Jiang, R. J. Schoelkopf. Quantum memory with millisecond coherence in circuit QED. \textit{Phys. Rev. B.} 94, 014506 (2016).

\bibitem{Romanenko2018}
A. Romanenko, R. Pilipenko, S. Zorzetti, D. Frolov, M. Awida, S. Posen, A. Grassellino. Three-dimensional superconducting resonators at T < 20 mK with the photon lifetime up to tau=2 seconds. \textit{arXiv:1810.03703} (2018).

\bibitem{Frattini2017}
N. E. Frattini, U. Vool, S. Shankar, A. Narla, K. M. Sliwa, M. H. Devoret. 3-wave mixing Josephson dipole element. \textit{Appl. Phys. Lett.} 110, 222603 (2017).

\bibitem{Lloyd1999}
S. Lloyd and S. L. Braunstein. Quantum Computation over Continuous Variables. \textit{Phys. Rev. Lett.} 82, 1784-1787 (1999).

\bibitem{Ma2019}
R. Ma, B. Saxberg, C. Owens, N. Leung, Y. Lu, J. Simon, D. I. Schuster. A dissipatively stabilized Mott insulator of photons. \textit{Nature}. 566, 51-57 (2019).

\bibitem{Yanay2019}
Y. Yanay, J. Braumuller, S. Gustavsson, W. D. Oliver, C. Tahan. Realizing the two-dimensional hard-core Bose-Hubbard model with superconducting qubits. \textit{arXiv:1910.00933 [quant-ph]} (2019) (available at http://arxiv.org/abs/1910.00933)

\end{thebibliography}
\end{document}


\preprint{}

\title{Supplemental Material: Efficient multiphoton sampling of molecular vibronic spectra on a superconducting bosonic processor}

\author{Christopher S. Wang}
\email{christopher.wang@yale.edu}
\author{Jacob C. Curtis}
\author{Brian J. Lester}
\author{Yaxing Zhang}
\author{Yvonne Y. Gao}
\affiliation{Departments of Physics and Applied Physics, Yale University, New Haven, CT 06511, USA.}
\affiliation{Yale Quantum Institute, Yale University, New Haven, CT 06520, USA.}
\author{Jessica Freeze}
\author{Victor S. Batista}
\author{Patrick H. Vaccaro}
\affiliation{Department of Chemistry, Yale University, New Haven, CT 06511, USA.}
\author{Isaac L. Chuang}
\affiliation{Department of Physics, Center for Ultracold Atoms, and Research Laboratory of Electronics, Massachusetts Institute of Technology, Cambridge, MA 02139, USA.}
\author{Luigi Frunzio}
\author{Liang Jiang}
\altaffiliation{current address: Pritzker School of Molecular Engineering, University of Chicago, Chicago, IL 60637, USA.}
\author{S. M. Girvin}
\author{Robert J. Schoelkopf}
\email{robert.schoelkopf@yale.edu}
\affiliation{Departments of Physics and Applied Physics, Yale University, New Haven, CT 06511, USA.}
\affiliation{Yale Quantum Institute, Yale University, New Haven, CT 06520, USA.}

\maketitle

\section{Obtaining Doktorov parameters}
The Doktorov parameters originate from the physical properties of a given molecule in the two electronic states of interest. Specifically, it is the structural information of the molecular configurations and the relationship between the two that fully parametrize the problem.

\subsection{Description of Quantum-Chemical Analyses}
Theoretical predictions of optimized equilibrium geometries (with imposed $C_{2v}$ symmetry constraints), harmonic (normal-mode) vibrational displacements, and Franck-Condon parameters (Duschinsky rotation matrices and associated shift vectors) exploited the commercial (G16 rev. A.03) version of the Gaussian quantum-chemical suite (TABLE I), \cite{Gaussian2016} with canonical Franck-Condon matrix elements for specific vibronic bands being evaluated through use of the open-source ezSpectrum (ver. 3.0) package. \cite{ezSpectrum2014}  All analyses relied on the CCSD(T) coupled-cluster paradigm, which includes single and double excitations along with non-iterative correction for triples. Dunning’s correlation-consistent basis sets \cite{Dunning1989,Kendall1992,Woon1993} of triple-$\zeta$ quality augmented by supplementary diffuse functions (aug-cc-pVTZ $\equiv$ apVTZ) were deployed for all targeted molecules except water, where a larger doubly augmented, quadruple-$\zeta$ basis was employed (daug-cc-pVQZ $\equiv$ dapVQZ). 

The Duschinsky rotation matrices and associated shift vectors provided by the commercial package Gaussian are defined via:
\begin{equation}
\textbf{Q}\boldsymbol{'} = \textbf{JQ}\boldsymbol{''} + \textbf{K} \tag{S1}
\end{equation}
where $\textbf{Q}\boldsymbol{'}$ and $\textbf{Q}\boldsymbol{''}$ are mass-weighted normal coordinates of the pre- and post-transition molecular configurations, respectively. Because our simulation considers the transformation from a vibrational state in the pre-transition configuration to the post-transition configuration, we must redefine the Duschinsky rotation matrices and associated shift vectors accordingly:
\begin{gather*}
U = \left(\begin{array}{cc} \textrm{cos}\theta & -\textrm{sin}\theta \\ \textrm{sin}\theta & \textrm{cos}\theta \end{array}\right) = \textbf{J}^T \tag{S2} \\
\textbf{d} = -\textbf{J}^T\textbf{K} \tag{S3}
\end{gather*}

\renewcommand{\arraystretch}{1.6}
\begin{table*}[]
\setlength{\tabcolsep}{8pt}
\begin{tabular}{|c|c|c|c|c|c|c|}
\hline
\multicolumn{1}{|c|}{\begin{tabular}[c]{@{}c@{}}Molecular photoelectron\\ process\end{tabular}} & \multicolumn{1}{|c|}{\begin{tabular}[c]{@{}c@{}}$\tilde{\nu}_\textrm{stretch}$ \\ (cm$^{-1}$)\end{tabular}} & \multicolumn{1}{|c|}{\begin{tabular}[c]{@{}c@{}}$\tilde{\nu}_\textrm{bend}$ \\ (cm$^{-1}$)\end{tabular}} & \multicolumn{1}{|c|}{\begin{tabular}[c]{@{}c@{}}$\tilde{\nu}'_\textrm{stretch}$ \\ (cm$^{-1}$)\end{tabular}} & \multicolumn{1}{|c|}{\begin{tabular}[c]{@{}c@{}}$\tilde{\nu}'_\textrm{bend}$ \\ (cm$^{-1}$)\end{tabular}} & \multicolumn{1}{|c|}{$\theta$ (deg)} & \multicolumn{1}{|c|}{\begin{tabular}[c]{@{}c@{}}\textbf{K} \\ $ (a_0\sqrt{m_e})$\end{tabular}} \\ \hline
\multicolumn{1}{|c|}{H$_2$O $\xrightarrow{h\nu}$ H$_2$O$^+ (\tilde{B}^2$B$_2)$ + e$^-$} & 3830.91 & 1649.27 & 2619.09 & 1602.85 & $-0.16598$ & (5.05, 49.47) \\ \hline
\multicolumn{1}{|c|}{O$^-_3 \xrightarrow{h\nu}$ O$_3$ + e$^-$} &  1031.10 & 582.58	& 1147.04 & 713.39 &	$-0.0417$ &	(27.36, 14.33) \\ \hline
\multicolumn{1}{|c|}{NO$^-_2 \xrightarrow{h\nu}$ NO$_2$ + e$^-$} & 1297.27 & 783.55 & 2633.34 & 796.94 & 2.40146 & (35.67, $-38.01$) \\ \hline
\multicolumn{1}{|c|}{SO$_2 \xrightarrow{h\nu}$ SO$^+_2$ + e$^-$} & 1136.38 & 506.27 & 1056.79 & 396.11 & 0.19012 & ($-8.86$, $-58.34$) \\ \hline
\end{tabular}
\caption{$\textbf{Theoretically optimized molecular parameters.}$ Vibrational frequencies for the symmetric-stretching and bending modes of each molecule in pre-($\tilde{\nu}$) and post-transition ($\tilde{\nu'}$) states are provided in wavenumbers (cm$^{-1}$), which is related to angular frequency $\omega$ via $\tilde{\nu} = \omega / 2\pi c$ , where $c$ is the speed of light. The rotation angle corresponding to the Duschinsky rotation matrix is defined in Eq. (S2). The shift vector $\textbf{K} = (k_1,k_2)$ is provided in mass weighted normal coordinates (where $a_0$ is the Bohr radius and $m_e$ is the electron mass) and reflects the relative displacement of equilibrium geometries between the two molecular configurations.}
\end{table*}

\subsection{Conversion from molecular parameters to Doktorov parameters}
The Doktorov transformation as given in Eq. (2) of the main text is:
\begin{equation}
\hat{U}_\textrm{Dok} = \hat{\boldsymbol{D}}(\boldsymbol{\alpha})\hat{\boldsymbol{S}}^\dagger(\boldsymbol{\zeta'})\hat{\boldsymbol{R}}(U)\hat{\boldsymbol{S}}(\boldsymbol{\zeta}) \tag{S4}
\end{equation}
where for $N$ = 2 modes, the squeezing and displacement operations are defined as:
\begin{align}
\hat{\textbf{S}}^{(\dagger)}(\boldsymbol{\zeta}^{(')}) & = \hat{S}^{(\dagger)}_A(\zeta^{(')}_1) \otimes \hat{S}^{(\dagger)}_B(\zeta^{(')}_2) \nonumber \\
& = \textrm{exp}\big(\frac{1}{2}(\zeta^{*(')}_1\hat{c}^2_A - \zeta^{(')}_1\hat{c}^{\dagger 2}_A)\big) \otimes \textrm{exp}\big(\frac{1}{2}(\zeta^{*(')}_2\hat{c}^2_B - \zeta^{(')}_2\hat{c}^{\dagger 2}_B)\big) \tag{S5} \\
\hat{\textbf{D}}(\boldsymbol{\alpha}) & = \hat{D}_A(\alpha_1) \otimes \hat{D}_B(\alpha_2) \nonumber \\
& = \textrm{exp}(\alpha_1\hat{c}^\dagger_A - \alpha^*_1\hat{c}_A) \otimes \textrm{exp}(\alpha_2\hat{c}^\dagger_B - \alpha^*_2\hat{c}_B) \tag{S6}
\end{align}
where $\zeta^{(')}_i = \textrm{ln}\bigg(\sqrt{\tilde{\nu}^{(')}_i}\bigg)$ and $\tilde{\nu}^{(')}_i$ is the vibrational frequency of mode $i$ in the pre- (post-) transition configuration and $\alpha_i = \sqrt{\frac{\omega^{(')}_i}{2\hbar}}d_i$ where $\{d_i\}$ are the vector elements of $\textbf{d}$ in Eq. (S3).

The Duschinsky rotation matrix $U$ generates the $N$-mode rotation operator $\hat{\textbf{R}}(U)$. The multi-mode mixing elements implemented in this experiment are two-mode beamsplitters, necessitating a decomposition of $U$ into nearest-neighbor rotations, and thus $\hat{\textbf{R}}$ into nearest-neighbor beamsplitters. $\hat{\textbf{R}}(U)$ becomes a product of two mode beamsplitters parametrized by $\{\theta_k\}$ and $\{i_k,j_k\}$, a sequence of angles and rotation axes derived from the decomposition of $U = \prod_k R_{i_k,j_k}(\theta_k)$. We can then write:
\begin{equation}
\hat{\textbf{R}}(U) = \prod_k \textrm{exp}\big(\theta_k (\hat{c}_{i_k}\hat{c}^\dagger_{j_k} - \hat{c}^\dagger_{i_k}\hat{c}_{j_k})\big) \tag{S7}
\end{equation}
The decomposition of $U$ is analogous to generalizing Euler angles to SO($N$); any rotation in $\mathbb{R}^N$ can be written as a product of rotations in a plane $R_{i_k,j_k}(\theta_k)$, known as Givens rotations. Following an algorithm similar to that in \cite{Reck1994, Cybenko2001}, but simplified to real orthogonal matrices, produces a decomposition of $U$ as a product of nearest-neighbor rotation matrices. The Duschinsky matrix for $N$ = 2 is a single Givens rotation parametrized by an angle $\theta$ which is enacted with one beamsplitter:
\begin{equation}
\hat{\textbf{R}}(\hat{U}(\theta)) = \textrm{exp}\big(\theta(\hat{c}^\dagger_A\hat{c}_B - \hat{c}_A\hat{c}^\dagger_B)\big) \tag{S8}
\end{equation}

\subsection{Optimization of squeezing parameters}
The modification of the creation and annihilation operators under the mode transformation is given in \cite{Huh2015}:
\begin{equation}
\hat{\textbf{a}}^{'\dagger} = \frac{1}{2}(L-(L^T)^{-1})\hat{\textbf{a}} + \frac{1}{2}(L+(L^T)^{-1})\hat{\textbf{a}}^\dagger +\vec{\boldsymbol{\alpha}} \tag{S9}
\end{equation}
where
\begin{gather}
L = \Omega'U\Omega^{-1} \nonumber \\
\Omega = \left( \begin{array}{ccc}
\sqrt{\tilde{\nu}_1} &  & 0 \\
 & \ddots &  \\
0 &  & \sqrt{\tilde{\nu}_N} \end{array} \right) \quad \Omega' = \left( \begin{array}{ccc}
\sqrt{\tilde{\nu}'_1} &  & 0 \\
 & \ddots &  \\
0 &  & \sqrt{\tilde{\nu}'_N} \end{array} \right) \tag{S10}
\end{gather}
The structure of $L$ allows for a free scaling parameter $\eta$ which leaves $L$ invarant, namely:
\begin{align*}
\tilde{\Omega}^{(')} & = \Omega^{(')} / \eta \nonumber \\
L(\Omega, \Omega') & = L(\tilde{\Omega}, \tilde{\Omega}') \tag{S11}
\end{align*}
Given that the squeezing operations of the Doktorov transformation take $\boldsymbol{\zeta} = \textrm{ln}(\vec{\Omega})$ as inputs, an optimization may be performed, as done in \cite{Shen2018}, that minimizes the total amount of squeezing while leaving the unitary invariant. This is desirable as less squeezing corresponds to shorter gate times in the simulation, which reduces the overall error rate. TABLE II lists the final set of dimensionless Doktorov parameters used in the experiment.

\renewcommand{\arraystretch}{1.2}
\begin{table*}[]
\setlength{\tabcolsep}{8pt}
\begin{tabular}{|c|c|c|c|c|}
\hline
 &    \begin{tabular}[c]{@{}c@{}}H$_2$O $\xrightarrow{h\nu}$ \\ H$_2$O$^+ (\tilde{B}^2$B$_2)$ + e$^-$\end{tabular}     &    \begin{tabular}[c]{@{}c@{}}O$^-_3 \xrightarrow{h\nu}$ \\ O$_3$ + e$^-$\end{tabular}      &    \begin{tabular}[c]{@{}c@{}}NO$^-_2 \xrightarrow{h\nu}$ \\ NO$_2$ + e$^-$\end{tabular}     &    \begin{tabular}[c]{@{}c@{}}SO$_2 \xrightarrow{h\nu}$ \\ SO$^+_2$ + e$^-$ \end{tabular}     \\ \hline
$\zeta_1$ & 0.262   & 0.104   & 0.035   & 0.242   \\ \hline
$\zeta_2$ & $-0.160$  & $-0.181$  & $-0.217$  & $-0.162$  \\ \hline
$\theta$ & $-0.166$  & $-0.042$  & 2.402   & 0.19    \\ \hline
$\zeta'_1$ & 0.072   & 0.157   & 0.389   & 0.206   \\ \hline
$\zeta'_2$ & $-0.174$  & $-0.080$  & $-0.208$  & $-0.285$  \\ \hline
$\alpha_1$ & $-1.0162$ & $-1.4278$ & 0.0546  & $-0.1140$ \\ \hline
$\alpha_2$ & $-2.8977$ & $-0.5311$ & $-2.2207$ & 1.7713  \\ \hline
$\eta$ & 47.6381 & 28.9364 & 34.7639 & 26.4676 \\ \hline
\end{tabular}
\caption{$\textbf{Dimensionless Doktorov parameters after converting from molecular parameters.}$ All values are truncated to the precision that the operations are able to be implemented experimentally.}
\end{table*}

\section{Theoretically predicted Hamiltonian terms}

\subsection{Derivation of ancilla-mediated operations}
In this section, we derive the ancilla-mediated beamsplitter and single-mode squeezing interactions as shown in the main text as well as the associated ancilla-induced cavity frequency shifts. Derivations based on the perturbative four wave frequency mixing enabled by a weak ancilla nonlinearity have been presented previously in \cite{Gao2018}. Here, we follow the formalism used in \cite{Zhang2019} and sketch the general results without assuming weak ancilla nonlinearity or weak pumps. We also give explicit expressions for the strength of the engineered interactions in the case of weak pumps.

We start from the Hamiltonian of the two bare cavity modes A and B coupled to the coupler transmon in module C:
\begin{equation}
\hat{H}/\hbar = \omega_A\hat{c}^\dagger_A\hat{c}_A + \omega_B\hat{c}^\dagger_B\hat{c}_B + \hat{H}_C + \hat{H}_I + \hat{H}_\textrm{pump}(t) \tag{S12}
\end{equation}
We emphasize that here the operators $\hat{c}_A$ and $\hat{c}_B$ are the annihilation operators for the bare cavity modes whereas in the main text the operators correspond to the dressed cavity modes that are weakly hybridized with the ancilla transmons.

$\hat{H}_C$ is the Hamiltonian of the bare coupler transmon in module C. After expanding the transmon potential energy to fourth order in the phase across the Josephson junction and neglecting counter rotating terms, we obtain \cite{Koch2007}:
\begin{equation}
\hat{H}_C / \hbar = \omega_C\hat{t}^\dagger_C\hat{t}_C - \frac{K_C}{2}\hat{t}^{\dagger 2}_C\hat{t}^2_C \tag{S13}
\end{equation}
where $\hat{t}^{(\dagger)}_C$ again is the bare annihilation (creation) operator for the coupler transmon with frequency $\omega_C$ and anharmonicity $K_C$.

$\hat{H}_I$ is the interaction energy between the coupler transmon and the two cavity modes. Neglecting counter rotating terms, it may be written as:
\begin{equation}
\hat{H}_I / \hbar = (g_A\hat{c}_A + g_B\hat{c}_B)\hat{t}^\dagger_C + \textrm{h.c.} \tag{S14}
\end{equation}

$\hat{H}_\textrm{pump}(t)$ represents two pumps on the coupler transmon:
\begin{equation}
\hat{H}_\textrm{pump}(t) / \hbar = (\Omega_1 e^{-i\omega_1 t} + \Omega_2 e^{-i\omega_2 t})\hat{t}^\dagger_C + \textrm{h.c.} \tag{S15}
\end{equation}

Of primary interest to us is the dispersive regime where the cavity-transmon coupling strengths are much smaller than their detuning: $|g_{A,B}| \ll |\omega_{A,B} - \omega_C|$\footnote{The condition for the dispersive approximation should be modified in the presence of pumps on the transmon. The cavity frequencies should not only be far away from the transition frequency of the transmon from the ground to the first excited state, but also transitions to higher excited states which are possible via the absorption of pump photons; see \cite{Zhang2019}.}. In this regime, we can treat the cavity-transmon interaction as a perturbation (while treating the remaining parts of the Hamiltonian exactly), and to second order in the interaction strength, we obtain an effective Hamiltonian:
\begin{equation}
\hat{H}_\textrm{eff} / \hbar = \sum_m (\delta\omega_{A,m}\hat{c}^\dagger_A\hat{c}_A + \delta\omega_{B,m}\hat{c}^\dagger_B\hat{c}_B) \otimes \ket{\Psi_m}\bra{\Psi_m} + \hat{V} \tag{S16}
\end{equation}
where $\ket{\Psi_m}$ is the $m^\textrm{th}$ Floquet state that quasi-adiabatically connects to the $m^\textrm{th}$ Fock state $\ket{m}$ of the bare transmon as the pumps are ramped up or down \cite{Zhang2019}. At zero pump amplitudes, $\ket{\Psi_m} = \ket{m}$. The first term in $\hat{H}_\textrm{eff}$ thus represents transmon-induced cavity frequency shifts $\delta\omega_{A,m}$ and $\delta\omega_{B,m}$ when the transmon is in $\ket{\Psi_m}$.

The difference between $\delta\omega_{A,m}$ or $\delta\omega_{B,m}$ with different $m$ leads to cavity-photon-number dependent transmon transition frequencies. In particular, at zero pump amplitudes, the transmon’s transition frequency from the ground to the first excited state decreases linearly with the cavity photon number with a proportionality constant:
\begin{equation}
\chi_{iC} = \delta\omega_{i,0} - \delta\omega_{i,1} = 2K_C\abs{\frac{g_i}{\delta_i}}^2\frac{\delta_i}{\delta_i + K_C}, \quad i \in \{A, B\} \tag{S17}
\end{equation}
where $\delta_i = \omega_i - \omega_C$. Physically, the factor $\abs{g_i/\delta_i}^2$ quantifies the participation ratio of cavity A or B in the coupler transmon. In the experiment, this factor is 0.3$\%$ for cavity A and 0.2$\%$ for cavity B.

Pumps on the coupler transmon can induce effective inter- or intra-cavity interactions (single-mode squeezing or beamsplitter) denoted as $\hat{V}$ in $\hat{H}_\textrm{eff}$. For the case of the beamsplitter interaction $(\omega_2 - \omega_1 = \omega_B - \omega_A)$, we have:
\begin{equation}
\hat{V} / \hbar = \hat{H}_\textrm{BS} / \hbar = \sum_m g_{\textrm{BS},m}(e^{i\varphi^{(m)}}\hat{c}_A\hat{c}^\dagger_B + e^{-i\varphi^{(m)}}\hat{c}^\dagger_A\hat{c}_B) \otimes \ket{\Psi_m}\bra{\Psi_m} \tag{S18}
\end{equation}
For the case of single-mode squeezing ($\omega_1 + \omega_2 = 2\omega_A$ or $2\omega_B$), we have:
\begin{equation}
\hat{V} / \hbar = \hat{H}_{\textrm{sq},i} / \hbar = \sum_m g_{\textrm{sq},i,m}(e^{i\phi^{(m)}_{\textrm{sq},i}}\hat{c}^2_i +e^{-i\phi^{(m)}_{\textrm{sq},i}}\hat{c}^{\dagger 2}_i) \otimes \ket{\Psi_m}\bra{\Psi_m}, \quad i \in \{A, B\} \tag{S19}
\end{equation}

Similar to the transmon-induced frequency shifts on the cavities, here both the strength and phase of the transmon-mediated interactions depend on the state of the transmon. Of primary interest to us is the strengths of the transmon-mediated interactions when the transmon is in the Floquet state $\ket{\Psi_0}$. For weak drives, these strengths are:
\begin{gather}
g_{\textrm{BS},0} \approx 2K_C \abs{\frac{g_A}{\delta_A}\frac{g_B}{\delta_B}\frac{\Omega_1}{\delta_1}\frac{\Omega_2}{\delta_2}\frac{\delta_A + \delta_2}{\delta_A + \delta_2 + K_C}} = \sqrt{\chi_{AC}\chi_{BC}}\sqrt{\abs{\frac{(\delta_A + K_C)(\delta_B + K_C)}{\delta_A \delta_B}}}\abs{\frac{\Omega_1}{\delta_1}\frac{\Omega_2}{\delta_2}\frac{\delta_A + \delta_2}{\delta_A + \delta_2 + K_C}} \tag{S20} \\
g_{\textrm{sq},i,0} \approx 2K_C\abs{(\frac{g_i}{\delta_A})^2\frac{\Omega_1}{\delta_1}\frac{\Omega_2}{\delta_2}\frac{\delta_i}{2\delta_i + K_C}} = \chi_{iC}\abs{\frac{\Omega_1}{\delta_1}\frac{\Omega_2}{\delta_2}\frac{\delta_i + K_C}{2\delta_i + K_C}}, \quad i \in \{A, B\} \tag{S21}
\end{gather}
where $\delta_{1,2} = \omega_{1,2} - \omega_C$. In the case where the transmon anharmonicity $K_C$ is small compared to the detunings $\abs{\delta_{1,2,A,B}}$, the expressions above reduce to those obtained based on perturbative multiwave frequency mixing \cite{Gao2018}. Note that the interactions strengths presented in the main text and the rest of the supplementary material refer to the values of $g_{\textrm{BS},0}$ and $g_{\textrm{sq},i,0}$.

For weak drives, Eqs. (S20, S21) show that the strengths of the engineered beamsplitter and single-mode squeezing increase linearly with both drive amplitudes $\Omega_{1,2}$. For strong drives, this dependence becomes nonlinear in $\Omega_{1,2}$ and can be accurately captured using Floquet theory for the driven transmon \cite{Zhang2019}. We have verified that the experimentally measured beamsplitter and single-mode squeezing rates match the expressions (S20, S21) for weak drives and the full Floquet analysis at strong drives.

\subsection{Transmon-induced cavity Kerr}
Another important effect and a source of infidelity is the cavity nonlinearity induced by the transmons. To fourth order in the cavity-transmon coupling, this nonlinearity is a Kerr nonlinearity and has the following form:
\begin{equation}
\hat{H}_\textrm{Kerr} / \hbar = \sum_m (-\frac{K_{A,m}}{2}\hat{c}^{\dagger 2}_A\hat{c}^2_A - \frac{K_{B,m}}{2}\hat{c}^{\dagger 2}_B\hat{c}^2_B - K_{AB,m}\hat{c}^\dagger_A\hat{c}_A\hat{c}^\dagger_B\hat{c}_B) \otimes \ket{\Psi_m}\bra{\Psi_m} \tag{S22}
\end{equation}
where $K_{A,m}$ and $K_{B,m}$ are the self-Kerr of cavities A and B and $K_{AB,m}$ is the cross-Kerr between cavities A and B when the transmon is in the state $\ket{\Psi_m}$.

First, we consider the case in the absence of pumps. Of interest to us is the cavity Kerr when the transmon is in the ground state $\ket{0}$:
\begin{gather}
K_{i,0} = 2K_C\abs{\frac{g_i}{\delta_i}}^4\frac{\delta_i}{2\delta_i + K_C} = \frac{\chi^2_{iC}}{2K_C}\frac{(\delta_i + K_C)^2}{\delta_i(2\delta_i + K_C)}, \quad i \in \{A, B\} \tag{S23} \\
K_{AB,0} = 2\abs{\frac{g_A}{\delta_A}\frac{g_B}{\delta_B}}^2\frac{K_C(\delta_A + \delta_B)}{\delta_A + \delta_B + K_C} = \frac{\chi_{AC}\chi_{BC}}{2K_C}\frac{(\delta_A + K_C)(\delta_B + K_C)}{\delta_A \delta_B}\frac{\delta_A + \delta+B}{\delta_A + \delta_B + K_C} \tag{S24}
\end{gather}

Also of interest to us is the difference between   and $K_{i,0}$ and $K_{i,1}$. This difference leads to a nonlinear dependence of the transmon transition frequency on the cavity photon number. This difference is usually denoted as:
\begin{equation}
\chi'_{iC} = \frac{K_{i,0}-K_{i,1}}{2} = \frac{\chi^2_{iC}}{\delta_i}f(\delta_1/K_C), \quad i \in \{A, B\} \tag{S25}
\end{equation}
where $f(x) = (18x^3 + 30x^2 + 22x + 6)/(4(x+1)(4x^2+8x+3))$. We note that there is also a contribution to $\chi'_{iC}$ from a term in the sixth order expansion of the transmon cosine potential, but for $\omega_C \gg \abs{\delta_i}$ this correction is negligible.

Here we have only considered the cavity Kerr induced by the coupler transmon. In general, the transmon ancillas in modules A and B also induce Kerr in their respective cavities. The total Kerr of each cavity will then be the sum of all contributions. 

In the presence of pumps on the coupler transmon, the cavity Kerr can be strongly modified due to a relatively strong hybridization between cavity photons and excitations of the coupler transmon. To illustrate this effect, we consider as an example the pumps used in generating the beamsplitter interaction between the two cavities. For the choice of pumps used in the experiment, the sum of the frequency of cavity A and the higher-frequency pump is close to the frequency of transition from transmon ground to the second excited state: $\omega_A + \omega_2 \approx \omega_{02}$. As a result, the cavity photons become relatively strongly hybridized with the second excited state of the transmon, thus modifying their nonlinearity. Using a sixth-order perturbation theory (fourth-order in $g_A$ and second order in $\Omega_2$), we find that the modification to the cavity Kerr is:
\begin{equation}
\delta K_{A,0} \approx 2K_C\abs{\frac{\Omega_2}{\delta_2}}^2\frac{\chi^2_{AC}}{\Delta^2}\frac{(2\delta_2 + K_C)\delta_2}{(\delta_2 + K_C)(\delta_2 - K_C)} \tag{S26}
\end{equation}
where $\Delta = \omega_A + \omega_2 - \omega_{02}$, and $\omega_{02}$ is the Stark shifted transmon transition frequency from the ground to the second excited state. The above expression, which applies for small $\abs{\Delta}$, qualitatively captures the observed enhanced self Kerr of cavity A in the experiment during the beamsplitter operation. Comparing this expression with that of the bare cavity Kerr $K_{A,0}$ without pumps, we see that $\delta K_{A,0}$ becomes comparable to $K_{A,0}$ when $K_C\abs{\Omega_2/\delta_2} \sim \abs{\Delta}$. We note that such dependence of the cavity Kerr on the drive parameters also potentially provides a knob to control the cavity Kerr for the purpose of simulating nonlinear bosonic modes.

\section{System characterization}

\subsection{Calibration of Gaussian operations}
In the dispersive regime, the transition frequency $\omega^l_{t_i}$ of ancilla $\hat{t}_i$ depends on the photon number $l$ in the respective cavity:
\begin{equation}
\omega^l_{t_i} = \omega^0_{t_i}-l\chi_i+(l^2-l)\frac{\chi'_i}{2} \tag{S27}
\end{equation}
where $\omega^0_{t_i}$ is the ancilla frequency when there are no photons in its respective cavity and $\chi_i$ and $\chi'_i$ are the dispersive shifts originating from fourth and sixth order Hamiltonian terms, respectively, as introduced in the main text. Using this, the photon number population of each cavity can be extracted via $\pi$ pulses selective on each photon number after applying various strengths of each operation (FIG. 1). These populations are then fit to the corresponding expected models, including an overall offset and scaling factor to take into account errors due to ancilla relaxation and readout imperfections (TABLE III). For the beamsplitter, we assume an effective detuning   between cavities A and B in a frame where $\delta_\textrm{BS} = 0$ if the beamsplitter resonance condition is satisfied. Transmon heating leads to fluctuations in $\delta_\textrm{BS}$, which dephases the beamsplitter operation with a dephasing rate: $\kappa^\textrm{BS}_{ph} = \int_0^\infty \langle\delta_\textrm{BS}(t)\delta_\textrm{BS}(0)\rangle dt$. Thus, the oscillating populations of a single photon in each cavity $P_{10/01}$ is given to leading order in $\kappa_{A,B}/g_\textrm{BS}$ and $\kappa^\textrm{BS}_{ph} / g_\textrm{BS}$ by the expression in TABLE III, where $\bar{\kappa} = (\kappa^\textrm{BS}_A + \kappa^\textrm{BS}_B)/2$ and $\kappa^\textrm{BS}_{A,B}$ are the effective linewidths of cavities A and B during the beamsplitter operation.

\subsection{Measurement of system parameters}
Static and pump-induced self-Kerr Hamiltonian terms, $-\frac{K_A}{2}\hat{c}^{\dagger 2}_A\hat{c}^2_A$ and $-\frac{K_B}{2}\hat{c}^{\dagger 2}_B\hat{c}^2_B$ , are estimated using the protocol detailed in \cite{ChouThesis} (FIG. 2). For the pump-induced cases, the one of the two pumps are detuned by $\delta$ = 20 kHz and 50 kHz for squeezing and beamsplitter operations, respectively, to make the engineered interaction off-resonant. We assume that the induced self-Kerr is not a strong function of this detuning.

Static and pump-induced cavity decay rates are measured via $T_1$ experiments (FIG. 3). A single photon is prepared in each cavity, followed by either a delay or an off-resonant pumped operation (with the same detunings as above). Again, we assume that the pumped-induced decay rates are not a strong function of the pump detuning. The ancillas are then flipped via selective $\pi$ rotations conditioned on $n$ = 1 photon. In both cases, the data is post-selected on the ancilla being in the ground state before the selective $\pi$ rotation. We attribute the higher decay rate to the hybridization of the cavities with the shorter-lived coupler transmon. Measured cavity Kerr and $T_1$ values are given in TABLE IV.

\clearpage
\begin{figure}
	\includegraphics[scale=0.75]{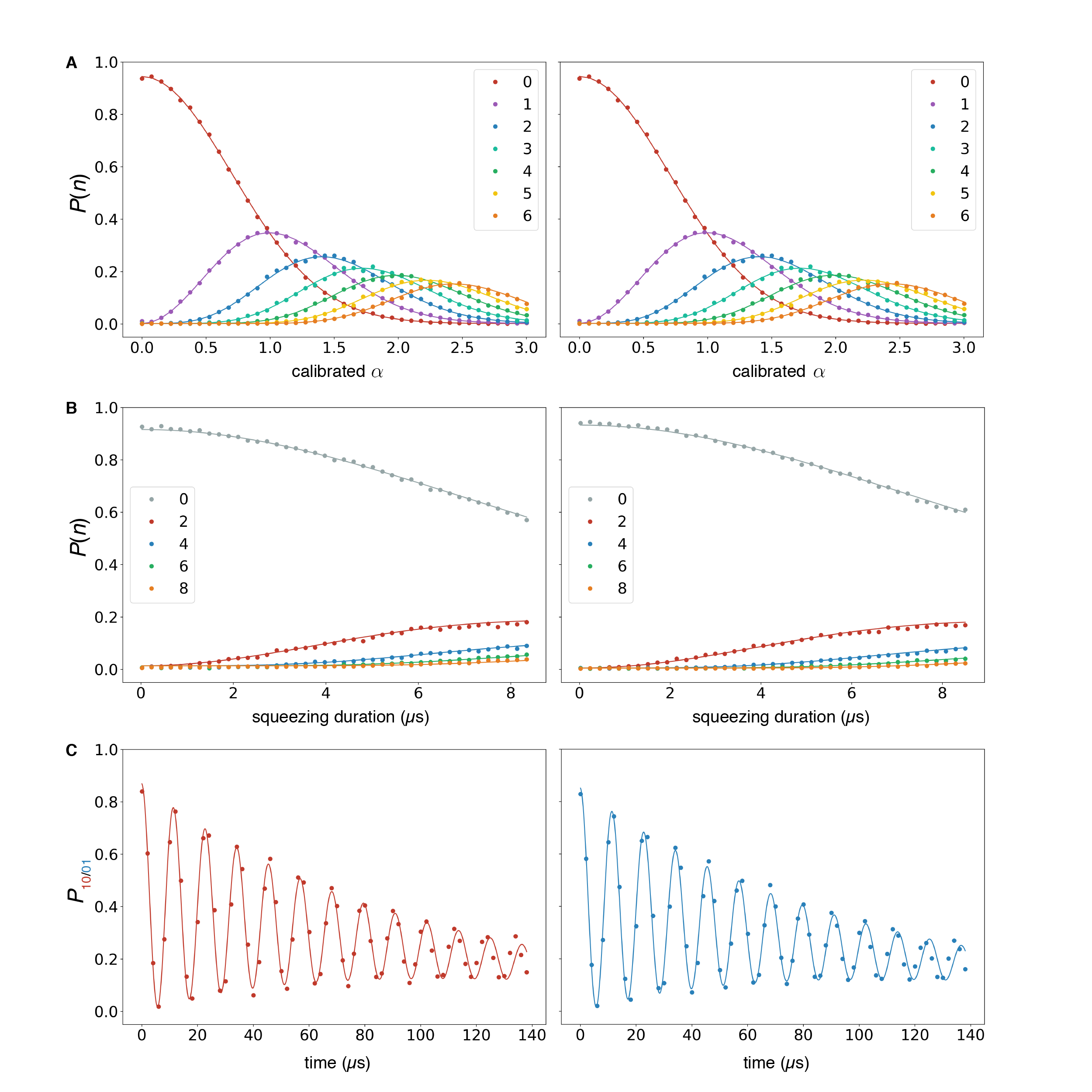}
	\caption{\textbf{Calibrations for Gaussian operations.} A) Displacement calibrations for cavities A (left) and B (right) starting in vacuum. Here, the amplitude of a resonant pulse is varied. B) Squeezing calibration for cavities A (left) and B (right) starting in vacuum. Here, the length of two squeezing pump tones is varied. Legends indicate population in photon number $n$. C) Beamsplitter calibration for beginning with a single photon in cavities A (left, $\ket{\psi_0} = \ket{1,0}$) and B (right, $\ket{\psi_0} = \ket{0,1}$). The length of two beamsplitter pump tones is varied, and the probability that the photon remains in the cavity that it started in is plotted over time.}
\end{figure}

\clearpage
\begin{figure}
	\includegraphics[scale=0.75]{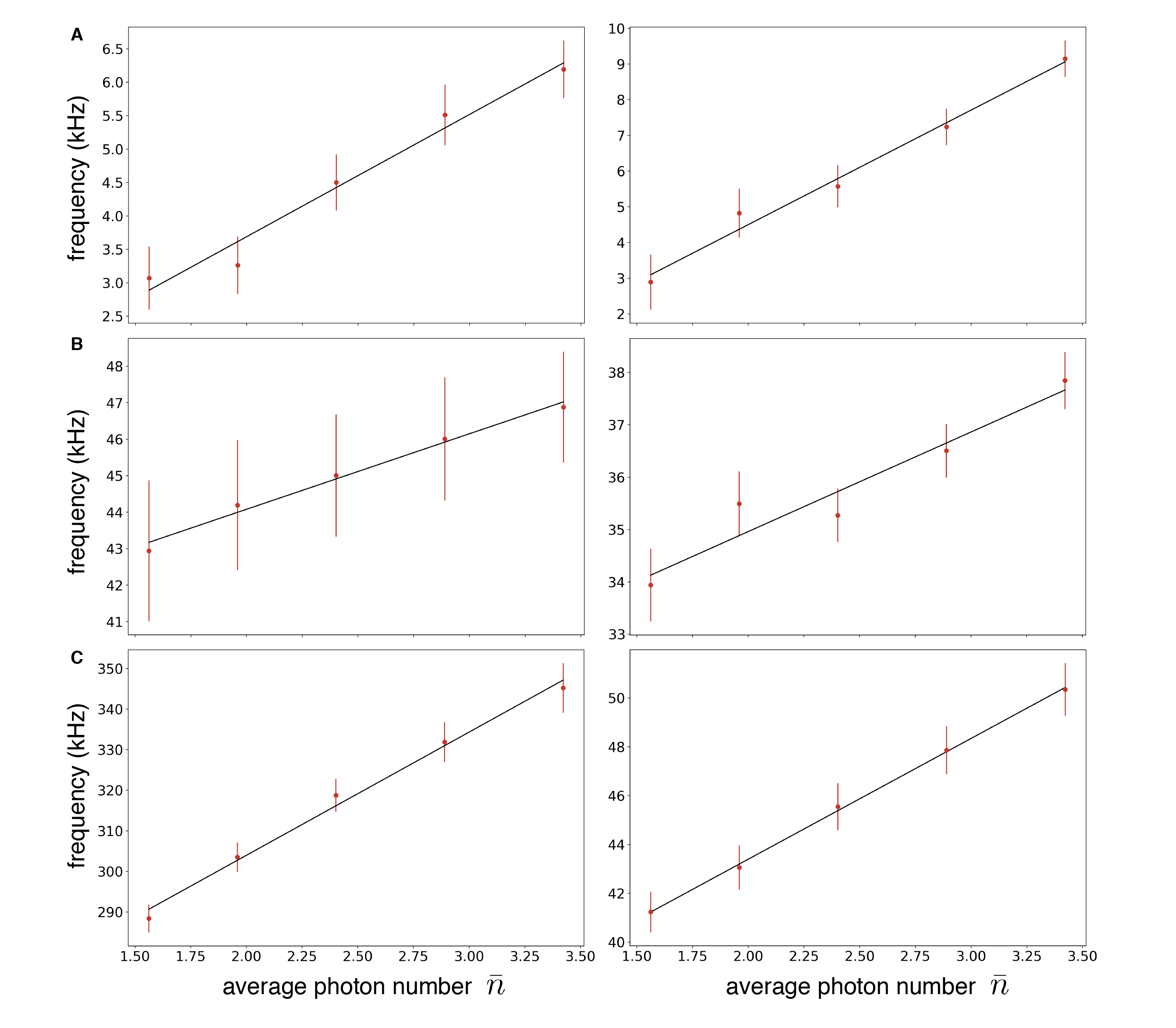}
	\caption{\textbf{Estimation of intrinsic $\&$ pump-induced self-Kerr.} In all plots, the y-coordinate corresponds to the effective frequency of a coherent state with average photon number $\bar{n}$. The slope determines the self-Kerr, and the offsets reflect pump-induced stark shifts. Experiments for estimating the self-Kerr in the A) absence of pumps, B) presence of off-resonant squeezing pumps, and C) presence of off-resonant beamsplitter pumps for cavity A (left panels) and B (right panels).}
\end{figure}

\clearpage
\begin{figure}
	\includegraphics[scale=0.75]{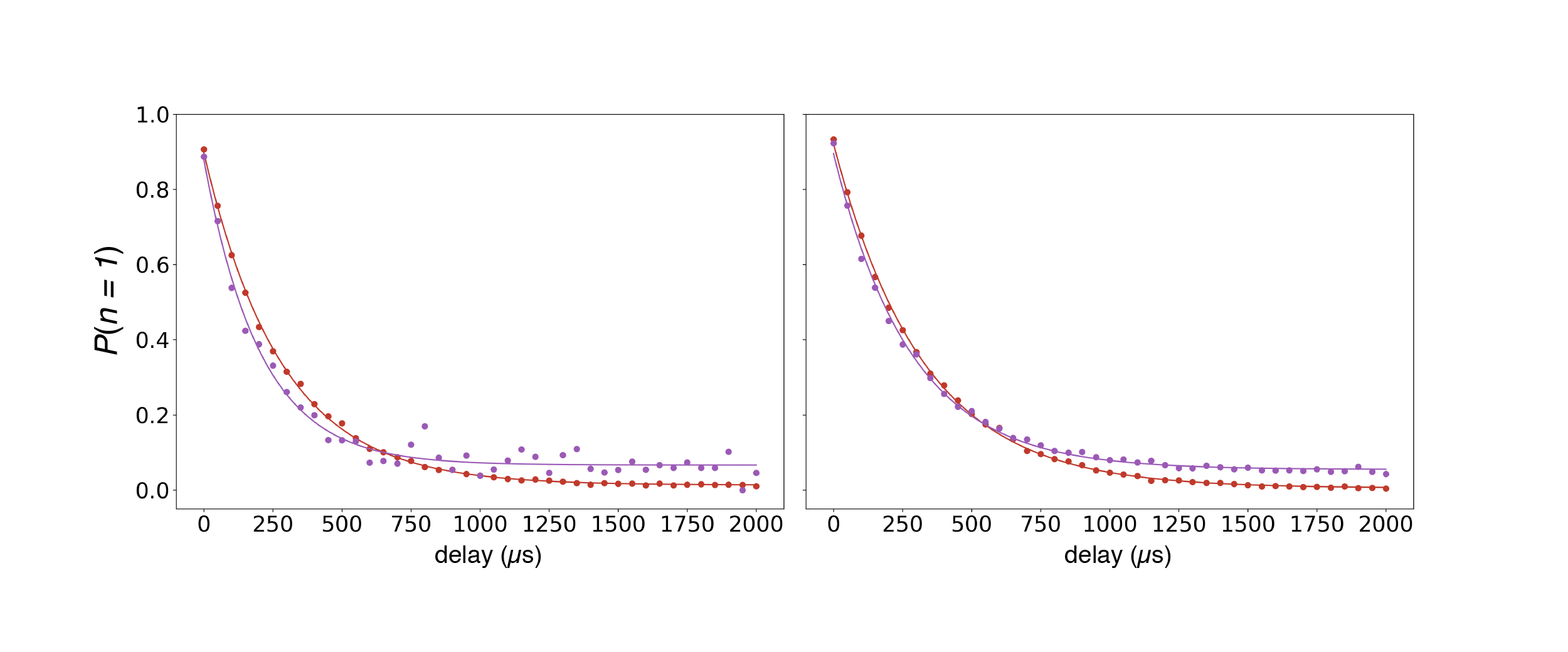}
	\caption{\textbf{Measurement of intrinsic $\&$ pump-induced decay rates.} Cavity $T_1$ experiments with either a varying delay (red) or the application of an off-resonant squeezing operation (purple) during the delay for cavities A (left) and B (right).}
\end{figure}

\begin{table*}
\begin{tabular}{|c|c|c|c|}
        \hline 
        \textbf{Operation} & \textbf{Model} & \textbf{Cavity} &  \textbf{Calibrated rate}\\ \hline
\multirow{2}{*}{Displacement} & \multirow{2}{*}{$P(l) = \frac{\abs{\alpha}^{2l}}{l!}\textrm{exp}(-\abs{\alpha}^2)$} & A & $\tau_{\alpha = 1} = 72 \,  \textrm{ns}$ \\ \cline{3-4}
      &  & B & $\tau_{\alpha = 1} = 72 \, \textrm{ns}$ \\ \hline
\multirow{2}{*}{Squeezing} & \multirow{2}{*}{$P(2l) = \frac{(2l)!}{2^{2l}(l!)^2}\frac{\textrm{tanh}^{2l}(2g_\textrm{sq}t)}{\textrm{cosh}(2g_\textrm{sq}t)}$} & A & $g_\textrm{sq} \approx 60 \, \textrm{kHz}$ \\ \cline{3-4}
      &  & B & $g_\textrm{sq} \approx 60 \, \textrm{kHz}$\\ \hline
\multirow{2}{*}{Beamsplitter} & $P_{10/01} = \frac{1}{2}\textrm{exp}\big(-\bar{\kappa}(t-t_0)\big)\times$ & A & \multirow{2}{*}{$g_\textrm{BS} \approx 2\pi \times 44 \, \textrm{kHz}$} \\ \cline{3-3}
      & $\big(1 + \textrm{exp}(-\kappa^\textrm{BS}_{ph} (t - t_0)/2)\textrm{cos}(2g_\textrm{BS}(t-t_0))\big)$ & B & \\ \hline
\end{tabular}
\caption{\textbf{Calibrated rates of Gaussian operations.} The amplitude of a displacement operation of fixed length $\tau$ = 72 ns is calibrated for generating a coherent state with   (FIG. 1(A)). The rates for the squeezing and beamsplitter operations are extracted from the fits (FIG. 1(B) $\&$ (C)).}
\end{table*}

\begin{table*}
\begin{tabular}{|c|c|c|c|}
        \hline 
        \textbf{Cavity} & \textbf{Operation} & $K/2\pi$ (kHz) &  $T_1$ ($\mu$s) \\ \hline
\multirow{3}{*}{A} & Native & 1.8 & 280 \\ \cline{2-4}
      & Squeezing & 2 & 200 \\ \cline{2-4}
      & Beamsplitter & 30 & 170 \\ \hline
\multirow{3}{*}{B} & Native & 3.2 & 320 \\ \cline{2-4}
      & Squeezing & 1.9 & 280 \\ \cline{2-4}
      & Beamsplitter & 5 & 170 \\ \hline
\end{tabular}
\caption{\textbf{Estimated cavity Kerr and} $T_1$  \textbf{values.} The beamsplitter decay rates are extracted from the fit performed in the calibration of the operation in FIG. 1(C) assuming that $\kappa^\textrm{BS}_A = \kappa^\textrm{BS}_B = \bar{\kappa}$.}
\end{table*}

\clearpage
\section{Circuit Implementation}

\begin{figure}
	\includegraphics[scale=1.0]{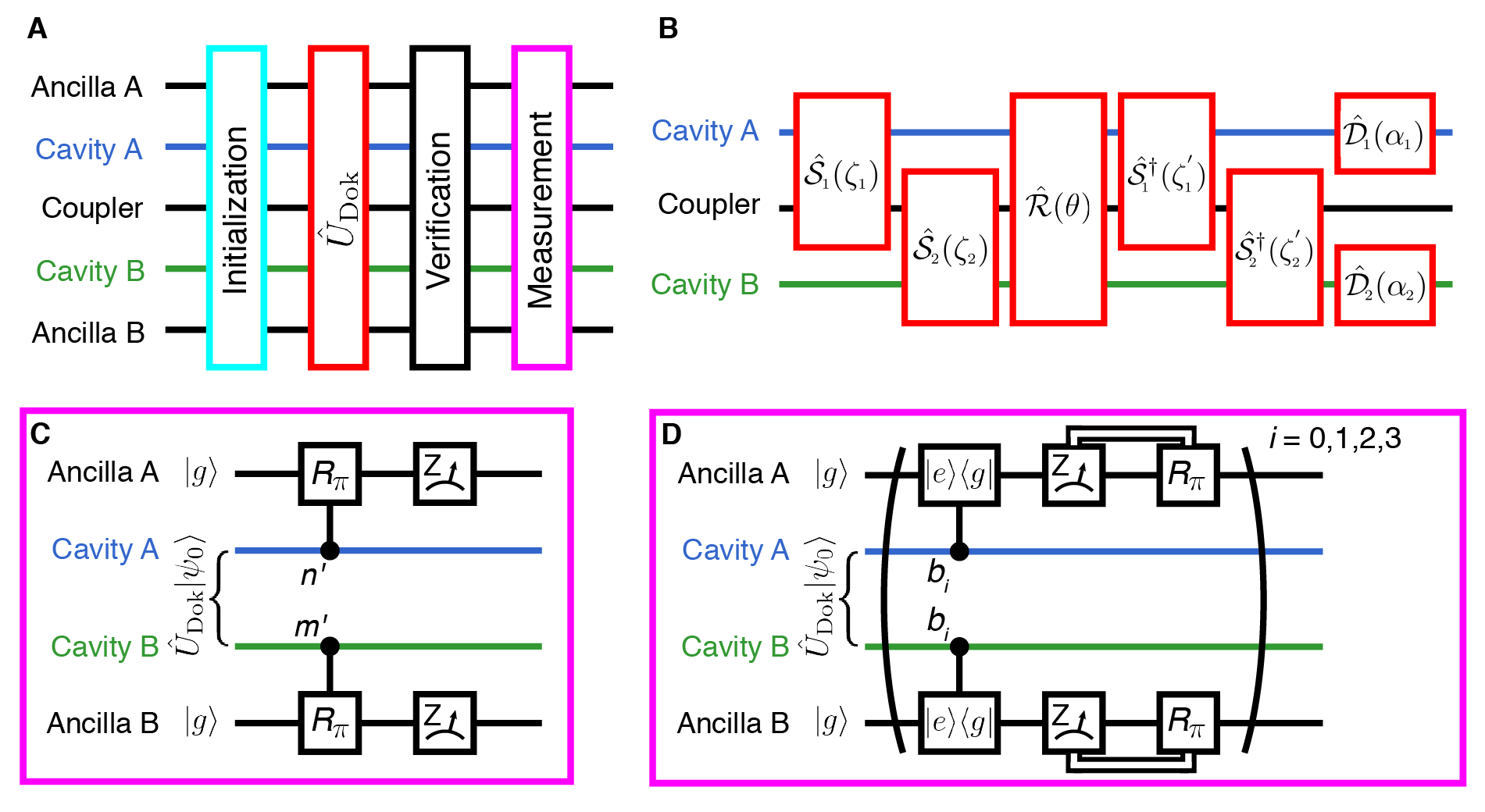}
	\caption{\textbf{Circuit implementation of the Franck-Condon simulation.} A) Overview of the quantum simulation algorithm, consisting of state preparation, unitary Doktorov transformation, and measurement. A set of verification measurements is performed after the unitary Doktorov transformation for the purpose of post-selecting the final data on measuring the transmons in their ground state. B) The two-mode circuit decomposition of the Doktorov transformation used in this experiment. The nonlinearity of the coupler transmon is primarily utilized to perform all three pumped operations, though in principle that of the ancilla transmons could have been used as well. C) \textit{Single-bit extraction.} Selective $\pi$-pulses $(R_\pi)$ flip each ancilla transmon conditioned on having $n'$ and $m'$ photons in cavities A and B, respectively, for a given run of the experiment. The ancillas are then simultaneously read out using standard dispersive techniques. Subsequent runs of the experiment thus need to scan $n'$ and $m'$ over the photon number range of interest up to the desired $n_\textrm{max}$. D) \textit{Sampling.} Optimal control pulses are designed to excite each ancilla transmon from $\ket{g}$ to $\ket{e}$ conditioned on the value of the binary bits $b_i$ of each cavity state, followed by dispersive readouts. Here, we measure the first 4 bits on a given run of the experiment, thus resolving the first 16 Fock states for each cavity. Real-time feedforward control is used to dynamically reset the state of the ancilla in between bit measurements to minimize errors due to ancilla relaxation.}
\end{figure}

The full quantum circuit implemented in our experiment is shown in FIG. 4. State preparation in our experiment (FIG. 5) consists of first performing measurement-based feedback cooling of all modes to their ground state (this protocol is described in full detail in the supplement of \cite{Elder2019}). For preparing Fock states, optimal control pulses are then played that perform the following state transfers:
\begin{align}
\ket{0}_A \otimes \ket{g}_{t_A} & \rightarrow \ket{n}_A \otimes \ket{g}_{t_A} \nonumber \\
\ket{0}_B \otimes \ket{g}_{t_B} & \rightarrow \ket{m}_B \otimes \ket{g}_{t_B} \tag{S28}
\end{align}
These state transfers, however, suffer a finite error probability on the order of a few percent due to decoherence during the operation. This error is suppressed by performing a series of QND measurements of each cavity photon number and post-selecting on outcomes that verify that the correct state was prepared. This is done via $k$ selective $\pi$ rotations on the ancilla transmons conditioned on the desired photon numbers in the cavities following by measurements, even if the desired state is joint vacuum. The final data is post-selected on the ancilla measurement outcomes being $(``e",``g")^{\otimes k/2}$ for both modules, where $k$ is chosen to be even. In our experiment, we choose $k$ = 2 for the ``single-bit extraction" measurement scheme and $k$ = 6 for the ``sampling" measurement scheme.

\begin{figure}
	\includegraphics[scale=0.75]{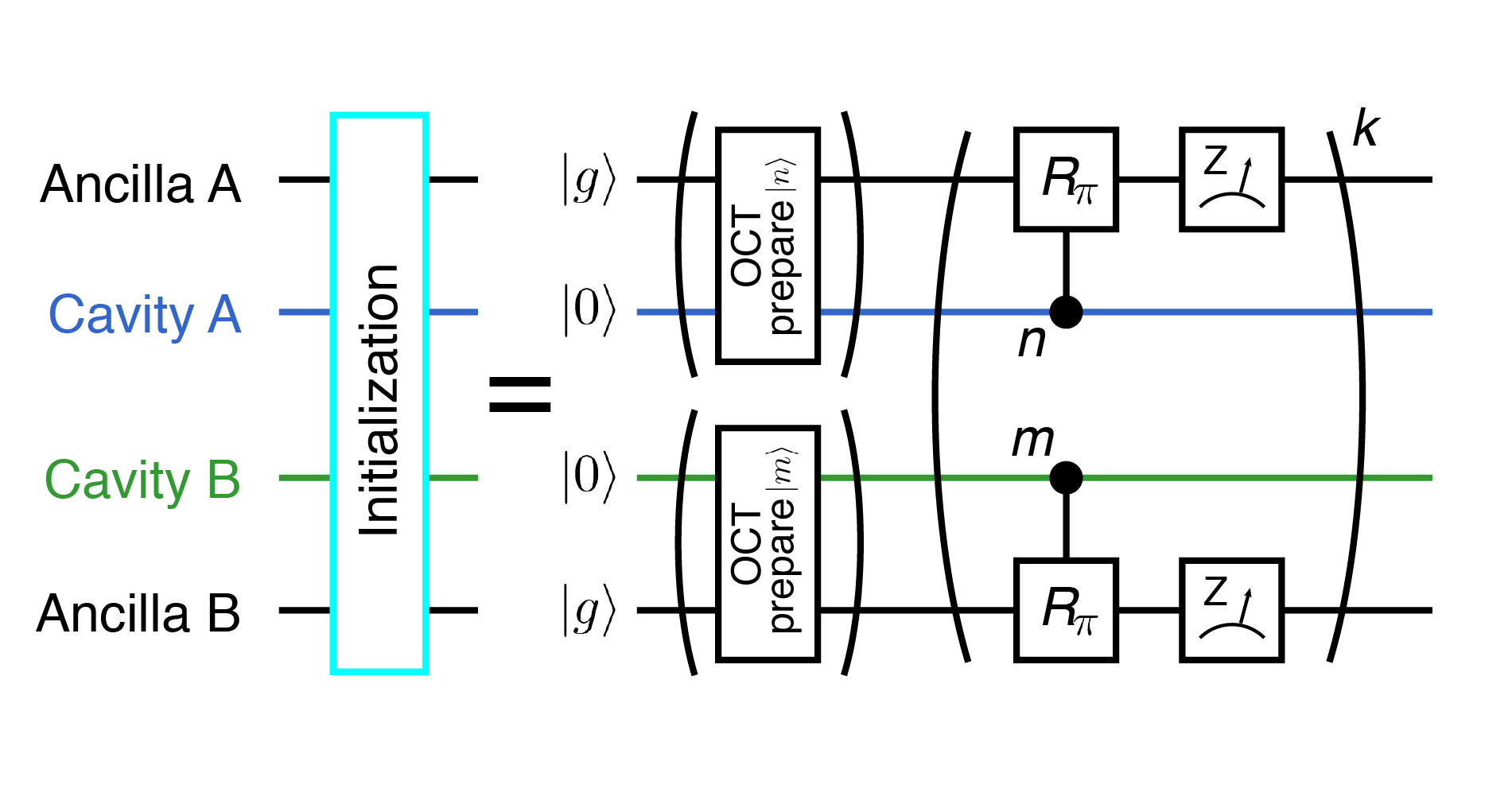}
	\caption{\textbf{Circuit implementation of heralded state preparation.} Measurement-based feedback cooling techniques prepare the full system in its ground state (i.e. both cavities in $\ket{0}$ and all transmons in $\ket{g}$). Optimal control pulses of the form listed in Table 1 of the main text are played simultaneously on each module to prepare a desired photon number state, followed by a set of $k$ check measurements.}
\end{figure}

\clearpage
\section{Correcting systematic errors due to transmon decoherence during single-bit extraction}
Errors due to ancilla decoherence during the ``single-bit extraction" measurement scheme may be systematically calibrated out. Specifically, decay and heating events during selective $\pi$ rotations and readout errors result in a systematic bias in the final estimate of the photon number population. For the case of a single ancilla qubit coupled to a single cavity, these effects result in a reduction of contrast for a Rabi experiment when both the ancilla and the cavity are prepared in their ground state (FIG. 6).

\begin{figure}
	\includegraphics[scale=0.3]{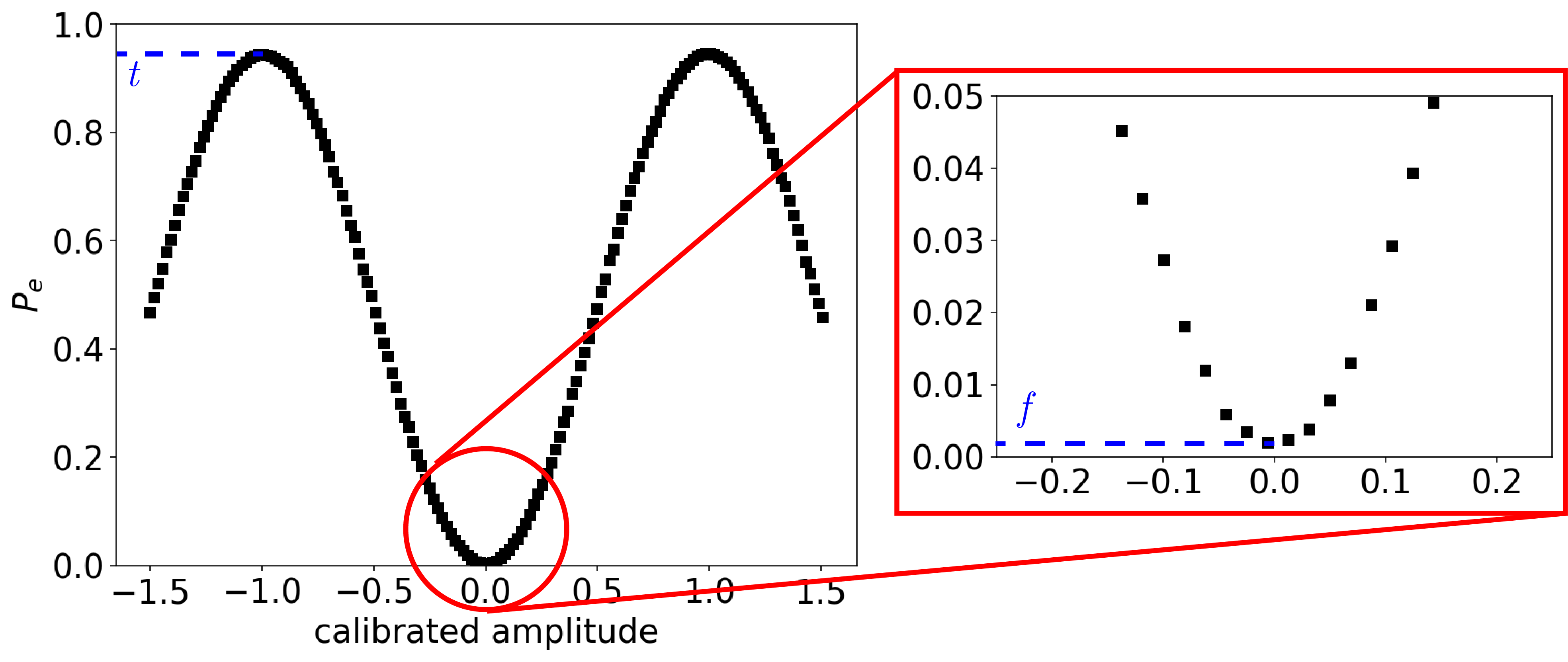}
	\caption{\textbf{Calibration of systematic measurement errors using selective ancilla pulses.} A standard Rabi calibration experiment of a selective pulse used for measurement (here shown for ancilla B). The maximum probability $t$ is limited by decoherence of the transmon during both the pulse and the readout. The floor $f$ is set by the probability of heating out of the ground state during both the pulse and the readout.}
\end{figure}

When using this pulse to infer cavity photon number populations, we assume that there is no photon number dependence to either the Rabi or decoherence rates of the ancilla. Under this model, we can relate the measured probabilities $\vec{Q}$ to the true probabilities $\vec{P}$ via:
\begin{equation}
\vec{P} = \frac{\vec{Q} - f}{t-f} \tag{S29}
\end{equation}
where $f$ and $t$ are the probabilities of assigning the ancilla measurement to the excited state when it is prepared in the ground and excited states, respectively. Thus, inferring the true probabilities from the measured probabilities is a relatively straightforward task.

For two modes, however, the problem becomes more complicated as a measurement of a joint probability relies on shot-by-shot correlations of the individual ancilla outcomes. Thus, false positive counts due to heating and readout errors lead to misassignment in a nonlinear fashion. We can again write what a given joint measured probability $Q_{nm}$ is in terms of the true distribution $P_{nm}$:
\begin{equation}
Q_{nm} = t_A t_B P_{nm} + t_A f_B P_{n\bar{m}} + f_A t_B P_{\bar{n}m} + f_A f_B P_{\overline{nm}} \tag{S30}
\end{equation}

Eq. (S30) may be solved for $P_{nm}$ by noting that:
\begin{align}
P_{\bar{n}m} & = \sum_k(1-\delta_{nk})P_{km} \nonumber \\
P_{n\bar{m}} & = \sum_l(1-\delta_{lm})P_{nl} \nonumber \\
P_{\overline{nm}} & = 1 - P_{nm} \tag{S31}
\end{align}
It is worth noting that this requires $Q_{nm}$ to be a square matrix, which translates to measuring both $n'$ and $m'$ up to a pre-specified $n_\textrm{max}$.

\clearpage
\section{Numerical Franck-Condon data}
Additional experimental data is provided in this section. TABLE V provides an overview of the different molecular processes that are simulated and corresponding information regarding post-selection, systematic offsets (see supplementary text V), and distance metrics. 

The data for each molecular process in the following tables is calculated as follows. For the ``single-bit extraction" scheme, the probability and standard error for a given joint photon number of interest   is:
\begin{align*}
q^\textrm{meas}_{n',m'} & = \frac{n^{ee}_{n',m'}}{N^\textrm{runs}_{n',m'}} \tag{S33} \\
\sigma_{n',m'} & = \sqrt{\frac{q^\textrm{meas}_{n',m'}(1 - q^\textrm{meas}_{n',m'})}{N^\textrm{runs}_{n',m'}}} \tag{S34}
\end{align*}
where $n^{ee}_{n',m'}$ is the number of counts where both ancillas are measured in their excited state, indicating a measure for population in $\ket{n',m'}$ (see Eq. 4 in the main text), and $N^\textrm{runs}_{n',m'}$ is the total number of runs of the experiment for probing $\ket{n',m'}$. The number of runs varies slightly among different final states due to varying post-selection probabilities. The correction protocol outlined in the supplementary text V is then applied to $q^\textrm{meas}_{n',m'}$ to retrieve a new probability distribution $p^\textrm{meas}_{n',m'}$. The standard error $\sigma_{n',m'}$ is truncated to one significant digit and $p^\textrm{meas}_{n',m'}$ is then rounded to the precision set by $\sigma_{n',m'}$. The data reported is $p^\textrm{meas}_{n',m'} \pm \sigma_{n',m'}$ only for probabilities with significant support relative to the precision of the experiment $(p^\textrm{ideal}_{n',m'} \gtrsim 10^{-4})$. 

The same method (sans the correction protocol) is applied to the data for the ``sampling" scheme, except there the probabilities and standard error are given by:
\begin{align*}
q^\textrm{meas}_{n',m'} & = \frac{n_{n',m'}}{N_\textrm{runs}} \tag{S35} \\
\sigma_{n',m'} & = \sqrt{\frac{q^\textrm{meas}_{n',m'}(1 - q^\textrm{meas}_{n',m'})}{N_\textrm{runs}}} \tag{S36}
\end{align*}
where $n_{n',m'}$ is the number of times the joint photon number $\ket{n',m'}$ is sampled from the total number of runs of the experiment $N_\textrm{runs}$.

Sometimes, there will be no counts reported for a given $\ket{n',m'}$ (i.e., $n^{ee}_{n',m'}$ or $n_{n',m'}$ = 0). In this case, we simply report a probability of zero. Furthermore, sometimes the correction protocol will return negative elements in the probability distribution due to statistical noise; these unphysical cases are also nulled and a zero is reported. All distances $D = \frac{1}{2}\sum_{i=0}^{n_\textrm{max}}\sum_{j=0}^{n_\textrm{max}}|p^\textrm{meas}_{ij} - p^\textrm{ideal}_{ij}|$ are calculated after this correction process, with the corresponding values for $n_\textrm{max}$ specified in TABLE V.

Full time-domain master equation simulations are performed using QuTiP and consider only the Hilbert space of the two cavities with $n_\textrm{max} = 30$. Each Gaussian operation is simulated by evolving the associated Hamiltonian term, while also including the corresponding self-Kerr terms and photon loss for each operation. While squeezing cavity A, for instance, the native self-Kerr and photon loss rates for cavity B are used, assuming that the pumped process for squeezing cavity A does not change the participation of cavity B in any nonlinear lossy modes (and vice-versa). The simulation also takes into account photon loss during the verification measurement, which takes 2.5 $\mu$s. The simulation does not consider imperfect state preparation and systematic errors in calibrations, which we believe to account for the remaining difference between the measured distances for the ``single-bit extraction" scheme and predicted distances from the master equation simulations.

\renewcommand{\arraystretch}{1.4}
\begin{table*}
\setlength{\tabcolsep}{5pt}
\begin{tabular}{|c|c|c|c|c|c|c|c|c|}
\hline
\multirow{3}{*}{\begin{tabular}[c]{@{}c@{}}Molecular \\ transition\end{tabular}} & \multirow{3}{*}{\begin{tabular}[c]{@{}c@{}}Initial state \\ $(n, m)$\end{tabular}} & \multirow{3}{*}{\begin{tabular}[c]{@{}c@{}}Percentage of \\ data kept\end{tabular}} & \multirow{3}{*}{$n_\textrm{max}$} & \multirow{3}{*}{\begin{tabular}[c]{@{}c@{}}$t_A, t_B$ \\ $f_A, f_B$\end{tabular}} & \multicolumn{3}{c|}{Distance to ideal distribution} & \multirow{3}{*}{Figure} \\ \cline{6-8}
 &  & & & & Single-bit & \multirow{2}{*}{Sampling} & Master & \\ 
 &  & & & & extraction &  & Equation & \\ \hline
 \begin{tabular}[c]{@{}c@{}}H$_2$O $\xrightarrow{h\nu}$ \\ H$_2$O$^+ (\tilde{B}^2$B$_2)$ + e$^-$\end{tabular} & $(0, 0)$ & $\sim 95\%$ & 16 & \begin{tabular}[c]{@{}c@{}} 0.937, 0.946 \\ 0.005, 0.002\end{tabular} & 0.049(1) & 0.151(9) & 0.0123 & FIG. 7 \\ \hline
 \multirow{3}{*}{O$^-_3 \xrightarrow{h\nu}$ O$_3$ + e$^-$}& $(0, 0)$ & $\sim 96\%$ & 12 & \begin{tabular}[c]{@{}c@{}} 0.937, 0.948 \\ 0.005, 0.002\end{tabular} & 0.039(9) & 0.075(2) & 0.0052 & FIG. 8 \\ \cline{2-9}
 & $(1, 0)$ & $\sim 95\%$ & 10 & \begin{tabular}[c]{@{}c@{}} 0.937, 0.950 \\ 0.004, 0.002\end{tabular} & 0.057(5) & 0.085(5) & 0.0131 & FIG. 9 \\ \cline{2-9}
 & $(1, 2)$ & $\sim 93\%$ & 12 & \begin{tabular}[c]{@{}c@{}} 0.938, 0.950 \\ 0.004, 0.001\end{tabular} & 0.105(3) & 0.148(4) & 0.0217 & FIG. 10 \\ \hline
 \multirow{2}{*}{NO$^-_2 \xrightarrow{h\nu}$ NO$_2$ + e$^-$}& $(0, 0)$ & $\sim 94\%$ & 12 & \begin{tabular}[c]{@{}c@{}} 0.935, 0.943 \\ 0.005, 0.003\end{tabular} & 0.034(0) & 0.110(9) & 0.0331 & FIG. 11 \\ \cline{2-9}
 & $(1, 0)$ & $\sim 92\%$ & 14 & \begin{tabular}[c]{@{}c@{}} 0.934, 0.951 \\ 0.004, 0.002\end{tabular} & 0.202(2) & 0.209(7) & 0.1269 & FIG. 12 \\ \hline
 \multirow{2}{*}{SO$_2 \xrightarrow{h\nu}$ SO$^+_2$ + e$^-$}& $(0, 0)$ & $\sim 96\%$ & 12 & \begin{tabular}[c]{@{}c@{}} 0.938, 0.950 \\ 0.004, 0.002\end{tabular} & 0.019(6) & 0.095(3) & 0.0065 & FIG. 13 \\ \cline{2-9}
 & $(0, 1)$ & $\sim 94\%$ & 12 & \begin{tabular}[c]{@{}c@{}} 0.931, 0.951 \\ 0.004, 0.001\end{tabular} & 0.063(7) & 0.136(6) & 0.0213 & FIG. 14 \\ \hline
 \end{tabular}
 \caption{\textbf{Summary of experimental data.} List of molecular processes simulated and corresponding post-selection probabilities, maximum photon number probed with the ``single-bit extraction" measurement scheme, and distances. Transmon offsets are independently measured after each dataset is taken.}
\end{table*}
 
\clearpage
\begin{figure}
	\includegraphics[scale=0.5]{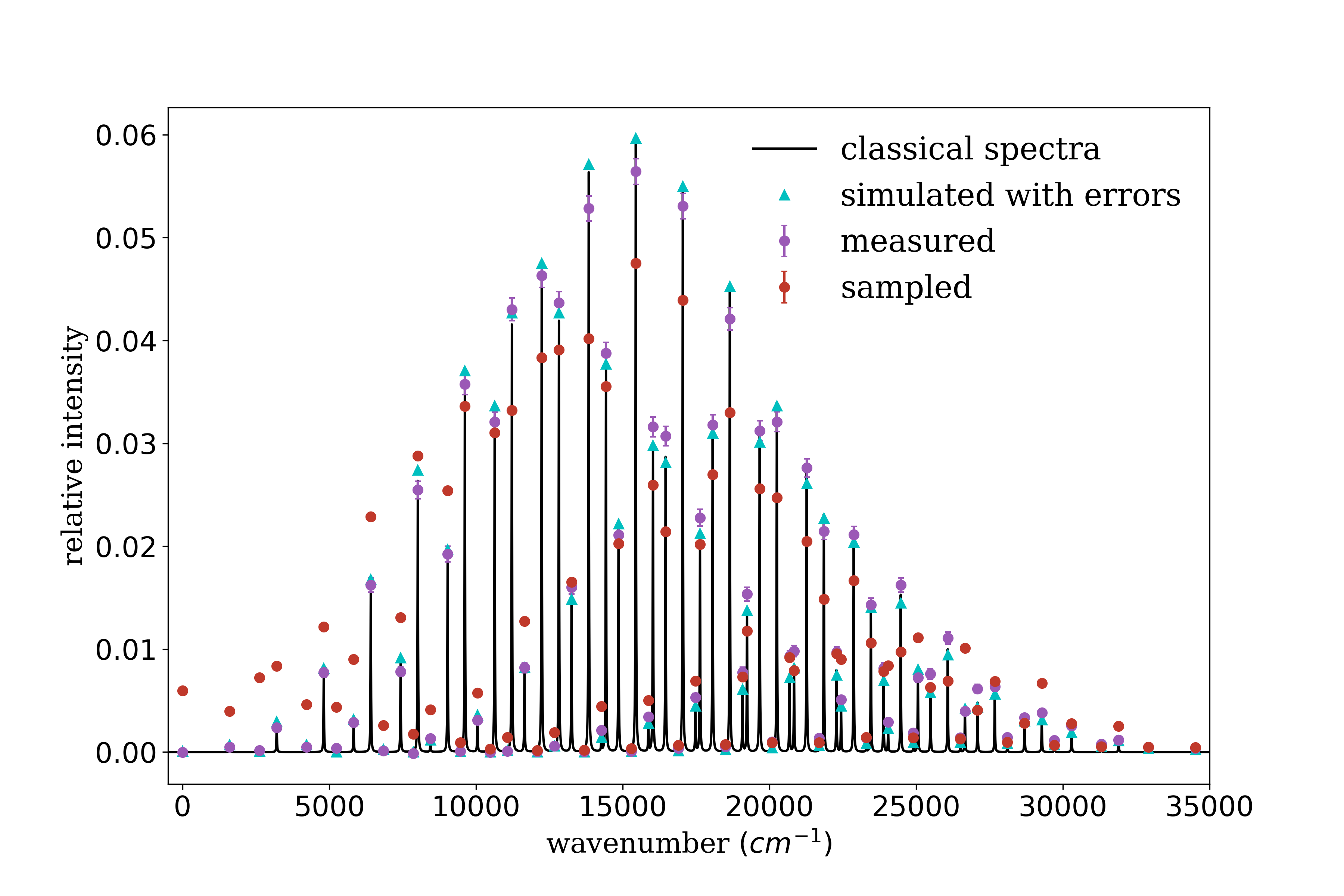}
	\caption{Photoionization of neutral water to the $(\tilde{B}^2\textrm{B}_2)$ excited state of the cation starting in the vibrationless state $n$ = 0, $m$ = 0.}
\end{figure}

\setlength{\tabcolsep}{8pt}
\renewcommand{\arraystretch}{1.5}
\begin{longtable}{|c|c|c|c|c|}
\hline
& \multicolumn{4}{c|}{H$_2$O $\xrightarrow{h\nu}$ H$_2$O$^+ (\tilde{B}^2$B$_2)$ + e$^-$ starting in ($n$ = 0, $m$ = 0)} \\ \hline
$(n',m')$ & \begin{tabular}[c]{@{}c@{}}Classically \\ calculated \end{tabular} & \begin{tabular}[c]{@{}c@{}}Master equation \\ simulation \end{tabular} & \begin{tabular}[c]{@{}c@{}}Single-bit \\ extraction \end{tabular} & Sampling \\ \hline
(0,0)   & 7.92E-05    & 8.26E-05   & 0               & 0.006 $\pm$ 0.0001   \\ \hline
(0,1)   & 6.67E-04    & 7.01E-04   & 0.0005 $\pm$ 0.0002 & 0.00396 $\pm$ 0.0001 \\ \hline
(0,2)   & 2.80E-03    & 2.94E-03   & 0.0024 $\pm$ 0.0003 & 0.0084 $\pm$ 0.0001  \\ \hline
(0,3)   & 7.76E-03    & 8.15E-03   & 0.0077 $\pm$ 0.0005 & 0.0122 $\pm$ 0.0002  \\ \hline
(0,4)   & 1.60E-02    & 1.68E-02   & 0.0162 $\pm$ 0.0007 & 0.0229 $\pm$ 0.0002  \\ \hline
(0,5)   & 2.64E-02    & 2.74E-02   & 0.0255 $\pm$ 0.0008 & 0.0288 $\pm$ 0.0003  \\ \hline
(0,6)   & 3.59E-02    & 3.70E-02   & 0.0357 $\pm$ 0.001  & 0.0336 $\pm$ 0.0003  \\ \hline
(0,7)   & 4.16E-02    & 4.26E-02   & 0.043 $\pm$ 0.001   & 0.0332 $\pm$ 0.0003  \\ \hline
(0,8)   & 4.19E-02    & 4.27E-02   & 0.044 $\pm$ 0.001   & 0.0391 $\pm$ 0.0003  \\ \hline
(0,9)   & 3.73E-02    & 3.77E-02   & 0.039 $\pm$ 0.001   & 0.0355 $\pm$ 0.0003  \\ \hline
(0,10)  & 2.96E-02    & 2.98E-02   & 0.0316 $\pm$ 0.0009 & 0.026 $\pm$ 0.0003   \\ \hline
(0,11)  & 2.13E-02    & 2.12E-02   & 0.0228 $\pm$ 0.0008 & 0.0202 $\pm$ 0.0002  \\ \hline
(0,12)  & 1.39E-02    & 1.38E-02   & 0.0154 $\pm$ 0.0006 & 0.0118 $\pm$ 0.0002  \\ \hline
(0,13)  & 8.33E-03    & 8.20E-03   & 0.0098 $\pm$ 0.0005 & 0.0079 $\pm$ 0.0001  \\ \hline
(0,14)  & 4.60E-03    & 4.50E-03   & 0.0051 $\pm$ 0.0004 & 0.009 $\pm$ 0.0001   \\ \hline
(0,15)  & 2.36E-03    & 2.29E-03   & 0.0029 $\pm$ 0.0003 & 0.0084 $\pm$ 0.0001  \\ \hline
(1,0)   & 7.82E-05    & 8.11E-05   & 0.0002 $\pm$ 0.0002 & 0.0072 $\pm$ 0.0001  \\ \hline
(1,1)   & 6.91E-04    & 7.19E-04   & 0.0005 $\pm$ 0.0002 & 0.0046 $\pm$ 0.0001  \\ \hline
(1,2)   & 3.03E-03    & 3.16E-03   & 0.0029 $\pm$ 0.0003 & 0.009 $\pm$ 0.0001   \\ \hline
(1,3)   & 8.81E-03    & 9.14E-03   & 0.0078 $\pm$ 0.0005 & 0.0131 $\pm$ 0.0002  \\ \hline
(1,4)   & 1.91E-02    & 1.97E-02   & 0.0192 $\pm$ 0.0007 & 0.0254 $\pm$ 0.0002  \\ \hline
(1,5)   & 3.28E-02    & 3.36E-02   & 0.0321 $\pm$ 0.0009 & 0.031 $\pm$ 0.0003   \\ \hline
(1,6)   & 4.66E-02    & 4.75E-02   & 0.046 $\pm$ 0.001   & 0.0383 $\pm$ 0.0003  \\ \hline
(1,7)   & 5.63E-02    & 5.71E-02   & 0.053 $\pm$ 0.001   & 0.0402 $\pm$ 0.0003  \\ \hline
(1,8)   & 5.92E-02    & 5.97E-02   & 0.056 $\pm$ 0.001   & 0.0475 $\pm$ 0.0003  \\ \hline
(1,9)   & 5.49E-02    & 5.50E-02   & 0.053 $\pm$ 0.001   & 0.0439 $\pm$ 0.0003  \\ \hline
(1,10)  & 4.55E-02    & 4.53E-02   & 0.042 $\pm$ 0.001   & 0.033 $\pm$ 0.0003   \\ \hline
(1,11)  & 3.40E-02    & 3.37E-02   & 0.0321 $\pm$ 0.0009 & 0.0247 $\pm$ 0.0002  \\ \hline
(1,12)  & 2.32E-02    & 2.28E-02   & 0.0214 $\pm$ 0.0008 & 0.0149 $\pm$ 0.0002  \\ \hline
(1,13)  & 1.44E-02    & 1.41E-02   & 0.0143 $\pm$ 0.0006 & 0.0106 $\pm$ 0.0002  \\ \hline
(1,14)  & 8.30E-03    & 8.04E-03   & 0.0073 $\pm$ 0.0005 & 0.0111 $\pm$ 0.0002  \\ \hline
(1,15)  & 4.42E-03    & 4.25E-03   & 0.004 $\pm$ 0.0004  & 0.0101 $\pm$ 0.0002  \\ \hline
(2,0)   & 2.61E-05    & 2.68E-05   & 0.0004 $\pm$ 0.0001 & 0.0044 $\pm$ 0.0001  \\ \hline
(2,1)   & 2.47E-04    & 2.52E-04   & 0.0001 $\pm$ 0.0001 & 0.00261 $\pm$ 8E-05  \\ \hline
(2,2)   & 1.15E-03    & 1.17E-03   & 0.0013 $\pm$ 0.0002 & 0.0041 $\pm$ 0.0001  \\ \hline
(2,3)   & 3.57E-03    & 3.60E-03   & 0.0031 $\pm$ 0.0003 & 0.0057 $\pm$ 0.0001  \\ \hline
(2,4)   & 8.19E-03    & 8.21E-03   & 0.0082 $\pm$ 0.0005 & 0.0127 $\pm$ 0.0002  \\ \hline
(2,5)   & 1.49E-02    & 1.48E-02   & 0.016 $\pm$ 0.0007  & 0.0165 $\pm$ 0.0002  \\ \hline
(2,6)   & 2.25E-02    & 2.22E-02   & 0.0211 $\pm$ 0.0008 & 0.0203 $\pm$ 0.0002  \\ \hline
(2,7)   & 2.87E-02    & 2.81E-02   & 0.0307 $\pm$ 0.0009 & 0.0214 $\pm$ 0.0002  \\ \hline
(2,8)   & 3.19E-02    & 3.10E-02   & 0.0318 $\pm$ 0.0009 & 0.027 $\pm$ 0.0003   \\ \hline
(2,9)   & 3.12E-02    & 3.01E-02   & 0.0312 $\pm$ 0.0009 & 0.0256 $\pm$ 0.0002  \\ \hline
(2,10)  & 2.72E-02    & 2.61E-02   & 0.0276 $\pm$ 0.0009 & 0.0205 $\pm$ 0.0002  \\ \hline
(2,11)  & 2.14E-02    & 2.04E-02   & 0.0211 $\pm$ 0.0008 & 0.0167 $\pm$ 0.0002  \\ \hline
(2,12)  & 1.53E-02    & 1.45E-02   & 0.0162 $\pm$ 0.0007 & 0.0097 $\pm$ 0.0002  \\ \hline
(2,13)  & 1.00E-02    & 9.46E-03   & 0.0111 $\pm$ 0.0006 & 0.0069 $\pm$ 0.0001  \\ \hline
(2,14)  & 6.02E-03    & 5.67E-03   & 0.0064 $\pm$ 0.0004 & 0.0069 $\pm$ 0.0001  \\ \hline
(2,15)  & 3.36E-03    & 3.14E-03   & 0.0038 $\pm$ 0.0003 & 0.0067 $\pm$ 0.0001  \\ \hline
(3,0)   & 2.78E-06    & 3.35E-06   & 0               & 0.00174 $\pm$ 7E-05  \\ \hline
(3,1)   & 2.97E-05    & 3.43E-05   & 8E-05 $\pm$ 8E-05   & 0.00091 $\pm$ 5E-05  \\ \hline
(3,2)   & 1.56E-04    & 1.74E-04   & 9E-05 $\pm$ 8E-05   & 0.00142 $\pm$ 6E-05  \\ \hline
(3,3)   & 5.37E-04    & 5.79E-04   & 0.0006 $\pm$ 0.0002 & 0.00189 $\pm$ 7E-05  \\ \hline
(3,4)   & 1.36E-03    & 1.43E-03   & 0.0021 $\pm$ 0.0003 & 0.0044 $\pm$ 0.0001  \\ \hline
(3,5)   & 2.73E-03    & 2.79E-03   & 0.0034 $\pm$ 0.0003 & 0.005 $\pm$ 0.0001   \\ \hline
(3,6)   & 4.50E-03    & 4.49E-03   & 0.0053 $\pm$ 0.0004 & 0.0069 $\pm$ 0.0001  \\ \hline
(3,7)   & 6.26E-03    & 6.12E-03   & 0.0078 $\pm$ 0.0005 & 0.0073 $\pm$ 0.0001  \\ \hline
(3,8)   & 7.52E-03    & 7.22E-03   & 0.0093 $\pm$ 0.0005 & 0.0092 $\pm$ 0.0002  \\ \hline
(3,9)   & 7.94E-03    & 7.50E-03   & 0.0097 $\pm$ 0.0005 & 0.0095 $\pm$ 0.0002  \\ \hline
(3,10)  & 7.46E-03    & 6.94E-03   & 0.0082 $\pm$ 0.0005 & 0.0078 $\pm$ 0.0001  \\ \hline
(3,11)  & 6.29E-03    & 5.78E-03   & 0.0076 $\pm$ 0.0005 & 0.0063 $\pm$ 0.0001  \\ \hline
(3,12)  & 4.81E-03    & 4.37E-03   & 0.0062 $\pm$ 0.0004 & 0.0041 $\pm$ 0.0001  \\ \hline
(3,13)  & 3.36E-03    & 3.01E-03   & 0.0033 $\pm$ 0.0003 & 0.00279 $\pm$ 8E-05  \\ \hline
(3,14)  & 2.16E-03    & 1.91E-03   & 0.0026 $\pm$ 0.0003 & 0.00277 $\pm$ 8E-05  \\ \hline
(3,15)  & 1.28E-03    & 1.12E-03   & 0.0012 $\pm$ 0.0002 & 0.00253 $\pm$ 8E-05  \\ \hline
(4,0)   & 2.66E-09    & 1.53E-07   & 0               & 0.00029 $\pm$ 3E-05  \\ \hline
(4,1)   & 1.45E-07    & 1.74E-06   & 6E-05 $\pm$ 5E-05   & 0.00017 $\pm$ 2E-05  \\ \hline
(4,2)   & 1.73E-06    & 9.99E-06   & 6E-05 $\pm$ 6E-05   & 0.0002 $\pm$ 2E-05   \\ \hline
(4,3)   & 1.02E-05    & 3.79E-05   & 0.0001 $\pm$ 8E-05  & 0.00033 $\pm$ 3E-05  \\ \hline
(4,4)   & 3.86E-05    & 1.06E-04   & 0.0004 $\pm$ 0.0001 & 0.00065 $\pm$ 4E-05  \\ \hline
(4,5)   & 1.06E-04    & 2.33E-04   & 0.0005 $\pm$ 0.0002 & 0.00073 $\pm$ 4E-05  \\ \hline
(4,6)   & 2.25E-04    & 4.19E-04   & 0.001 $\pm$ 0.0002  & 0.0009 $\pm$ 5E-05   \\ \hline
(4,7)   & 3.88E-04    & 6.34E-04   & 0.0014 $\pm$ 0.0002 & 0.00092 $\pm$ 5E-05  \\ \hline
(4,8)   & 5.63E-04    & 8.24E-04   & 0.0014 $\pm$ 0.0002 & 0.00144 $\pm$ 6E-05  \\ \hline
(4,9)   & 7.01E-04    & 9.36E-04   & 0.0019 $\pm$ 0.0003 & 0.00139 $\pm$ 6E-05  \\ \hline
(4,10)  & 7.63E-04    & 9.43E-04   & 0.0014 $\pm$ 0.0002 & 0.00127 $\pm$ 6E-05  \\ \hline
(4,11)  & 7.36E-04    & 8.50E-04   & 0.0014 $\pm$ 0.0002 & 0.00094 $\pm$ 5E-05  \\ \hline
(4,12)  & 6.36E-04    & 6.92E-04   & 0.0011 $\pm$ 0.0002 & 0.00067 $\pm$ 4E-05  \\ \hline
(4,13)  & 4.97E-04    & 5.13E-04   & 0.0008 $\pm$ 0.0002 & 0.00051 $\pm$ 4E-05  \\ \hline
(4,14)  & 3.54E-04    & 3.49E-04   & 0.0005 $\pm$ 0.0001 & 0.00049 $\pm$ 3E-05  \\ \hline
(4,15)  & 2.31E-04    & 2.18E-04   & 0.0004 $\pm$ 0.0001 & 0.00046 $\pm$ 3E-05  \\ \hline
\end{longtable}

\newpage
\begin{figure}
	\includegraphics[scale=0.5]{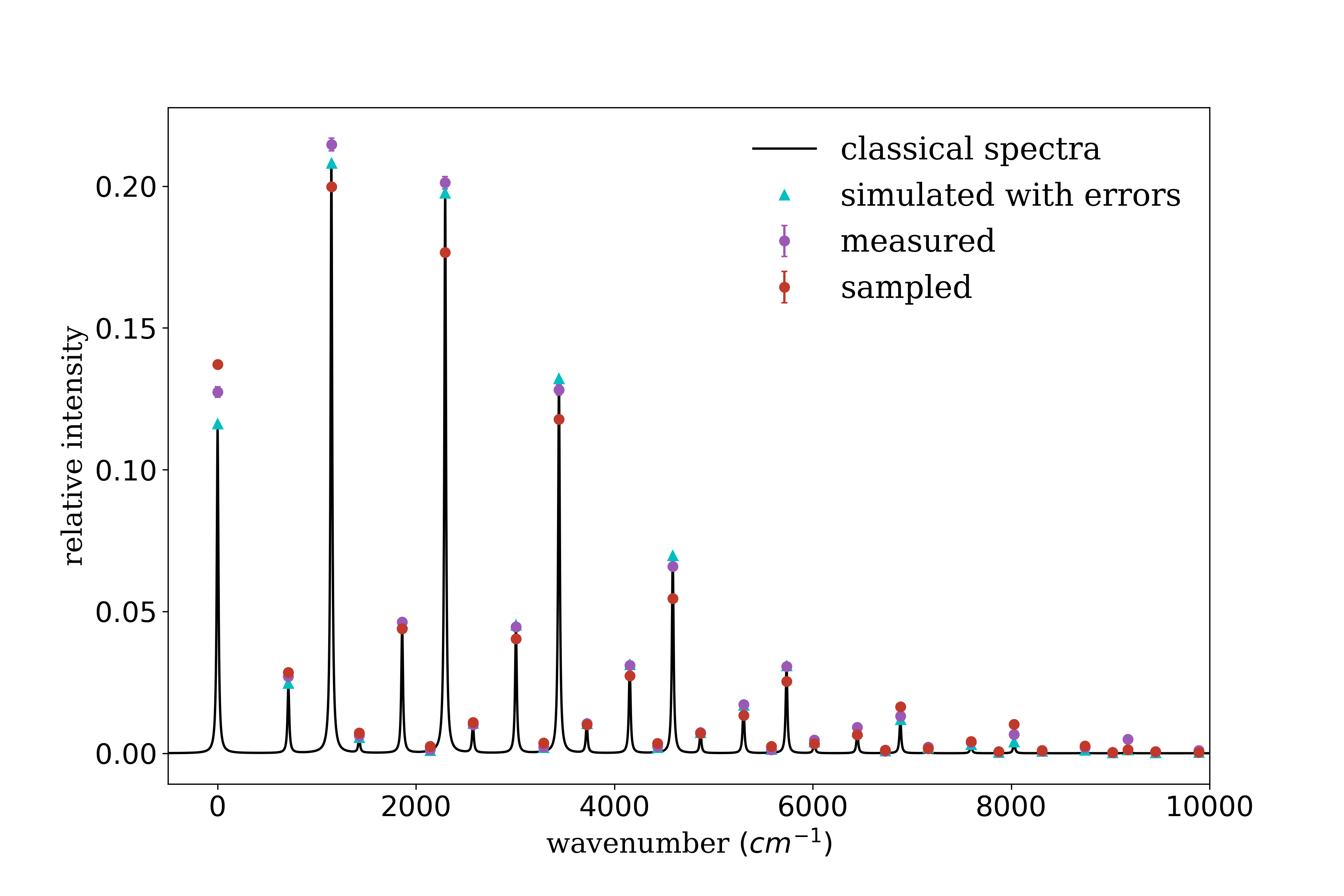}
	\caption{Photoionization of the ozone anion to neutral ozone starting in the vibrationless state $n$ = 0, $m$ = 0.}
\end{figure}

\setlength{\tabcolsep}{8pt}
\renewcommand{\arraystretch}{1.5}
\begin{longtable}{|c|c|c|c|c|}
\hline
& \multicolumn{4}{c|}{O$^-_3 \xrightarrow{h\nu}$ O$_3$ + e$^-$ starting in ($n$ = 0, $m$ = 0)} \\ \hline
$(n',m')$ & \begin{tabular}[c]{@{}c@{}}Classically \\ calculated \end{tabular} & \begin{tabular}[c]{@{}c@{}}Master equation \\ simulation \end{tabular} & \begin{tabular}[c]{@{}c@{}}Single-bit \\ extraction \end{tabular} & Sampling \\ \hline
(0,0) & 1.14E-01 & 1.16E-01 & 0.127 $\pm$ 0.002   & 0.1372 $\pm$ 0.0005 \\ \hline
(0,1) & 2.42E-02 & 2.47E-02 & 0.0271 $\pm$ 0.0008 & 0.0286 $\pm$ 0.0003 \\ \hline
(0,2) & 5.57E-03 & 5.60E-03 & 0.0063 $\pm$ 0.0004 & 0.0071 $\pm$ 0.0001 \\ \hline
(0,3) & 1.06E-03 & 1.06E-03 & 0.0013 $\pm$ 0.0002 & 0.0024 $\pm$ 8E-05  \\ \hline
(1,0) & 2.06E-01 & 2.08E-01 & 0.215 $\pm$ 0.002   & 0.1998 $\pm$ 0.0006 \\ \hline
(1,1) & 4.55E-02 & 4.60E-02 & 0.046 $\pm$ 0.001   & 0.0439 $\pm$ 0.0003 \\ \hline
(1,2) & 1.06E-02 & 1.06E-02 & 0.01 $\pm$ 0.0005   & 0.0109 $\pm$ 0.0002 \\ \hline
(1,3) & 2.06E-03 & 2.04E-03 & 0.0022 $\pm$ 0.0003 & 0.0036 $\pm$ 0.0001 \\ \hline
(2,0) & 1.98E-01 & 1.98E-01 & 0.201 $\pm$ 0.002   & 0.1766 $\pm$ 0.0006 \\ \hline
(2,1) & 4.52E-02 & 4.52E-02 & 0.045 $\pm$ 0.001   & 0.0403 $\pm$ 0.0003 \\ \hline
(2,2) & 1.07E-02 & 1.05E-02 & 0.0104 $\pm$ 0.0005 & 0.0101 $\pm$ 0.0002 \\ \hline
(2,3) & 2.11E-03 & 2.06E-03 & 0.0025 $\pm$ 0.0003 & 0.00349 $\pm$ 9E-05 \\ \hline
(3,0) & 1.33E-01 & 1.32E-01 & 0.128 $\pm$ 0.002   & 0.1177 $\pm$ 0.0005 \\ \hline
(3,1) & 3.15E-02 & 3.12E-02 & 0.031 $\pm$ 0.0009  & 0.0274 $\pm$ 0.0003 \\ \hline
(3,2) & 7.54E-03 & 7.37E-03 & 0.0073 $\pm$ 0.0005 & 0.0071 $\pm$ 0.0001 \\ \hline
(3,3) & 1.51E-03 & 1.47E-03 & 0.0012 $\pm$ 0.0002 & 0.00248 $\pm$ 8E-05 \\ \hline
(4,0) & 7.08E-02 & 6.97E-02 & 0.066 $\pm$ 0.001   & 0.0546 $\pm$ 0.0004 \\ \hline
(4,1) & 1.73E-02 & 1.70E-02 & 0.0173 $\pm$ 0.0007 & 0.0133 $\pm$ 0.0002 \\ \hline
(4,2) & 4.19E-03 & 4.06E-03 & 0.0047 $\pm$ 0.0004 & 0.00342 $\pm$ 9E-05 \\ \hline
(4,3) & 8.51E-04 & 8.19E-04 & 0.0008 $\pm$ 0.0002 & 0.0012 $\pm$ 6E-05  \\ \hline
(5,0) & 3.15E-02 & 3.09E-02 & 0.0307 $\pm$ 0.0009 & 0.0253 $\pm$ 0.0003 \\ \hline
(5,1) & 7.91E-03 & 7.74E-03 & 0.0092 $\pm$ 0.0005 & 0.0065 $\pm$ 0.0001 \\ \hline
(5,2) & 1.94E-03 & 1.87E-03 & 0.0022 $\pm$ 0.0003 & 0.00173 $\pm$ 7E-05 \\ \hline
(5,3) & 4.01E-04 & 3.83E-04 & 0.0004 $\pm$ 0.0001 & 0.00064 $\pm$ 4E-05 \\ \hline
(6,0) & 1.22E-02 & 1.19E-02 & 0.0132 $\pm$ 0.0007 & 0.0165 $\pm$ 0.0002 \\ \hline
(6,1) & 3.15E-03 & 3.07E-03 & 0.0039 $\pm$ 0.0004 & 0.0041 $\pm$ 0.0001 \\ \hline
(6,2) & 7.83E-04 & 7.52E-04 & 0.0007 $\pm$ 0.0002 & 0.00105 $\pm$ 5E-05 \\ \hline
(6,3) & 1.64E-04 & 1.56E-04 & 0.00024 $\pm$ 9E-05 & 0.00036 $\pm$ 3E-05 \\ \hline
(7,0) & 4.23E-03 & 4.10E-03 & 0.0066 $\pm$ 0.0005 & 0.0102 $\pm$ 0.0002 \\ \hline
(7,1) & 1.12E-03 & 1.09E-03 & 0.0021 $\pm$ 0.0003 & 0.00252 $\pm$ 8E-05 \\ \hline
(7,2) & 2.81E-04 & 2.69E-04 & 0.0005 $\pm$ 0.0001 & 0.00068 $\pm$ 4E-05 \\ \hline
(7,3) & 5.96E-05 & 5.64E-05 & 0.00019 $\pm$ 8E-05 & 0.00017 $\pm$ 2E-05 \\ \hline
(8,0) & 1.33E-03 & 1.28E-03 & 0.0049 $\pm$ 0.0005 & 0.00132 $\pm$ 6E-05 \\ \hline
(8,1) & 3.61E-04 & 3.48E-04 & 0.001 $\pm$ 0.0002  & 0.00035 $\pm$ 3E-05 \\ \hline
(8,2) & 9.18E-05 & 8.73E-05 & 0.0002 $\pm$ 0.0001 & 9E-05 $\pm$ 2E-05   \\ \hline
(8,3) & 1.97E-05 & 1.85E-05 & 0.00013 $\pm$ 7E-05 & 2.8E-05 $\pm$ 8E-06 \\ \hline
(9,0) & 3.86E-04 & 3.68E-04 & 0.0047 $\pm$ 0.0005 & 0.00156 $\pm$ 6E-05 \\ \hline
(9,1) & 1.07E-04 & 1.03E-04 & 0.0012 $\pm$ 0.0002 & 0.00038 $\pm$ 3E-05 \\ \hline
(9,2) & 2.76E-05 & 2.60E-05 & 0.0004 $\pm$ 0.0001 & 8E-05 $\pm$ 1E-05   \\ \hline
(9,3) & 5.98E-06 & 5.59E-06 & 0.00013 $\pm$ 7E-05 & 3.6E-05 $\pm$ 1E-05 \\ \hline
\end{longtable}

\newpage
\begin{figure}
	\includegraphics[scale=0.5]{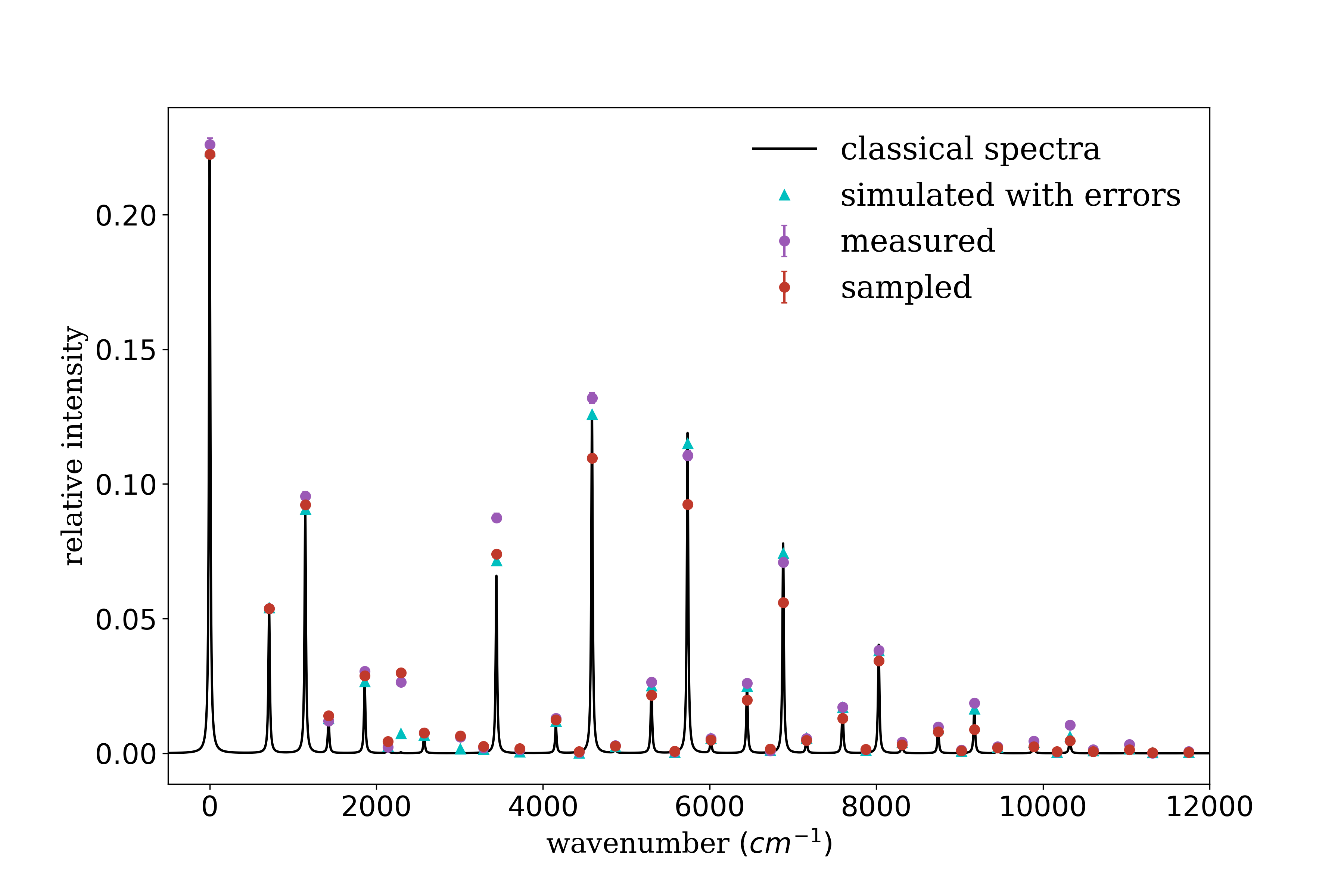}
	\caption{Photoionization of the ozone anion to neutral ozone starting with one quanta in the symmetric-stretching mode and zero in the bending mode $n$ = 1, $m$ = 0.}
\end{figure}

\setlength{\tabcolsep}{8pt}
\renewcommand{\arraystretch}{1.5}
\begin{longtable}{|c|c|c|c|c|}
\hline
& \multicolumn{4}{c|}{O$^-_3 \xrightarrow{h\nu}$ O$_3$ + e$^-$ starting in ($n$ = 1, $m$ = 0)} \\ \hline
$(n',m')$ & \begin{tabular}[c]{@{}c@{}}Classically \\ calculated \end{tabular} & \begin{tabular}[c]{@{}c@{}}Master equation \\ simulation \end{tabular} & \begin{tabular}[c]{@{}c@{}}Single-bit \\ extraction \end{tabular} & Sampling \\ \hline
(0,0) & 2.24E-01 & 2.23E-01 & 0.226 $\pm$ 0.002   & 0.2226 $\pm$ 0.0007 \\ \hline
(0,1) & 5.42E-02 & 5.40E-02 & 0.054 $\pm$ 0.001   & 0.0537 $\pm$ 0.0004 \\ \hline
(0,2) & 1.31E-02 & 1.28E-02 & 0.0121 $\pm$ 0.0006 & 0.0139 $\pm$ 0.0002 \\ \hline
(0,3) & 2.64E-03 & 2.56E-03 & 0.0023 $\pm$ 0.0003 & 0.0044 $\pm$ 0.0001 \\ \hline
(1,0) & 8.96E-02 & 9.09E-02 & 0.095 $\pm$ 0.002   & 0.0923 $\pm$ 0.0005 \\ \hline
(1,1) & 2.67E-02 & 2.69E-02 & 0.0304 $\pm$ 0.0009 & 0.0288 $\pm$ 0.0003 \\ \hline
(1,2) & 6.90E-03 & 6.83E-03 & 0.0075 $\pm$ 0.0005 & 0.0075 $\pm$ 0.0001 \\ \hline
(1,3) & 1.51E-03 & 1.48E-03 & 0.002 $\pm$ 0.0003  & 0.00263 $\pm$ 9E-05 \\ \hline
(2,0) & 3.14E-04 & 5.76E-03 & 0.0265 $\pm$ 0.0009 & 0.0299 $\pm$ 0.0003 \\ \hline
(2,1) & 2.69E-04 & 1.39E-03 & 0.006 $\pm$ 0.0004  & 0.0064 $\pm$ 0.0001 \\ \hline
(2,2) & 1.59E-04 & 4.07E-04 & 0.0013 $\pm$ 0.0002 & 0.00174 $\pm$ 7E-05 \\ \hline
(2,3) & 6.61E-05 & 1.11E-04 & 0.0004 $\pm$ 0.0001 & 0.00059 $\pm$ 4E-05 \\ \hline
(3,0) & 6.59E-02 & 6.96E-02 & 0.088 $\pm$ 0.002   & 0.074 $\pm$ 0.0004  \\ \hline
(3,1) & 1.05E-02 & 1.14E-02 & 0.0129 $\pm$ 0.0006 & 0.0124 $\pm$ 0.0002 \\ \hline
(3,2) & 2.11E-03 & 2.27E-03 & 0.0028 $\pm$ 0.0003 & 0.00267 $\pm$ 9E-05 \\ \hline
(3,3) & 3.34E-04 & 3.67E-04 & 0.0005 $\pm$ 0.0001 & 0.00078 $\pm$ 5E-05 \\ \hline
(4,0) & 1.26E-01 & 1.25E-01 & 0.132 $\pm$ 0.002   & 0.1097 $\pm$ 0.0005 \\ \hline
(4,1) & 2.51E-02 & 2.49E-02 & 0.0264 $\pm$ 0.0009 & 0.0216 $\pm$ 0.0002 \\ \hline
(4,2) & 5.55E-03 & 5.42E-03 & 0.0054 $\pm$ 0.0004 & 0.0049 $\pm$ 0.0001 \\ \hline
(4,3) & 1.01E-03 & 9.81E-04 & 0.0009 $\pm$ 0.0002 & 0.00165 $\pm$ 7E-05 \\ \hline
(5,0) & 1.19E-01 & 1.16E-01 & 0.111 $\pm$ 0.002   & 0.0925 $\pm$ 0.0005 \\ \hline
(5,1) & 2.58E-02 & 2.51E-02 & 0.0261 $\pm$ 0.0009 & 0.0198 $\pm$ 0.0002 \\ \hline
(5,2) & 5.94E-03 & 5.68E-03 & 0.0054 $\pm$ 0.0004 & 0.0048 $\pm$ 0.0001 \\ \hline
(5,3) & 1.14E-03 & 1.08E-03 & 0.0014 $\pm$ 0.0002 & 0.00153 $\pm$ 7E-05 \\ \hline
(6,0) & 7.79E-02 & 7.54E-02 & 0.071 $\pm$ 0.001   & 0.056 $\pm$ 0.0004  \\ \hline
(6,1) & 1.79E-02 & 1.73E-02 & 0.0171 $\pm$ 0.0007 & 0.013 $\pm$ 0.0002  \\ \hline
(6,2) & 4.23E-03 & 4.01E-03 & 0.0041 $\pm$ 0.0004 & 0.00313 $\pm$ 9E-05 \\ \hline
(6,3) & 8.33E-04 & 7.82E-04 & 0.0012 $\pm$ 0.0002 & 0.00091 $\pm$ 5E-05 \\ \hline
(7,0) & 4.03E-02 & 3.89E-02 & 0.038 $\pm$ 0.001   & 0.0344 $\pm$ 0.0003 \\ \hline
(7,1) & 9.69E-03 & 9.34E-03 & 0.0098 $\pm$ 0.0005 & 0.008 $\pm$ 0.0001  \\ \hline
(7,2) & 2.33E-03 & 2.21E-03 & 0.0024 $\pm$ 0.0003 & 0.00207 $\pm$ 8E-05 \\ \hline
(7,3) & 4.70E-04 & 4.40E-04 & 0.0004 $\pm$ 0.0001 & 0.00065 $\pm$ 4E-05 \\ \hline
(8,0) & 1.75E-02 & 1.69E-02 & 0.0186 $\pm$ 0.0008 & 0.0089 $\pm$ 0.0002 \\ \hline
(8,1) & 4.38E-03 & 4.22E-03 & 0.0045 $\pm$ 0.0004 & 0.00239 $\pm$ 8E-05 \\ \hline
(8,2) & 1.07E-03 & 1.02E-03 & 0.0013 $\pm$ 0.0002 & 0.00059 $\pm$ 4E-05 \\ \hline
(8,3) & 2.20E-04 & 2.06E-04 & 0.00016 $\pm$ 8E-05 & 0.0002 $\pm$ 2E-05  \\ \hline
(9,0) & 6.66E-03 & 6.43E-03 & 0.0105 $\pm$ 0.0006 & 0.0046 $\pm$ 0.0001 \\ \hline
(9,1) & 1.72E-03 & 1.66E-03 & 0.0033 $\pm$ 0.0003 & 0.00132 $\pm$ 6E-05 \\ \hline
(9,2) & 4.28E-04 & 4.05E-04 & 0.0006 $\pm$ 0.0002 & 0.00034 $\pm$ 3E-05 \\ \hline
(9,3) & 8.93E-05 & 8.37E-05 & 3E-05 $\pm$ 6E-05   & 0.00013 $\pm$ 2E-05 \\ \hline
\end{longtable}

\newpage
\begin{figure}
	\includegraphics[scale=0.5]{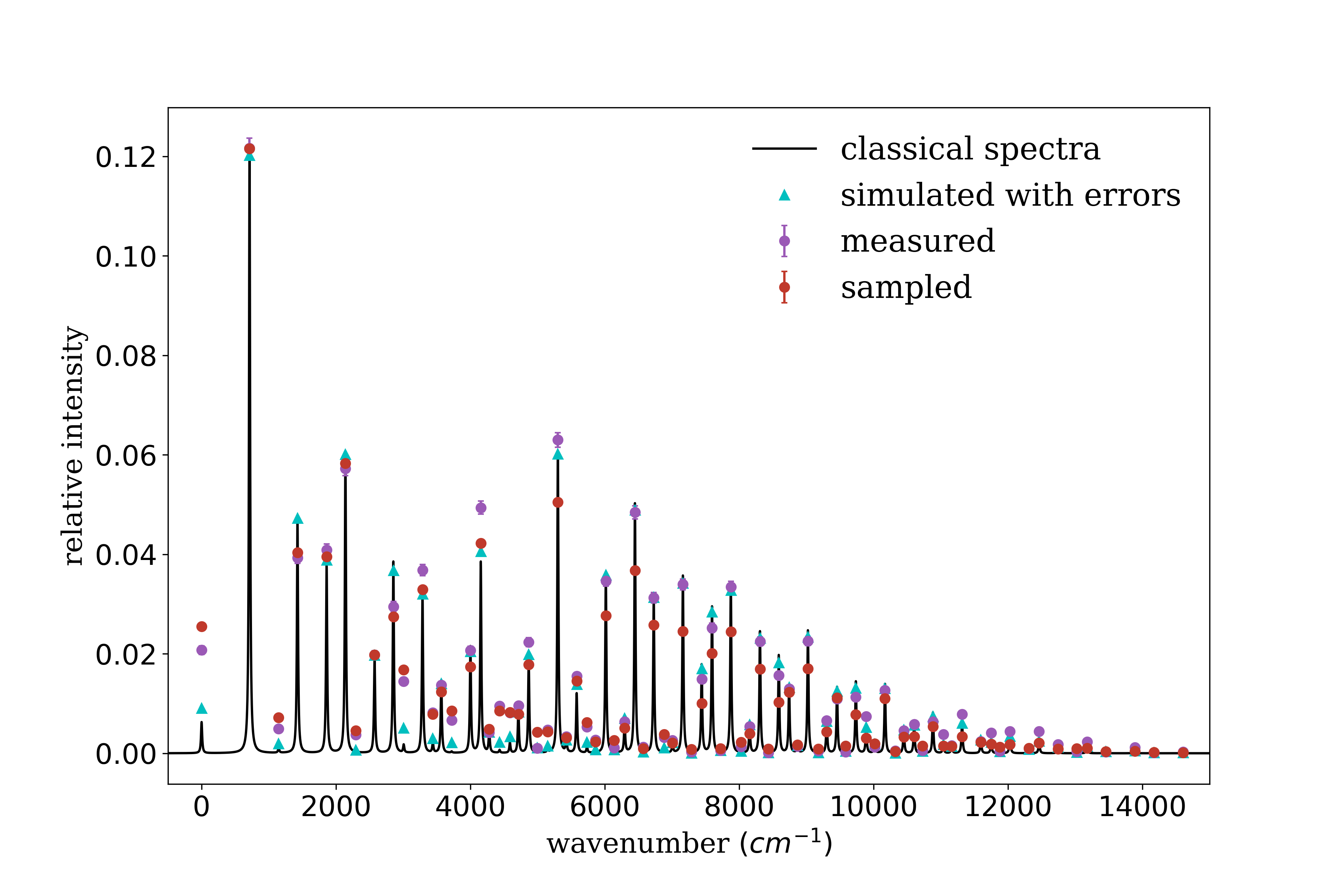}
	\caption{Photoionization of the ozone anion to neutral ozone starting with one quanta in the symmetric-stretching mode and two in the bending mode $n$ = 1, $m$ = 2.}
\end{figure}

\setlength{\tabcolsep}{8pt}
\renewcommand{\arraystretch}{1.5}
\begin{longtable}{|c|c|c|c|c|}
\hline
& \multicolumn{4}{c|}{O$^-_3 \xrightarrow{h\nu}$ O$_3$ + e$^-$ starting in ($n$ = 1, $m$ = 2)} \\ \hline
$(n',m')$ & \begin{tabular}[c]{@{}c@{}}Classically \\ calculated \end{tabular} & \begin{tabular}[c]{@{}c@{}}Master equation \\ simulation \end{tabular} & \begin{tabular}[c]{@{}c@{}}Single-bit \\ extraction \end{tabular} & Sampling \\ \hline
(0,0) & 6.24E-03 & 9.10E-03 & 0.0208 $\pm$ 0.0008 & 0.0255 $\pm$ 0.0003  \\ \hline
(0,1) & 1.21E-01 & 1.20E-01 & 0.122 $\pm$ 0.002   & 0.1215 $\pm$ 0.0006  \\ \hline
(0,2) & 4.73E-02 & 4.75E-02 & 0.039 $\pm$ 0.001   & 0.0403 $\pm$ 0.0004  \\ \hline
(0,3) & 6.01E-02 & 6.02E-02 & 0.057 $\pm$ 0.001   & 0.0583 $\pm$ 0.0004  \\ \hline
(0,4) & 3.85E-02 & 3.65E-02 & 0.0295 $\pm$ 0.001  & 0.0274 $\pm$ 0.0003  \\ \hline
(0,5) & 1.46E-02 & 1.39E-02 & 0.0137 $\pm$ 0.0007 & 0.0124 $\pm$ 0.0002  \\ \hline
(0,6) & 4.57E-03 & 4.18E-03 & 0.0042 $\pm$ 0.0004 & 0.0049 $\pm$ 0.0001  \\ \hline
(0,7) & 1.21E-03 & 1.10E-03 & 0.001 $\pm$ 0.0002  & 0.0042 $\pm$ 0.0001  \\ \hline
(1,0) & 9.99E-04 & 1.99E-03 & 0.0049 $\pm$ 0.0004 & 0.0072 $\pm$ 0.0002  \\ \hline
(1,1) & 3.79E-02 & 3.89E-02 & 0.041 $\pm$ 0.001   & 0.0395 $\pm$ 0.0004  \\ \hline
(1,2) & 1.90E-02 & 1.99E-02 & 0.0198 $\pm$ 0.0008 & 0.0198 $\pm$ 0.0003  \\ \hline
(1,3) & 3.22E-02 & 3.22E-02 & 0.037 $\pm$ 0.001   & 0.0329 $\pm$ 0.0003  \\ \hline
(1,4) & 2.14E-02 & 2.04E-02 & 0.0207 $\pm$ 0.0008 & 0.0174 $\pm$ 0.0002  \\ \hline
(1,5) & 8.78E-03 & 8.31E-03 & 0.0096 $\pm$ 0.0006 & 0.0079 $\pm$ 0.0002  \\ \hline
(1,6) & 2.87E-03 & 2.63E-03 & 0.0033 $\pm$ 0.0003 & 0.0031 $\pm$ 0.0001  \\ \hline
(1,7) & 7.99E-04 & 7.23E-04 & 0.0011 $\pm$ 0.0002 & 0.00258 $\pm$ 0.0001 \\ \hline
(2,0) & 4.14E-04 & 6.10E-04 & 0.0037 $\pm$ 0.0004 & 0.0045 $\pm$ 0.0001  \\ \hline
(2,1) & 1.56E-03 & 4.26E-03 & 0.0145 $\pm$ 0.0007 & 0.0168 $\pm$ 0.0002  \\ \hline
(2,2) & 2.44E-04 & 1.72E-03 & 0.0066 $\pm$ 0.0005 & 0.0085 $\pm$ 0.0002  \\ \hline
(2,3) & 5.64E-04 & 1.90E-03 & 0.0095 $\pm$ 0.0006 & 0.0085 $\pm$ 0.0002  \\ \hline
(2,4) & 6.09E-04 & 1.33E-03 & 0.0047 $\pm$ 0.0004 & 0.0043 $\pm$ 0.0001  \\ \hline
(2,5) & 4.53E-04 & 6.66E-04 & 0.0027 $\pm$ 0.0003 & 0.00229 $\pm$ 9E-05  \\ \hline
(2,6) & 1.95E-04 & 2.51E-04 & 0.0012 $\pm$ 0.0002 & 0.00098 $\pm$ 6E-05  \\ \hline
(2,7) & 6.97E-05 & 7.85E-05 & 0.0003 $\pm$ 0.0001 & 0.0008 $\pm$ 5E-05   \\ \hline
(3,0) & 2.02E-03 & 2.95E-03 & 0.0082 $\pm$ 0.0005 & 0.0079 $\pm$ 0.0002  \\ \hline
(3,1) & 3.84E-02 & 3.97E-02 & 0.049 $\pm$ 0.001   & 0.0422 $\pm$ 0.0004  \\ \hline
(3,2) & 1.87E-02 & 1.94E-02 & 0.0224 $\pm$ 0.0009 & 0.0179 $\pm$ 0.0003  \\ \hline
(3,3) & 1.20E-02 & 1.34E-02 & 0.0155 $\pm$ 0.0007 & 0.0145 $\pm$ 0.0002  \\ \hline
(3,4) & 6.36E-03 & 6.72E-03 & 0.0064 $\pm$ 0.0005 & 0.0051 $\pm$ 0.0001  \\ \hline
(3,5) & 1.83E-03 & 2.05E-03 & 0.0026 $\pm$ 0.0003 & 0.00206 $\pm$ 9E-05  \\ \hline
(3,6) & 4.70E-04 & 5.06E-04 & 0.0006 $\pm$ 0.0002 & 0.00093 $\pm$ 6E-05  \\ \hline
(3,7) & 9.93E-05 & 1.14E-04 & 0.0002 $\pm$ 0.0001 & 0.00086 $\pm$ 6E-05  \\ \hline
(4,0) & 2.05E-03 & 3.41E-03 & 0.0082 $\pm$ 0.0005 & 0.0082 $\pm$ 0.0002  \\ \hline
(4,1) & 6.06E-02 & 6.01E-02 & 0.063 $\pm$ 0.001   & 0.0505 $\pm$ 0.0004  \\ \hline
(4,2) & 3.64E-02 & 3.58E-02 & 0.035 $\pm$ 0.001   & 0.0277 $\pm$ 0.0003  \\ \hline
(4,3) & 3.11E-02 & 3.13E-02 & 0.031 $\pm$ 0.001   & 0.0258 $\pm$ 0.0003  \\ \hline
(4,4) & 1.77E-02 & 1.68E-02 & 0.0149 $\pm$ 0.0007 & 0.01 $\pm$ 0.0002    \\ \hline
(4,5) & 5.91E-03 & 5.69E-03 & 0.0054 $\pm$ 0.0004 & 0.0039 $\pm$ 0.0001  \\ \hline
(4,6) & 1.69E-03 & 1.55E-03 & 0.0014 $\pm$ 0.0002 & 0.00167 $\pm$ 8E-05  \\ \hline
(4,7) & 4.09E-04 & 3.80E-04 & 0.0003 $\pm$ 0.0001 & 0.00151 $\pm$ 7E-05  \\ \hline
(5,0) & 1.18E-03 & 2.29E-03 & 0.0053 $\pm$ 0.0004 & 0.0062 $\pm$ 0.0002  \\ \hline
(5,1) & 5.02E-02 & 4.93E-02 & 0.048 $\pm$ 0.001   & 0.0367 $\pm$ 0.0004  \\ \hline
(5,2) & 3.57E-02 & 3.46E-02 & 0.034 $\pm$ 0.001   & 0.0245 $\pm$ 0.0003  \\ \hline
(5,3) & 3.39E-02 & 3.31E-02 & 0.033 $\pm$ 0.001   & 0.0244 $\pm$ 0.0003  \\ \hline
(5,4) & 1.96E-02 & 1.82E-02 & 0.0156 $\pm$ 0.0007 & 0.0103 $\pm$ 0.0002  \\ \hline
(5,5) & 6.87E-03 & 6.40E-03 & 0.0066 $\pm$ 0.0005 & 0.0043 $\pm$ 0.0001  \\ \hline
(5,6) & 2.03E-03 & 1.81E-03 & 0.0012 $\pm$ 0.0002 & 0.00196 $\pm$ 8E-05  \\ \hline
(5,7) & 5.10E-04 & 4.57E-04 & 0.0005 $\pm$ 0.0002 & 0.00148 $\pm$ 7E-05  \\ \hline
(6,0) & 4.73E-04 & 1.13E-03 & 0.0032 $\pm$ 0.0003 & 0.0038 $\pm$ 0.0001  \\ \hline
(6,1) & 2.94E-02 & 2.89E-02 & 0.0252 $\pm$ 0.0009 & 0.0201 $\pm$ 0.0003  \\ \hline
(6,2) & 2.45E-02 & 2.36E-02 & 0.0225 $\pm$ 0.0009 & 0.0169 $\pm$ 0.0002  \\ \hline
(6,3) & 2.47E-02 & 2.37E-02 & 0.0226 $\pm$ 0.0009 & 0.017 $\pm$ 0.0002   \\ \hline
(6,4) & 1.44E-02 & 1.32E-02 & 0.0113 $\pm$ 0.0006 & 0.0078 $\pm$ 0.0002  \\ \hline
(6,5) & 5.18E-03 & 4.76E-03 & 0.0046 $\pm$ 0.0004 & 0.0033 $\pm$ 0.0001  \\ \hline
(6,6) & 1.56E-03 & 1.38E-03 & 0.0015 $\pm$ 0.0002 & 0.00143 $\pm$ 7E-05  \\ \hline
(6,7) & 4.00E-04 & 3.53E-04 & 0.0005 $\pm$ 0.0001 & 0.00122 $\pm$ 7E-05  \\ \hline
(7,0) & 1.44E-04 & 4.53E-04 & 0.0013 $\pm$ 0.0002 & 0.00221 $\pm$ 9E-05  \\ \hline
(7,1) & 1.37E-02 & 1.35E-02 & 0.013 $\pm$ 0.0007  & 0.0122 $\pm$ 0.0002  \\ \hline
(7,2) & 1.33E-02 & 1.28E-02 & 0.0109 $\pm$ 0.0006 & 0.0112 $\pm$ 0.0002  \\ \hline
(7,3) & 1.39E-02 & 1.33E-02 & 0.0126 $\pm$ 0.0007 & 0.011 $\pm$ 0.0002   \\ \hline
(7,4) & 8.18E-03 & 7.46E-03 & 0.0064 $\pm$ 0.0005 & 0.0054 $\pm$ 0.0001  \\ \hline
(7,5) & 3.00E-03 & 2.73E-03 & 0.0024 $\pm$ 0.0003 & 0.00221 $\pm$ 9E-05  \\ \hline
(7,6) & 9.14E-04 & 8.02E-04 & 0.001 $\pm$ 0.0002  & 0.00102 $\pm$ 6E-05  \\ \hline
(7,7) & 2.38E-04 & 2.08E-04 & 0.0004 $\pm$ 0.0001 & 0.00092 $\pm$ 6E-05  \\ \hline
(8,0) & 3.37E-05 & 1.55E-04 & 0.0005 $\pm$ 0.0002 & 0.00085 $\pm$ 6E-05  \\ \hline
(8,1) & 5.37E-03 & 5.35E-03 & 0.0074 $\pm$ 0.0005 & 0.003 $\pm$ 0.0001   \\ \hline
(8,2) & 6.07E-03 & 5.83E-03 & 0.0058 $\pm$ 0.0005 & 0.0033 $\pm$ 0.0001  \\ \hline
(8,3) & 6.54E-03 & 6.22E-03 & 0.0079 $\pm$ 0.0005 & 0.0034 $\pm$ 0.0001  \\ \hline
(8,4) & 3.86E-03 & 3.52E-03 & 0.0044 $\pm$ 0.0004 & 0.00176 $\pm$ 8E-05  \\ \hline
(8,5) & 1.44E-03 & 1.30E-03 & 0.0018 $\pm$ 0.0003 & 0.00085 $\pm$ 6E-05  \\ \hline
(8,6) & 4.42E-04 & 3.88E-04 & 0.0003 $\pm$ 0.0001 & 0.00033 $\pm$ 3E-05  \\ \hline
(8,7) & 1.17E-04 & 1.02E-04 & 0.0001 $\pm$ 7E-05  & 0.00023 $\pm$ 3E-05  \\ \hline
(9,0) & 5.90E-06 & 4.75E-05 & 0.0005 $\pm$ 0.0002 & 0.00035 $\pm$ 4E-05  \\ \hline
(9,1) & 1.84E-03 & 1.84E-03 & 0.0038 $\pm$ 0.0004 & 0.00149 $\pm$ 7E-05  \\ \hline
(9,2) & 2.42E-03 & 2.32E-03 & 0.0041 $\pm$ 0.0004 & 0.00184 $\pm$ 8E-05  \\ \hline
(9,3) & 2.67E-03 & 2.53E-03 & 0.0044 $\pm$ 0.0004 & 0.00208 $\pm$ 9E-05  \\ \hline
(9,4) & 1.58E-03 & 1.44E-03 & 0.0023 $\pm$ 0.0003 & 0.00105 $\pm$ 6E-05  \\ \hline
(9,5) & 5.95E-04 & 5.39E-04 & 0.0012 $\pm$ 0.0002 & 0.00044 $\pm$ 4E-05  \\ \hline
(9,6) & 1.85E-04 & 1.62E-04 & 0.0003 $\pm$ 0.0001 & 0.00016 $\pm$ 2E-05  \\ \hline
(9,7) & 4.94E-05 & 4.28E-05 & 0.00012 $\pm$ 7E-05 & 0.00015 $\pm$ 2E-05  \\ \hline
\end{longtable}

\newpage
\begin{figure}
	\includegraphics[scale=0.5]{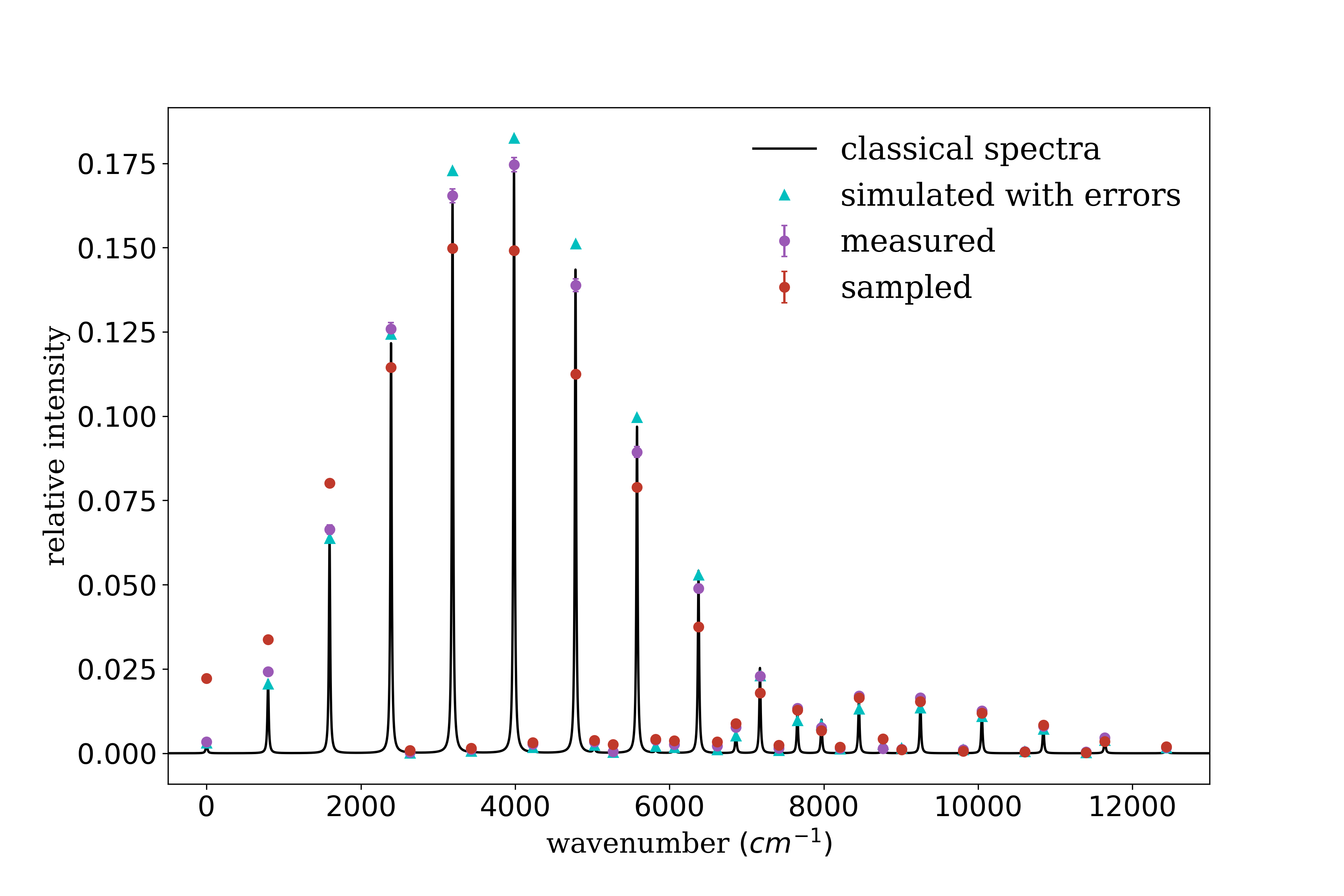}
	\caption{Photoionization of nitrite to nitrogen dioxide starting in the vibrationless state $n$ = 0, $m$ = 0.}
\end{figure}

\setlength{\tabcolsep}{8pt}
\renewcommand{\arraystretch}{1.5}
\begin{longtable}{|c|c|c|c|c|}
\hline
& \multicolumn{4}{c|}{NO$^-_2 \xrightarrow{h\nu}$ NO$_2$ + e$^-$ starting in ($n$ = 0, $m$ = 0)} \\ \hline
$(n',m')$ & \begin{tabular}[c]{@{}c@{}}Classically \\ calculated \end{tabular} & \begin{tabular}[c]{@{}c@{}}Master equation \\ simulation \end{tabular} & \begin{tabular}[c]{@{}c@{}}Single-bit \\ extraction \end{tabular} & Sampling \\ \hline
(0,0)  & 3.54E-03 & 3.14E-03 & 0.0033 $\pm$ 0.0004  & 0.0222 $\pm$ 0.0002  \\ \hline
(0,1)  & 2.17E-02 & 2.07E-02 & 0.0243 $\pm$ 0.0008  & 0.0337 $\pm$ 0.0003  \\ \hline
(0,2)  & 6.42E-02 & 6.40E-02 & 0.066 $\pm$ 0.001    & 0.0802 $\pm$ 0.0004  \\ \hline
(0,3)  & 1.22E-01 & 1.24E-01 & 0.126 $\pm$ 0.002    & 0.1145 $\pm$ 0.0005  \\ \hline
(0,4)  & 1.66E-01 & 1.73E-01 & 0.165 $\pm$ 0.002    & 0.1498 $\pm$ 0.0006  \\ \hline
(0,5)  & 1.73E-01 & 1.82E-01 & 0.175 $\pm$ 0.002    & 0.1492 $\pm$ 0.0005  \\ \hline
(0,6)  & 1.43E-01 & 1.51E-01 & 0.139 $\pm$ 0.002    & 0.1125 $\pm$ 0.0005  \\ \hline
(0,7)  & 9.68E-02 & 9.95E-02 & 0.089 $\pm$ 0.002    & 0.079 $\pm$ 0.0004   \\ \hline
(0,8)  & 5.41E-02 & 5.27E-02 & 0.049 $\pm$ 0.001    & 0.0375 $\pm$ 0.0003  \\ \hline
(0,9)  & 2.53E-02 & 2.29E-02 & 0.0229 $\pm$ 0.0008  & 0.0179 $\pm$ 0.0002  \\ \hline
(0,10) & 9.93E-03 & 8.49E-03 & 0.0076 $\pm$ 0.0005  & 0.0067 $\pm$ 0.0001  \\ \hline
(0,11) & 3.29E-03 & 2.75E-03 & 0.0014 $\pm$ 0.0003  & 0.0043 $\pm$ 0.0001  \\ \hline
(1,0)  & 2.74E-04 & 1.03E-04 & 0.00026 $\pm$ 0.0001 & 0.0008 $\pm$ 4E-05   \\ \hline
(1,1)  & 1.18E-03 & 6.52E-04 & 0.0012 $\pm$ 0.0002  & 0.00147 $\pm$ 6E-05  \\ \hline
(1,2)  & 2.24E-03 & 1.73E-03 & 0.0027 $\pm$ 0.0003  & 0.00316 $\pm$ 9E-05  \\ \hline
(1,3)  & 2.32E-03 & 2.48E-03 & 0.0033 $\pm$ 0.0003  & 0.00384 $\pm$ 0.0001 \\ \hline
(1,4)  & 1.26E-03 & 2.04E-03 & 0.0039 $\pm$ 0.0004  & 0.00414 $\pm$ 0.0001 \\ \hline
(1,5)  & 2.02E-04 & 1.04E-03 & 0.0022 $\pm$ 0.0003  & 0.00336 $\pm$ 9E-05  \\ \hline
(1,6)  & 6.79E-05 & 7.77E-04 & 0.0015 $\pm$ 0.0002  & 0.00245 $\pm$ 8E-05  \\ \hline
(1,7)  & 6.34E-04 & 1.27E-03 & 0.0015 $\pm$ 0.0002  & 0.0018 $\pm$ 7E-05   \\ \hline
(1,8)  & 1.12E-03 & 1.53E-03 & 0.0012 $\pm$ 0.0002  & 0.00107 $\pm$ 5E-05  \\ \hline
(1,9)  & 1.13E-03 & 1.15E-03 & 0.0011 $\pm$ 0.0002  & 0.00067 $\pm$ 4E-05  \\ \hline
(1,10) & 8.04E-04 & 5.44E-04 & 0.0005 $\pm$ 0.0001  & 0.00039 $\pm$ 3E-05  \\ \hline
(1,11) & 4.38E-04 & 1.56E-04 & 0.0004 $\pm$ 0.0001  & 0.00021 $\pm$ 2E-05  \\ \hline
(2,0)  & 4.71E-04 & 2.67E-04 & 0.0005 $\pm$ 0.0001  & 0.0026 $\pm$ 8E-05   \\ \hline
(2,1)  & 2.62E-03 & 1.71E-03 & 0.0026 $\pm$ 0.0003  & 0.00376 $\pm$ 9E-05  \\ \hline
(2,2)  & 7.06E-03 & 5.16E-03 & 0.0077 $\pm$ 0.0005  & 0.0088 $\pm$ 0.0001  \\ \hline
(2,3)  & 1.24E-02 & 9.77E-03 & 0.0134 $\pm$ 0.0006  & 0.0128 $\pm$ 0.0002  \\ \hline
(2,4)  & 1.58E-02 & 1.31E-02 & 0.017 $\pm$ 0.0007   & 0.0164 $\pm$ 0.0002  \\ \hline
(2,5)  & 1.58E-02 & 1.35E-02 & 0.0165 $\pm$ 0.0007  & 0.0154 $\pm$ 0.0002  \\ \hline
(2,6)  & 1.28E-02 & 1.09E-02 & 0.0126 $\pm$ 0.0006  & 0.0119 $\pm$ 0.0002  \\ \hline
(2,7)  & 8.70E-03 & 7.12E-03 & 0.008 $\pm$ 0.0005   & 0.0084 $\pm$ 0.0001  \\ \hline
(2,8)  & 5.07E-03 & 3.79E-03 & 0.0046 $\pm$ 0.0004  & 0.00354 $\pm$ 9E-05  \\ \hline
(2,9)  & 2.56E-03 & 1.67E-03 & 0.0017 $\pm$ 0.0002  & 0.00191 $\pm$ 7E-05  \\ \hline
(2,10) & 1.14E-03 & 6.20E-04 & 0.0008 $\pm$ 0.0002  & 0.00064 $\pm$ 4E-05  \\ \hline
(2,11) & 4.45E-04 & 1.96E-04 & 0.0005 $\pm$ 0.0001  & 0.00039 $\pm$ 3E-05  \\ \hline
\end{longtable}

\newpage
\begin{figure}
	\includegraphics[scale=0.5]{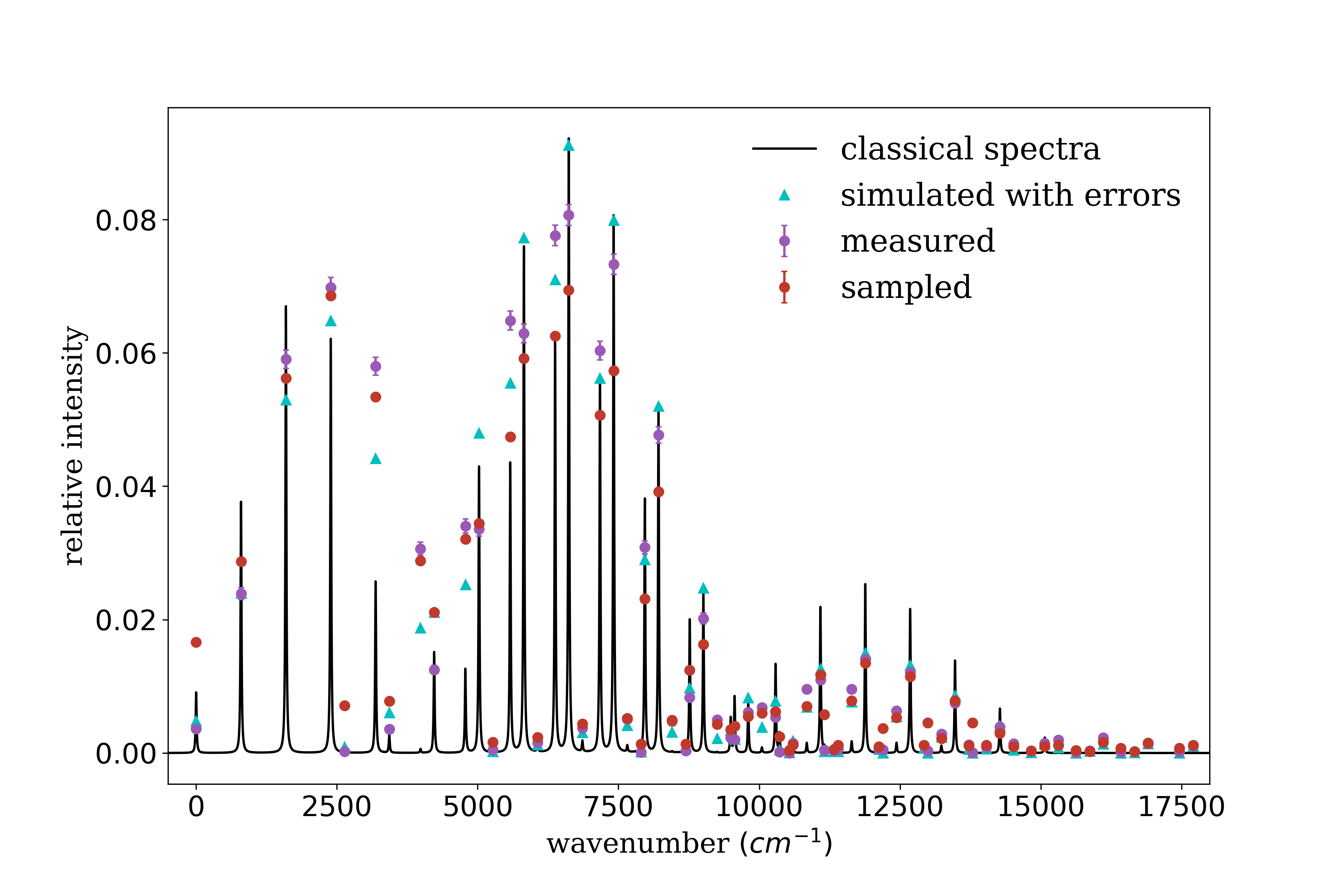}
	\caption{Photoionization of nitrite to nitrogen dioxide starting with one quanta in the symmetric-stretching mode and zero in the bending mode $n$ = 1, $m$ = 0. The more significant errors are primarily due to having a large self-Kerr on cavity A ($\sim$30 kHz) during the beamsplitter operation after starting in a state with higher photon number.}
\end{figure}

\setlength{\tabcolsep}{8pt}
\renewcommand{\arraystretch}{1.5}
\begin{longtable}{|c|c|c|c|c|}
\hline
& \multicolumn{4}{c|}{NO$^-_2 \xrightarrow{h\nu}$ NO$_2$ + e$^-$ starting in ($n$ = 1, $m$ = 0)} \\ \hline
$(n',m')$ & \begin{tabular}[c]{@{}c@{}}Classically \\ calculated \end{tabular} & \begin{tabular}[c]{@{}c@{}}Master equation \\ simulation \end{tabular} & \begin{tabular}[c]{@{}c@{}}Single-bit \\ extraction \end{tabular} & Sampling \\ \hline
(0,0)  & 9.08E-03 & 4.87E-03 & 0.0036 $\pm$ 0.0004 & 0.0166 $\pm$ 0.0002 \\ \hline
(0,1)  & 3.77E-02 & 2.40E-02 & 0.0239 $\pm$ 0.0008 & 0.0287 $\pm$ 0.0003 \\ \hline
(0,2)  & 6.70E-02 & 5.28E-02 & 0.059 $\pm$ 0.001   & 0.0562 $\pm$ 0.0004 \\ \hline
(0,3)  & 6.21E-02 & 6.36E-02 & 0.07 $\pm$ 0.001    & 0.0686 $\pm$ 0.0004 \\ \hline
(0,4)  & 2.57E-02 & 4.12E-02 & 0.058 $\pm$ 0.001   & 0.0534 $\pm$ 0.0004 \\ \hline
(0,5)  & 5.78E-04 & 1.47E-02 & 0.0306 $\pm$ 0.001  & 0.0288 $\pm$ 0.0003 \\ \hline
(0,6)  & 1.25E-02 & 2.20E-02 & 0.034 $\pm$ 0.001   & 0.0321 $\pm$ 0.0003 \\ \hline
(0,7)  & 4.34E-02 & 5.43E-02 & 0.065 $\pm$ 0.001   & 0.0474 $\pm$ 0.0004 \\ \hline
(0,8)  & 6.17E-02 & 7.15E-02 & 0.078 $\pm$ 0.001   & 0.0625 $\pm$ 0.0004 \\ \hline
(0,9)  & 5.65E-02 & 5.71E-02 & 0.06 $\pm$ 0.001    & 0.0507 $\pm$ 0.0004 \\ \hline
(0,10) & 3.80E-02 & 2.95E-02 & 0.0308 $\pm$ 0.001  & 0.0231 $\pm$ 0.0003 \\ \hline
(0,11) & 2.00E-02 & 9.85E-03 & 0.0083 $\pm$ 0.0005 & 0.0124 $\pm$ 0.0002 \\ \hline
(0,12) & 8.41E-03 & 2.11E-03 & 0.002 $\pm$ 0.0003  & 0.004 $\pm$ 0.0001  \\ \hline
(0,13) & 2.87E-03 & 4.62E-04 & 0.0002 $\pm$ 0.0002 & 0.00248 $\pm$ 9E-05 \\ \hline
(0,14) & 8.00E-04 & 2.41E-04 & 0.0004 $\pm$ 0.0002 & 0.0058 $\pm$ 0.0001 \\ \hline
(1,0)  & 1.46E-04 & 9.10E-04 & 0.0002 $\pm$ 0.0002 & 0.0071 $\pm$ 0.0001 \\ \hline
(1,1)  & 2.73E-03 & 6.17E-03 & 0.0036 $\pm$ 0.0004 & 0.0078 $\pm$ 0.0002 \\ \hline
(1,2)  & 1.51E-02 & 2.15E-02 & 0.0125 $\pm$ 0.0006 & 0.0211 $\pm$ 0.0002 \\ \hline
(1,3)  & 4.29E-02 & 4.89E-02 & 0.0336 $\pm$ 0.001  & 0.0344 $\pm$ 0.0003 \\ \hline
(1,4)  & 7.59E-02 & 7.90E-02 & 0.063 $\pm$ 0.001   & 0.0592 $\pm$ 0.0004 \\ \hline
(1,5)  & 9.21E-02 & 9.34E-02 & 0.081 $\pm$ 0.001   & 0.0694 $\pm$ 0.0004 \\ \hline
(1,6)  & 8.05E-02 & 8.20E-02 & 0.073 $\pm$ 0.001   & 0.0573 $\pm$ 0.0004 \\ \hline
(1,7)  & 5.16E-02 & 5.33E-02 & 0.048 $\pm$ 0.001   & 0.0392 $\pm$ 0.0003 \\ \hline
(1,8)  & 2.39E-02 & 2.52E-02 & 0.0202 $\pm$ 0.0008 & 0.0163 $\pm$ 0.0002 \\ \hline
(1,9)  & 7.45E-03 & 8.37E-03 & 0.0061 $\pm$ 0.0005 & 0.0055 $\pm$ 0.0001 \\ \hline
(1,10) & 1.24E-03 & 1.78E-03 & 0.0014 $\pm$ 0.0003 & 0.00124 $\pm$ 6E-05 \\ \hline
(1,11) & 1.95E-05 & 1.99E-04 & 0.0006 $\pm$ 0.0002 & 0.00118 $\pm$ 6E-05 \\ \hline
(1,12) & 8.94E-05 & 1.33E-05 & 0.0005 $\pm$ 0.0002 & 0.0037 $\pm$ 0.0001 \\ \hline
(1,13) & 1.56E-04 & 1.69E-05 & 0.0003 $\pm$ 0.0002 & 0.0046 $\pm$ 0.0001 \\ \hline
(1,14) & 1.08E-04 & 2.40E-05 & 0               & 0.0045 $\pm$ 0.0001 \\ \hline
(2,0)  & 3.83E-04 & 1.99E-04 & 0.0005 $\pm$ 0.0001 & 0.00161 $\pm$ 7E-05 \\ \hline
(2,1)  & 1.23E-03 & 1.20E-03 & 0.0016 $\pm$ 0.0002 & 0.00233 $\pm$ 8E-05 \\ \hline
(2,2)  & 1.61E-03 & 3.06E-03 & 0.0037 $\pm$ 0.0004 & 0.0044 $\pm$ 0.0001 \\ \hline
(2,3)  & 9.74E-04 & 4.06E-03 & 0.0051 $\pm$ 0.0004 & 0.0052 $\pm$ 0.0001 \\ \hline
(2,4)  & 1.41E-04 & 2.92E-03 & 0.0048 $\pm$ 0.0004 & 0.005 $\pm$ 0.0001  \\ \hline
(2,5)  & 1.05E-04 & 1.84E-03 & 0.005 $\pm$ 0.0004  & 0.0043 $\pm$ 0.0001 \\ \hline
(2,6)  & 7.93E-04 & 3.63E-03 & 0.0068 $\pm$ 0.0005 & 0.006 $\pm$ 0.0001  \\ \hline
(2,7)  & 1.49E-03 & 6.90E-03 & 0.0096 $\pm$ 0.0005 & 0.007 $\pm$ 0.0001  \\ \hline
(2,8)  & 1.71E-03 & 7.86E-03 & 0.0096 $\pm$ 0.0005 & 0.0078 $\pm$ 0.0002 \\ \hline
(2,9)  & 1.50E-03 & 5.59E-03 & 0.0063 $\pm$ 0.0004 & 0.0053 $\pm$ 0.0001 \\ \hline
(2,10) & 1.08E-03 & 2.48E-03 & 0.0028 $\pm$ 0.0003 & 0.00219 $\pm$ 8E-05 \\ \hline
(2,11) & 6.70E-04 & 6.10E-04 & 0.0009 $\pm$ 0.0002 & 0.0012 $\pm$ 6E-05  \\ \hline
(2,12) & 3.63E-04 & 5.24E-05 & 0.0003 $\pm$ 0.0001 & 0.00034 $\pm$ 3E-05 \\ \hline
(2,13) & 1.72E-04 & 1.32E-05 & 0.0001 $\pm$ 8E-05  & 0.00037 $\pm$ 3E-05 \\ \hline
(2,14) & 7.12E-05 & 2.40E-05 & 0.00014 $\pm$ 9E-05 & 0.00074 $\pm$ 5E-05 \\ \hline
(3,0)  & 8.70E-05 & 1.50E-04 & 0.00012 $\pm$ 9E-05 & 0.00133 $\pm$ 6E-05 \\ \hline
(3,1)  & 1.12E-03 & 9.95E-04 & 0.0003 $\pm$ 0.0001 & 0.00132 $\pm$ 6E-05 \\ \hline
(3,2)  & 5.23E-03 & 3.46E-03 & 0.0023 $\pm$ 0.0003 & 0.0036 $\pm$ 0.0001 \\ \hline
(3,3)  & 1.33E-02 & 7.92E-03 & 0.0054 $\pm$ 0.0004 & 0.0062 $\pm$ 0.0001 \\ \hline
(3,4)  & 2.19E-02 & 1.29E-02 & 0.011 $\pm$ 0.0006  & 0.0117 $\pm$ 0.0002 \\ \hline
(3,5)  & 2.53E-02 & 1.53E-02 & 0.014 $\pm$ 0.0007  & 0.0135 $\pm$ 0.0002 \\ \hline
(3,6)  & 2.16E-02 & 1.35E-02 & 0.0121 $\pm$ 0.0006 & 0.0115 $\pm$ 0.0002 \\ \hline
(3,7)  & 1.39E-02 & 8.77E-03 & 0.0075 $\pm$ 0.0005 & 0.0078 $\pm$ 0.0002 \\ \hline
(3,8)  & 6.67E-03 & 4.15E-03 & 0.0039 $\pm$ 0.0004 & 0.00296 $\pm$ 9E-05 \\ \hline
(3,9)  & 2.31E-03 & 1.37E-03 & 0.0014 $\pm$ 0.0002 & 0.00103 $\pm$ 6E-05 \\ \hline
(3,10) & 5.13E-04 & 2.86E-04 & 0.0004 $\pm$ 0.0001 & 0.00025 $\pm$ 3E-05 \\ \hline
(3,11) & 4.40E-05 & 3.18E-05 & 0.0002 $\pm$ 0.0001 & 0.00021 $\pm$ 2E-05 \\ \hline
(3,12) & 2.11E-06 & 7.28E-06 & 0.0002 $\pm$ 0.0001 & 0.00071 $\pm$ 5E-05 \\ \hline
(3,13) & 1.79E-05 & 8.77E-06 & 0.00018 $\pm$ 9E-05 & 0.00091 $\pm$ 5E-05 \\ \hline
(3,14) & 1.77E-05 & 7.78E-06 & 0.0001 $\pm$ 8E-05  & 0.00084 $\pm$ 5E-05 \\ \hline
(4,0)  & 9.46E-06 & 3.58E-05 & 0               & 0.0004 $\pm$ 3E-05  \\ \hline
(4,1)  & 8.74E-06 & 2.20E-04 & 0.0004 $\pm$ 0.0001 & 0.00055 $\pm$ 4E-05 \\ \hline
(4,2)  & 2.89E-07 & 5.52E-04 & 0.0006 $\pm$ 0.0002 & 0.00096 $\pm$ 5E-05 \\ \hline
(4,3)  & 3.10E-05 & 7.20E-04 & 0.0011 $\pm$ 0.0002 & 0.00115 $\pm$ 6E-05 \\ \hline
(4,4)  & 8.57E-05 & 5.26E-04 & 0.0012 $\pm$ 0.0002 & 0.00112 $\pm$ 6E-05 \\ \hline
(4,5)  & 9.82E-05 & 3.86E-04 & 0.0014 $\pm$ 0.0002 & 0.00095 $\pm$ 5E-05 \\ \hline
(4,6)  & 6.28E-05 & 7.62E-04 & 0.002 $\pm$ 0.0003  & 0.00118 $\pm$ 6E-05 \\ \hline
(4,7)  & 2.51E-05 & 1.35E-03 & 0.0023 $\pm$ 0.0003 & 0.0016 $\pm$ 7E-05  \\ \hline
(4,8)  & 7.91E-06 & 1.48E-03 & 0.0014 $\pm$ 0.0002 & 0.00149 $\pm$ 7E-05 \\ \hline
(4,9)  & 3.98E-06 & 1.03E-03 & 0.0011 $\pm$ 0.0002 & 0.00117 $\pm$ 6E-05 \\ \hline
(4,10) & 5.12E-06 & 4.58E-04 & 0.0004 $\pm$ 0.0001 & 0.00043 $\pm$ 4E-05 \\ \hline
(4,11) & 8.37E-06 & 1.20E-04 & 0.00015 $\pm$ 8E-05 & 0.00027 $\pm$ 3E-05 \\ \hline
(4,12) & 1.07E-05 & 1.67E-05 & 0               & 0.0001 $\pm$ 2E-05  \\ \hline
(4,13) & 1.00E-05 & 5.98E-06 & 0               & 7E-05 $\pm$ 1E-05   \\ \hline
(4,14) & 7.02E-06 & 5.32E-06 & 0.00015 $\pm$ 7E-05 & 0.00013 $\pm$ 2E-05 \\ \hline
\end{longtable}

\newpage
\begin{figure}
	\includegraphics[scale=0.5]{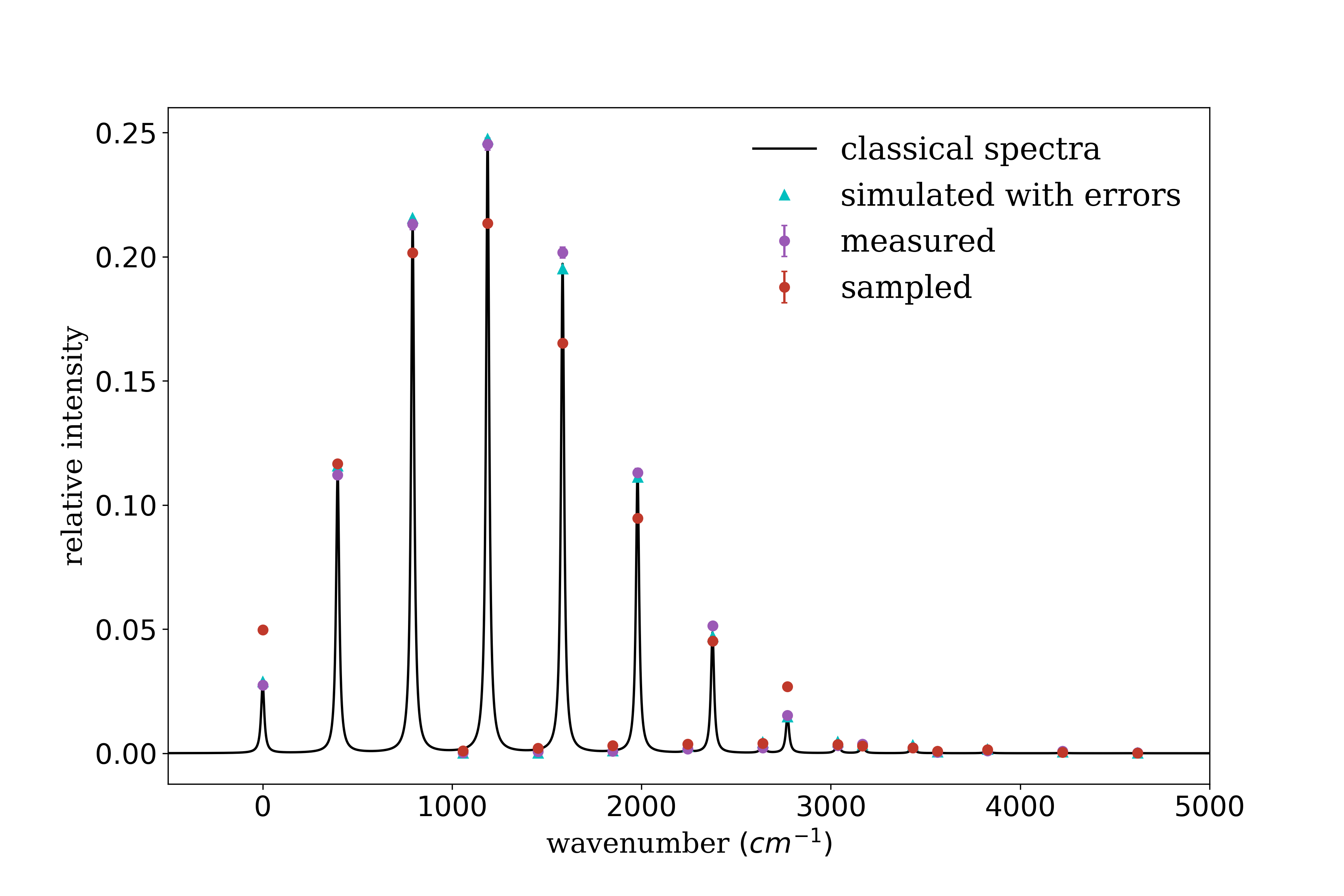}
	\caption{Photoionization of sulfur dioxide to the cation starting in the vibrationless state $n$ = 0, $m$ = 0.}
\end{figure}

\setlength{\tabcolsep}{8pt}
\renewcommand{\arraystretch}{1.5}
\begin{longtable}{|c|c|c|c|c|}
\hline
& \multicolumn{4}{c|}{SO$_2 \xrightarrow{h\nu}$ SO$^+_2$ + e$^-$ starting in ($n$ = 0, $m$ = 0)} \\ \hline
$(n',m')$ & \begin{tabular}[c]{@{}c@{}}Classically \\ calculated \end{tabular} & \begin{tabular}[c]{@{}c@{}}Master equation \\ simulation \end{tabular} & \begin{tabular}[c]{@{}c@{}}Single-bit \\ extraction \end{tabular} & Sampling \\ \hline
(0,0) & 2.82E-02 & 2.88E-02 & 0.0275 $\pm$ 0.0009 & 0.0497 $\pm$ 0.0003  \\ \hline
(0,1) & 1.14E-01 & 1.15E-01 & 0.112 $\pm$ 0.002   & 0.1167 $\pm$ 0.0005  \\ \hline
(0,2) & 2.13E-01 & 2.15E-01 & 0.213 $\pm$ 0.002   & 0.2017 $\pm$ 0.0006  \\ \hline
(0,3) & 2.47E-01 & 2.47E-01 & 0.245 $\pm$ 0.002   & 0.2136 $\pm$ 0.0006  \\ \hline
(0,4) & 1.97E-01 & 1.95E-01 & 0.202 $\pm$ 0.002   & 0.1652 $\pm$ 0.0006  \\ \hline
(0,5) & 1.14E-01 & 1.11E-01 & 0.113 $\pm$ 0.002   & 0.0947 $\pm$ 0.0005  \\ \hline
(0,6) & 4.85E-02 & 4.72E-02 & 0.051 $\pm$ 0.001   & 0.0452 $\pm$ 0.0003  \\ \hline
(0,7) & 1.54E-02 & 1.48E-02 & 0.0152 $\pm$ 0.0007 & 0.0268 $\pm$ 0.0003  \\ \hline
(0,8) & 3.57E-03 & 3.38E-03 & 0.0037 $\pm$ 0.0004 & 0.00302 $\pm$ 9E-05  \\ \hline
(0,9) & 5.68E-04 & 5.28E-04 & 0.0003 $\pm$ 0.0002 & 0.00091 $\pm$ 5E-05  \\ \hline
(1,0) & 9.40E-06 & 2.89E-05 & 0.0003 $\pm$ 0.0001 & 0.00108 $\pm$ 5E-05  \\ \hline
(1,1) & 4.35E-05 & 1.53E-04 & 0.0007 $\pm$ 0.0002 & 0.00206 $\pm$ 7E-05  \\ \hline
(1,2) & 7.70E-04 & 9.82E-04 & 0.0008 $\pm$ 0.0002 & 0.00318 $\pm$ 9E-05  \\ \hline
(1,3) & 2.67E-03 & 2.84E-03 & 0.0017 $\pm$ 0.0003 & 0.00374 $\pm$ 0.0001 \\ \hline
(1,4) & 4.54E-03 & 4.53E-03 & 0.0023 $\pm$ 0.0003 & 0.004 $\pm$ 0.0001   \\ \hline
(1,5) & 4.78E-03 & 4.60E-03 & 0.0031 $\pm$ 0.0003 & 0.0035 $\pm$ 9E-05   \\ \hline
(1,6) & 3.44E-03 & 3.22E-03 & 0.0022 $\pm$ 0.0003 & 0.00219 $\pm$ 7E-05  \\ \hline
(1,7) & 1.76E-03 & 1.62E-03 & 0.0009 $\pm$ 0.0002 & 0.0014 $\pm$ 6E-05   \\ \hline
(1,8) & 6.53E-04 & 5.89E-04 & 0.0008 $\pm$ 0.0002 & 0.00038 $\pm$ 3E-05  \\ \hline
(1,9) & 1.73E-04 & 1.54E-04 & 0.0003 $\pm$ 9E-05  & 0.00011 $\pm$ 2E-05  \\ \hline
\end{longtable}

\newpage
\begin{figure}
	\includegraphics[scale=0.5]{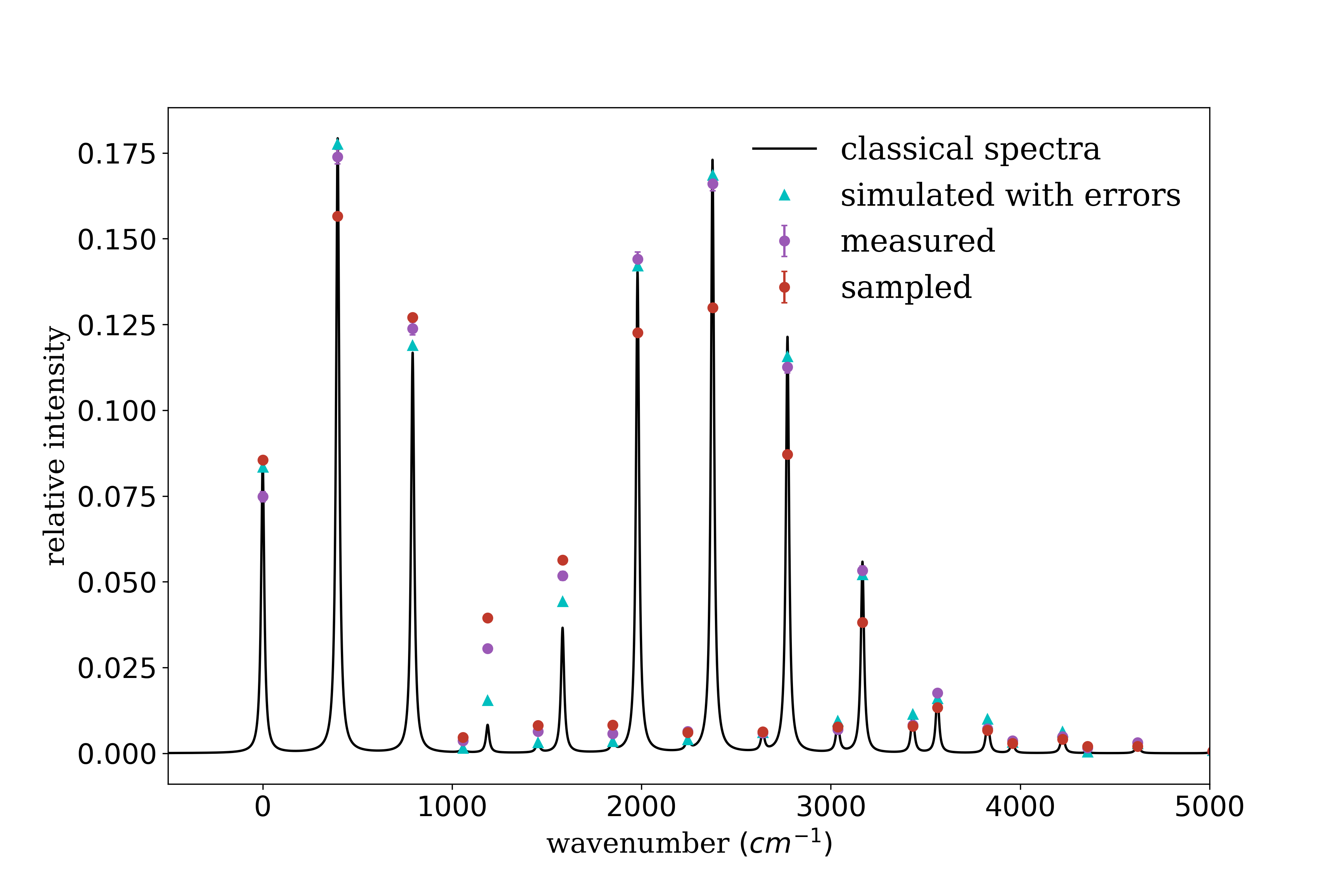}
	\caption{Photoionization of sulfur dioxide to the cation starting with zero quanta in the symmetric-stretching mode and one quantum in the bending mode  $n$ = 0, $m$ = 1.}
\end{figure}

\setlength{\tabcolsep}{8pt}
\renewcommand{\arraystretch}{1.5}
\begin{longtable}{|c|c|c|c|c|}
\hline
& \multicolumn{4}{c|}{SO$_2 \xrightarrow{h\nu}$ SO$^+_2$ + e$^-$ starting in ($n$ = 0, $m$ = 1)} \\ \hline
$(n',m')$ & \begin{tabular}[c]{@{}c@{}}Classically \\ calculated \end{tabular} & \begin{tabular}[c]{@{}c@{}}Master equation \\ simulation \end{tabular} & \begin{tabular}[c]{@{}c@{}}Single-bit \\ extraction \end{tabular} & Sampling \\ \hline
(0,0)  & 8.52E-02 & 8.36E-02 & 0.075 $\pm$ 0.001   & 0.0855 $\pm$ 0.0005 \\ \hline
(0,1)  & 1.79E-01 & 1.78E-01 & 0.174 $\pm$ 0.002   & 0.1566 $\pm$ 0.0006 \\ \hline
(0,2)  & 1.17E-01 & 1.19E-01 & 0.124 $\pm$ 0.002   & 0.1271 $\pm$ 0.0006 \\ \hline
(0,3)  & 8.11E-03 & 1.50E-02 & 0.0306 $\pm$ 0.0009 & 0.0395 $\pm$ 0.0003 \\ \hline
(0,4)  & 3.64E-02 & 4.42E-02 & 0.052 $\pm$ 0.001   & 0.0564 $\pm$ 0.0004 \\ \hline
(0,5)  & 1.40E-01 & 1.42E-01 & 0.144 $\pm$ 0.002   & 0.1226 $\pm$ 0.0006 \\ \hline
(0,6)  & 1.73E-01 & 1.69E-01 & 0.166 $\pm$ 0.002   & 0.1299 $\pm$ 0.0006 \\ \hline
(0,7)  & 1.21E-01 & 1.16E-01 & 0.113 $\pm$ 0.002   & 0.0872 $\pm$ 0.0005 \\ \hline
(0,8)  & 5.56E-02 & 5.22E-02 & 0.053 $\pm$ 0.001   & 0.0382 $\pm$ 0.0003 \\ \hline
(0,9)  & 1.74E-02 & 1.60E-02 & 0.0176 $\pm$ 0.0007 & 0.0133 $\pm$ 0.0002 \\ \hline
(0,10) & 3.62E-03 & 3.24E-03 & 0.0036 $\pm$ 0.0004 & 0.00296 $\pm$ 9E-05 \\ \hline
(0,11) & 4.51E-04 & 3.96E-04 & 0.0016 $\pm$ 0.0003 & 0.00206 $\pm$ 8E-05 \\ \hline
(1,0)  & 1.68E-03 & 1.55E-03 & 0.0037 $\pm$ 0.0003 & 0.0047 $\pm$ 0.0001 \\ \hline
(1,1)  & 3.27E-03 & 3.16E-03 & 0.0064 $\pm$ 0.0005 & 0.0082 $\pm$ 0.0002 \\ \hline
(1,2)  & 3.48E-03 & 3.48E-03 & 0.0058 $\pm$ 0.0004 & 0.0083 $\pm$ 0.0002 \\ \hline
(1,3)  & 3.89E-03 & 3.99E-03 & 0.0064 $\pm$ 0.0004 & 0.006 $\pm$ 0.0001  \\ \hline
(1,4)  & 5.95E-03 & 6.09E-03 & 0.006 $\pm$ 0.0004  & 0.0062 $\pm$ 0.0001 \\ \hline
(1,5)  & 9.44E-03 & 9.38E-03 & 0.007 $\pm$ 0.0005  & 0.0077 $\pm$ 0.0001 \\ \hline
(1,6)  & 1.18E-02 & 1.14E-02 & 0.0083 $\pm$ 0.0005 & 0.008 $\pm$ 0.0001  \\ \hline
(1,7)  & 1.07E-02 & 9.95E-03 & 0.0072 $\pm$ 0.0005 & 0.0067 $\pm$ 0.0001 \\ \hline
(1,8)  & 6.90E-03 & 6.24E-03 & 0.0048 $\pm$ 0.0004 & 0.0041 $\pm$ 0.0001 \\ \hline
(1,9)  & 3.18E-03 & 2.79E-03 & 0.0031 $\pm$ 0.0003 & 0.00203 $\pm$ 8E-05 \\ \hline
(1,10) & 1.03E-03 & 8.79E-04 & 0.0005 $\pm$ 0.0001 & 0.00061 $\pm$ 4E-05 \\ \hline
(1,11) & 2.26E-04 & 1.86E-04 & 4E-05 $\pm$ 6E-05   & 0.00035 $\pm$ 3E-05 \\ \hline
\end{longtable}